**Title:**

The Letter Pi

**Subtitle:**

Bayesian interpretation of p-values, Reproducibility and Considerations for Replication in the Generalized Linear Model

**Authors:**

Christos Argyropoulos[*,‡] Andy P Grieve[¶]

*Department of Internal Medicine, University of New Mexico School of Medicine, MSC10 5550

1 University of New Mexico Albuquerque, NM 87131, USA

¶ Centre for Excellence in Statistical Innovation of UCB Celltech

‡ Corresponding author: cargyropoulos@salud.unm.edu

**Abstract**

Significance testing based on p-values has been implicated in the reproducibility crisis in scientific research, with one of the proposals being to eliminate them in favor of Bayesian analyses. Defenders of the p-values have countered that it is the improper use and errors in interpretation, rather than the p-values themselves that are to blame. Similar exchanges about the role of p-values have occurred with some regularity every 10 to 15 years since their formal introduction in statistical practice. The apparent contradiction between the repeated failures in interpretation and continuous use of p-values suggest that there is an inferential value in the computation of these values. In this work we propose to attach a radical Bayesian interpretation to the number computed and reported as a p-value for the Generalized Linear Model. We show that under the mathematical assumptions that are used to justify non-Bayesian analyses the former will agree with the Bayesian posterior under locally uniform priors. We then propose a decision analytic framework for thresholding posterior tail areas (π-values) which for any given Bayesian analysis will be numerically identical to the non-Bayesian p-values. πvalues are non-controversial, but are relatively uninteresting posterior summaries of treatment effects. A predictive probability argument is then proposed to explore the stochastic variation (replication probability) of p and π-values to illustrate the severe shortcomings of interpreting findings as reproducible or not based on the statistical significance of p-values.

**Table of Contents**

# 1. Introduction

The recent literature has identified p-value significance testing as a culprit for the reproducibility crisis in scientific research(Siegfried 2010; Colquhoun 2014, 2017; Wasserstein et al. 2019; Gibson 2021). Bayesian methods have been proposed(Wagenmakers 2007; Morey et al. 2016a, b) as one remedy that will somehow address these shortcomings of traditional hypothesis testing. We feel that Bayesians should be concerned about the magical thinking that is indiscriminately applied to Bayesian methods as the savior of science from irreproducibility. Furthermore, the generalized criticism against p-values reported for many biomedical relevant models may be misplaced and counterproductive to the goals of a reproducible scientific enterprise unless properly qualified.

In this piece we illustrate several interrelated points that pertain to the role of p-values in the reproducibility crisis. First, we establish the numerical equivalence between non-Bayesian analyses for the commonly used Generalized Linear Models (GLMs) and Bayesian ones under the uniform and other "minimally informative", *locally uniform* priors. Such priors are (among other more mathematically sophisticated choices) used in Objective Bayesian Analysis, which can be viewed as a collection of methodologies for learning from data, when an analysis based on subjective priors is not tenable (Berger 2006) or will be contested by the scientific community. This numerical equivalency allows one to attach a Bayesian interpretation to non-Bayesian analysis of GLMs. Secondly, it is possible to provide a Bayesian justification for the calculation and the thresholding of p-values. These Bayesian analogues (designated as π-values in this text) are numerically identical to p-values and provide the machinery for directional statistical decisions (Kaiser 1960) and inference stated as ordinal claims about direction of an effect (Chow 1988; Harris 1997; Nickerson 2000; Lakens 2021). Third, we propose a general, predictive probability framework for the reproducibility of the findings of a given study. At its most elementary level, that of replicating a study with the exact same characteristics as the original one, this framework leads one to consider the posterior predictive distribution of p and π-values. Fourth, we derive the form of the "reproducibility probability density" (RPD) for p-values from GLMs thus extending previous results for binomial proportions, one and two sample comparisons and multiple parallel group designs (Lambert and Hall 1982; Shao and Chow 2002; De Martini 2008; Boos and Stefanski 2011; De Capitani and De Martini 2016) to GLM regressions. We demonstrate that for these widely used GLMs p and π-values exhibit considerable variability mirroring previous observations(Senn 2002; Boos and Stefanski 2011). Hence, inconsistencies in the results ("significant vs non-significant" from individual studies may be fairly common, stemming not from irreproducibility of the underlying science but rather from the noisy nature of the p and π-value.

The monograph is structured as follows: In Section 2 we provide a self-contained exposition of non-Bayesian (likelihood) based methods for GLMs. In section 3, we present the Bayesian analysis of GLMs under the (locally) uniform prior. We justify the use of this prior for GLMs under three different perspectives: inference for finite effects, non-informativeness for the value of bounded linear functional relationships and as a reference prior for community of investigators, in the absence of a prior elicitation activity. The corresponding posteriors have the exact same functional form as the Maximum Likelihood (ML) estimates for the Generalized Linear Models (GLMs) typically used to analyze biomedical data. This is done under the same conditions of asymptotic convergence of the likelihood function to a multivariate quadratic function, which underlies the ML estimators. At a technical level, we show that various non-Bayesian approaches for GLMs e.g., Extended Quasi-Likelihood (EQL), the modified profile likelihood or method of moment estimators have a Bayesian interpretation as approximate calculations to a full Bayesian analysis. We also establish a more general result, about the asymptotic equivalence of non-quadratic likelihood based approaches to GLMs, i.e. the p* formula of Barndorff-Nielsen (Barndorff-Nielsen 1983) to Bayesian posteriors under locally, rather than exactly, uniform priors. These analytical results establish the numerical equivalency of non-Bayesian and Bayesian computations for GLMs, allowing us to conduct further work from within the Bayesian paradigm.

In Section 4 we propose a utility function for the decision problem of reporting a directional effect for the coefficients in a GLM regression. This utility function is obtained by considering the interactions between the statistician analyst of the data and their scientist client who commissioned the relevant analyses. We show that the outcome of this directional decision problem is not only determined by the π-value but the latter also measures the strength of evidence as it determines the Expected Value of Perfect Information for that decision problem.

In Section 5 , we describe a proper Bayesian framework for the reproducibility of scientific findings. This framework allows the *consumer* (e.g. policy makers, regulatory bodies or even the general public) of scientific discovery to probe the robustness of the inferences from a given study if it were to be replicated from a predictive probability vantage point. A closed form of the RPDs of p (Goodman 1992; Greenwald et al. 1996; Senn 2002; Boos and Stefanski 2011) and π-values is derived in this ;section under the assumption that the likelihood converges to a quadratic polynomial. We also illustrate a straightforward Monte Carlo approach to derive the RPD of the p and π-values when the quadratic approximation of the log-likelihood is to be questioned. The mathematical investigation of the properties of the relevant probability densities allows us to criticize the use of replicate p and π-values to gauge reproducibility of scientific discovery. This criticism is grounded less in philosophy and more in the technical issue of the large variability of the p and π-values. In Section 6 we use the analytical machinery developed in the previous sections to undertake an investigation of the reproducibility of the kidney specific outcomes of Sodium Glucose Cotransporter 2 Inhibitors (SGLT2i), an emerging class of cardiometabolic drugs with broad kidney benefits(Johansen and Argyropoulos 2020), but also a unique profile of side effects. We use these trials to illustrate issues of reproducibility in clinical trials, the role of the priors and the limits of the approximations for GLM regressions. In the last section we attempt a synthesis of the material of sections 2-6, by proposing a hybrid Bayesian/likelihood approach to presenting and interpreting the results of scientific investigation that considers reproducibility.

Programming of the relevant calculations in the R language is illustrated in the Appendix.

# 2. Likelihood methods for GLMs

GLMs are widely applicable methods for the analysis of continuous (linear regression), binomial (logistic regression), discrete (multinomial regression), rate or time to event (Poisson regression) outcomes. These techniques are widely used to analyze data in the fields of the social sciences and medicine that are at the forefront of the current reproducibility crisis. Regression methods for GLMs provide *effect measures*, i.e. quantitative measures of association or treatment effects in observational and randomized studies. Our intention in this section is provide a succinct reference to likelihood-based approaches to GLMs to ground the Bayesian analyses that follow, referring the interested reader to the following textbooks (McCullagh and Nelder 1989; Lindsey 1997; Pawitan 2013; Wood 2017; Lee et al. 2017; Dobson and Barnett 2018) for a more extensive treatment of GLMs.

## 2.1. Estimation of treatment effects for GLMs

In essence GLM implements a hierarchical estimation problem which can be solved in stages as follows:

- ➢ Stage 1 : relating the observables, $y$, to parameters $\boldsymbol{\theta} = \theta_i, \ i = 1, \dots, n$
- ➢ Stage 2 : specifying the nature of the relationship existing among the parameters $\theta_i$ (sampling model with correlations if present) and any nuisance parameters ($\phi$)
- ➢ Stage 3 : incorporating further information, if any, about the general form of the relationship (link function – to be defined below), an additive regression structure and a reparameterization by projecting to a space of parameters ($\boldsymbol{\beta}$) which has fewer dimensions than the space of the observables and the parameters $\boldsymbol{\theta}$

In stages 1 and 2, we will assume that all data collected in a scientific investigation come from a distribution that belongs to the exponential family with the following density representation:

$$f(y_i|\theta,\phi) = \exp\left\{\frac{y_i\theta_i - b(\theta_i)}{\phi a_i^{-1}} + c(y_i,\phi)\right\}$$ Eq. 1.

The parameters $\theta_i, \phi$ are location and scale parameters which determine the mean and the variance of the response variable: $\mu_i = b'(\theta_i)$ and $\sigma_i^2 = b''(\theta_i)\phi a_i^{-1}$, while the *variance function* of the GLM is $V(\mu_i) = V_i = b''(\theta_i)$. The weights $a_i$ are assumed to be known and are taken equal to one for the remainder without sacrificing generality). Stage 2 is completed by specifying a sampling model that relates the individual parameters to each other indirectly through the correlation of observations. For independent, identically distributed data (i.i.d) the contribution of the individual observations to the *likelihood function* (L) of the problem is multiplicative. The likelihood is a mathematical function of the parameters of the problem and the data, $y = \{y_1, y_2, \ldots, y_n\}$ and for the i.i.d case assumes the form:

$$L(\theta,\phi; y|X) = \prod_{i=1}^{n} f(y_i|\theta,\phi) = \prod_{i=1}^{n} \exp\left\{\frac{y_i\theta_i - b(\theta_i)}{\phi a_i^{-1}} + c(y_i,\phi)\right\}$$

The GLM is completed at stage 3 by the introduction of one-to-one continuous differentiable transformations in the mean of the response which may assumed to follow a linear model for covariates:

$$\eta_i = g(\mu_i) = x_i^T \beta, \; i = 1,2,\ldots,n, \; \dim(\beta) = p, \; \theta_i = (b')^{-1}(g^{-1}(\eta_i))$$

The function $g(\cdot)$ is the link function of the GLM, and when $\theta_i = \eta_i$, i.e. $g(\cdot) = b^{-1}(\cdot)$, the link is called *canonical*.

For an i.i.d sample the log-likelihood (*ll*) function (the logarithm of the likelihood) viewed as a function of $\beta$, rather than $\theta$, is given by:

$$ll(\beta,\phi; y|X) = \sum_{i=1}^{n}\left\{\frac{y_i\theta_i - b(\theta_i)}{\phi} + c(y_i,\phi)\right\}$$ Eq. 2.

The estimate of the parameter vector $\beta$ is found by solving the score equations:

$$s(\beta_j) = \frac{\partial l}{\partial \beta_j} = 0 = \sum_{i=1}^{n} \frac{y_i - \mu_i}{\phi V(\mu_i)} \times \frac{x_{ij}}{g'(\mu_i)} = \sum_{i=1}^{n} W_i(y_i - \mu_i)x_{ij}G_i,$$

$$W_i = \frac{w_i(\mu_i)}{V(\mu_i)\,g'(\mu_i)^2}, \quad w_i(\mu_i) = 1 + (y_i - \mu_i)\{V'(\mu_i)/V(\mu_i) + g''(\mu_i)/g'(\mu_i)\}, \quad G_i = g'(\mu_i),$$

For the canonical link function, $V(\mu_i) = 1/g'(\mu_i)$, $V''(\mu_i) = -g''(\mu_i)/g'(\mu_i)^2$ and $w_i = 1$, $W_i = V(\mu_i)$, leading to considerable analytical simplification of the score equations. From a statistical perspective opting for the canonical link makes the *observed Fisher information matrix* (Hessian of the log-likelihood) equal to its expectation taken over the distribution $p(y|\beta,\phi)$. The Iteratively Reweighted Least Squares algorithm is used to solve the nonlinear score equations $\mathbf{X}^T \mathbf{W} \mathbf{G}(\mathbf{y} - \boldsymbol{\mu})$, with $\mathbf{W} = diag\{W_i\}$ and $\mathbf{G} = diag\{G_i\}$ to obtain the ML parameter estimates and their variance. Note that the scale parameter cancels out of the score equations when solving for the maximum of the likelihood over the parameters $\boldsymbol{\beta}$. If we denote by $\widehat{\boldsymbol{\beta}}$ the estimates of the IWLS at convergence, $\mathbf{X}$ the design matrix for the regression model (constructed by "stacking" column wise the row vectors $x_i^T$), $\mathbf{W}^k = diag\{W_i^k\}$, the value of the weight matrix at the kth (convergent) step of the IWLS, $I(\widehat{\boldsymbol{\beta}})$ the observed Fisher observation matrix, we obtain the multivariate normal (MVN) as the limiting distribution of the ML estimates:

$$\widehat{\boldsymbol{\beta}}|\mathbf{X},\boldsymbol{\beta},\phi \to MVN\left(\boldsymbol{\beta}, I(\widehat{\boldsymbol{\beta}})^{-1}\right) = MVN(\boldsymbol{\beta}, \boldsymbol{\Sigma}), \quad \boldsymbol{\Sigma} = \phi(\mathbf{X}^T \mathbf{W}^k \mathbf{X})^{-1} \quad \text{Eq. 3.}$$

To reduce clutter, we will write $\mathbf{W}_{\widehat{\boldsymbol{\beta}}}$ rather than $\mathbf{W}^k$, a notation which makes clear that the matrix of weights, and thus the observed information matrix is a function of $\widehat{\boldsymbol{\beta}}$. More generally, one can define the matrix function, $\mathbf{W}(\boldsymbol{\beta}) = \mathbf{W}_{\boldsymbol{\beta}} = diag\left\{1/V(\mu_i(\boldsymbol{\beta}))\,g'(\mu_i(\boldsymbol{\beta}))^2\right\}$ to express the *expected* Fisher information matrix associated with the log-likelihood at an arbitrary point $\boldsymbol{\beta}$ as:

$$\mathcal{I}(\boldsymbol{\beta}) \equiv \left(E\left[-\frac{\partial^2 ll(\boldsymbol{\beta},\phi|\mathbf{y})}{\partial \beta_j \partial \beta_l}\right]\right)_{j,k} = \phi^{-1}(\mathbf{X}^T \mathbf{W}_{\boldsymbol{\beta}} \mathbf{X}) \quad \text{Eq. 4.}$$

In the case of the canonical link, $\mathbf{W}_\beta = diag\{V(\mu_i(\boldsymbol{\beta}))\} = diag\{1/g'(\mu_i(\boldsymbol{\beta}))\} = diag\{b''(\theta_i)\}$, while for non-canonical links, $\mathbf{W}_\beta = diag\{b''(\theta_i)\}\, diag\{(d\theta_i/d\eta_i)^2\}$ contains an adjustment for the link function. The covariance matrix will retain the general form given in the right hand side of Eq.4 even if the observed, rather than the expected Fisher information, is used to form standard deviation under a suitable redefinition of the values of the weight matrix $\mathbf{W}_\beta$ (Wood 2017).

## 2.2. Estimation of the scale parameter

Several non-Bayesian calculations are available in the literature, if the value of the dispersion (nuisance) parameter is known as in logistic or Poisson regression. Scale parameters provide the simplest example of *nuisance* parameters and there have been numerous non-Bayesian approaches for the calculation of estimators. We will describe them below, before illustrating how these calculations can be derived as (modal) approximations to a Bayesian posterior in Section 3. The method of moments estimate for the scale is given by the equation:

$$\hat{\phi}_{MOM} = \frac{1}{n-p} \sum_{i=1}^{n} \frac{(y_i - \hat{\mu}_i)^2}{\hat{V}_i} \qquad \text{Eq. 5.}$$

where $\hat{\mu}_i, \hat{V}_i = V(\hat{\mu}_i)$ are the of the mean and the variance function of the GLM for the ith observation, with the values of the parameters set to their ML estimates. This estimator is identical to the *pseudolikelihood* approach estimator which assumes that the Pearson residuals, $r_{PL,i} = (y_i - \hat{\mu}_i)\sqrt{1/\hat{V}_i}$ are normally distributed.

Alternatively, one may use the GLM *deviance* in the Extended Quasi-Likelihood (EQL) approach (Nelder and Pregibon 1987) to estimate the scale parameter. The deviance is the (scaled) difference between the log-likelihood of a reduced the model under consideration and a saturated one, in which there is a single parameter $\theta_i$ for each observation $y_i$. If we denote by $\widetilde{\boldsymbol{\theta}} = \boldsymbol{\theta}(\mathbf{y}(\widetilde{\boldsymbol{\beta}}))$ and $\widehat{\boldsymbol{\theta}} = \boldsymbol{\theta}(\widehat{\boldsymbol{\mu}}(\widehat{\boldsymbol{\beta}}))$ the estimates of the canonical parameters under the saturated and the reduced models, and $\widetilde{\boldsymbol{\beta}}$ and $\widehat{\boldsymbol{\beta}}$ the corresponding regression parameters, the deviance is given by:

$$\hat{\phi}_{EQL} = \frac{2\phi\{ll(\widetilde{\boldsymbol{\beta}},\phi;\boldsymbol{y}|\mathbf{X}) - ll(\widehat{\boldsymbol{\beta}},\phi;\boldsymbol{y}|\mathbf{X})\}}{n} = \frac{D(\boldsymbol{y};\widehat{\boldsymbol{\mu}})}{n} = \phi \sum_{i=1}^{n} \frac{d(y_i,\hat{\mu}_i)}{n} \qquad \text{Eq. 6.}$$

$$d(y_i,\mu_i) = 2\{y_i(\widetilde{\theta}_i - \theta_i) - b(\widetilde{\theta}_i) + b(\theta_i)\} \qquad \text{Eq. 7.}$$

The quantity $d(y_i,\hat{\mu}_i)$, which is equal to the squared scaled deviance residual, is obtained by evaluating Eg.7 at the ML estimate, $\hat{\theta}_i$. The EQL approach is based on a saddlepoint approximation to the GLM family, Eq. 1. The starting point for this approximation is the rewriting of the GLM family as an exponential dispersion model (Jorgensen 1987, 1997):

$$f(y_i) = h(y_i,\phi) \exp\left\{-\frac{1}{2\phi} d(y_i,\mu_i)\right\} \qquad \text{Eq. 8.}$$

$$h(y_i,\phi) = \exp(c(y_i,\phi) + 2\phi\{y_i\widetilde{\theta}_i - b(\widetilde{\theta}_i)\})$$

As $\phi \to 0$ (small dispersion case), $h(y_i,\phi) \approx \{2\pi\phi V(y_i)\}^{-1/2}$ in the saddlepoint approximation (the $V(y_i)$ is the variance function of the GLM evaluated at $y_i$). The saddlepoint approximation may be rewritten in the GLM form as:

$$f(y_i) = \frac{\exp\left\{\frac{y_i\theta_i - b(\theta_i)}{\phi} - \frac{y_i\widetilde{\theta}_i - b(\widetilde{\theta}_i)}{\phi}\right\}}{\{2\pi\phi V(y_i)\}^{1/2}} \qquad \text{Eq. 9.}$$

This approximation is exact for the normal and the inverse-Gaussian distribution. The quality of this approximation is surprisingly very good for other distributions e.g. with relative error less than 3% when the number of Poisson counts is greater than 3, or the number of binomial successes and failures greater than 3 (Smyth and Verbyla 1999). For the Poisson and the gamma GLMs, the saddlepoint approximation consists of replacing the factorial/Gamma function appearing in the definition of the probability mass function of these models with their Stirling formula approximation. More generally, this approximation will approximately hold true if the mean of the response variable, is more than three standard deviations away from the boundary of the distribution. Informally, the EQL approximation decomposes the data into two components: "sufficient statistics" ($\widehat{\boldsymbol{\beta}}$) and "measures of model goodness-of-fit" ($D(\boldsymbol{y};\widehat{\boldsymbol{\mu}})$). It also orthogonalizes the parameters of the model, thus considerably simplifying both exposition but also the numerical calculations needed to fit the model and estimate the parameters.

A deviance estimate of the scale may also be computed for the saddlepoint approximation as: $\hat{\phi}_{dev} \approx D(\boldsymbol{y};\widehat{\boldsymbol{\mu}})/(n-p)$ (Jorgensen 1987; Jørgensen 1992). The final estimate for the scale parameter that has been proposed in the literature is based on the modified profile likelihood (Barndorff-Nielsen 1983; Jorgensen 1987):

$$\hat{\phi}_{mPL} = \max_{\phi} \left(\frac{p}{2}\log(\phi) + \max_{\boldsymbol{\beta}|\phi} ll(\boldsymbol{\beta},\phi;\boldsymbol{y}|\mathbf{X})\right) \qquad \text{Eq. 10.}$$

The relationship among these different estimators has been discussed previously (Jorgensen 1987; Jørgensen 1987, 1992): the $\hat{\phi}_{mPL}$ is equivalent to $\hat{\phi}_{dev}$ and similarly the $\hat{\phi}_{MOM}$ is equivalent to $\hat{\phi}_{dev}$ under the saddlepoint approximation. It has been suggested that the $\hat{\phi}_{mPL}$ is preferred to the moment estimator if numerical robustness is not a consideration (Jørgensen 1992). The deviance based estimator plays a key role for the EQL approaches to GLMs (Lee et al. 2017), yet its consistency as the dispersion parameters tends to zero or when the sample size tends to infinity is less clear(Jorgensen 1987; Jørgensen 1987, 1992).

### 2.3. P-values and Null Hypothesis Significance Testing in GLMs

Non-Bayesian analyses typically report tail area probabilities ("p-values") for testing parameter estimates against specific values in Null Hypothesis Significance Testing (NHST). The p-value is the probability of obtaining a result at least as extreme as the one obtained under the null hypothesis $H_0$. A small p-value, indicates that the data are unlikely under the null hypothesis, and thus may be taken as evidence against the null hypothesis(Senn 2002; Berger 2003). Such area probabilities are also subjected to a threshold type of decision analysis, and p-values that are smaller than the threshold of statistical significance (conventionally taken to be equal to $a = 0.05$) lead to the *rejection* of $H_0$. P – values are calculated by computing a test statistic, a random variable that is a function of the observed data observed, and then reading off the corresponding probability from the known cumulative distribution function of the statistic. In the NHST, decisions about $H_0$ are made by comparing the test statistic against $c_a$ the critical statistic value corresponding to the threshold of statistical significance.

In the context of GLMs $H_0$ states that the true value of a subset of the parameters,$\boldsymbol{\beta}$, is equal to some prespecified value. For example one may specify that all the elements of $\boldsymbol{\beta}$, except a single element $\beta_i$ is equal to $\beta_{i,0}$. The latter is usually taken to be equal to zero to signify the lack of a treatment effect upon the linear predictor and thus the response variable of the GLM. There are two major approaches in calculating test statistics for GLMs: the *likelihood ratio test* (*LRT*) which is the twice the logarithm of the likelihood under the alternative and the null hypotheses and the Wald test. These statistics are identical if the log-likelihood is exactly quadratic: in such a case the likelihood ratio test is based upon the "vertical" distance between the likelihood at the parameter value implied by the null v.s. the one at the ML, while the Wald on the distance between the two parameter values(Buse 1982; Engle 1984). The two tests are asymptotically equivalent for regular log-likelihoods that are well approximated by quadratic functions (one of the key implicit assumptions of likelihood methods of GLMs)(Hayakawa 1975; Buse 1982; Cordeiro 1983, 2004; Engle 1984; Vargas et al. 2014). Given the close relationship between the two tests, we will thus restrict attention to the simpler (two sided) Wald test. The p-value and the rejection decision for the Wald test for a single parameter may be written as:

$$p = 2\Phi(-|z_i|), reject\ H_0\ iff\ |z_i| > \Phi\left(1 - \frac{a}{2}\right), z_i = \frac{\hat{\beta}_i - \beta_{i,0}}{se(\hat{\beta}_i)} \qquad \text{Eq. 11.}$$

In this expression, $\Phi$ the cumulative density of the standard normal distribution and $z_i$ is the value of the square root of the Wald statistic. When the dispersion estimate is to be estimated from the data the underlying distribution of the $z_i$ is the Student t rather than the normal one. The standard error of $\hat{\beta}_i$, $se(\hat{\beta}_i) = \sqrt{\Sigma[i,i]}$, is the square root of the ith diagonal element in the covariance matrix $\Sigma = \phi(\mathbf{X}^T\mathbf{W}_{\hat{\beta}}\mathbf{X})^{-1}$. There are several observations that are worth noting about the univariate Wald test: First, the directionality of the effect, i.e. the sign of the difference $\hat{\beta}_i - \beta_{i,0}$ is effectively lost during the calculation of the p-value. Furthermore, even though the test itself concerns a precise null hypothesis, its value can be computed as a function of the probabilities of two complimentary directional events:

$$p = 2 \times min(\{P(\hat{\beta}_\iota \geq \beta_i | \beta_i = \beta_{i,0}), P(\hat{\beta}_\iota < \beta_i | \beta_i = \beta_{i,0})\}) \qquad \text{Eq. 12.}$$

Finally, rejection of $H_0$ follows if the probability of the *least likely* directional event is less than $a/2$. Equivalently, one would reject $H_0$ if the probability of the most likely directional event is greater than $1 - a/2$. In this very specific sense, one may interpret p-values as providing evidence for a *directional* covariate effect since the rejection of the null hypothesis is ultimately driven by the probabilities of directional effects. Even though this analysis of the p-value calculation would seem complicated, it provides an insight into what takes place, "under-the-hood" when this test is being reported.

# 3. Bayesian methods for GLMs

In the Bayesian analysis of GLMs one must specify priors for the unknown regression and dispersion parameter (if not known). In this section we will argue that the default prior for the Bayesian analysis of the generalized linear model should be a possibly improper distribution that is constant in a hypercube region over the parameter vector $\boldsymbol{\beta}$ and inversely related to the scale parameter $\phi$. Priors for $\boldsymbol{\beta}$ and the scale parameter are thus independent, a choice that can be justified because these parameters independently determine the location of the scale of the response (Box and Tiao 1973). While this proposal differs from the joint, noninformative prior(Jeffreys 1946), it pragmatically recognizes that the Jeffrey's prior for $\boldsymbol{\beta}, \phi$ is "quite cumbersome to work with analytically and may not be practical for numerical computations"(Ibrahim and Laud 1991), but a prior which factorizes $p(\boldsymbol{\beta}, \phi) = p(\boldsymbol{\beta}) \times p(\phi)$ may be quite tractable. To the extent that informative priors are then chosen separately for the parameters determining location ($\boldsymbol{\beta}$) and those affecting scale ($\phi$), we recover Jeffreys' modified proposal for an objective, minimally informative prior(Jeffreys 1946; Box and Tiao 1973). It is under this specification for $p(\boldsymbol{\beta})$ that one may obtain a numerical equivalence between Bayesian and frequentist analyses, with the various variance parameter estimators arising from different choices for the prior over $\phi$. We will thus consider this reference prior in some detail in the following sections.

## 3.1. Uniform, bounded priors for treatment effects

Our particular choice of the prior over $\boldsymbol{\beta}$ may be justified from three different perspectives: the finite world argument, as one inducing a uniform prior for functional relationships (Bornkamp 2012) and a suitable extension of the dominant likelihood/locally uniform prior approach (Box and Tiao 1973) for a community of investigators.

### 3.1.1. A finite world argument

GLMs are often used to analyze properties about real-world phenomena. This context of use immediately brings out a contradiction between the space of the statistical model (possibly infinite) and the finite nature of measurements obtained by sampling the physical world. As GLM parameters may be put in direct correspondence to the actual measurement, one can resolve this apparent contradiction by restricting the admissible range of each of the GLM parameters to a finite interval. The length of this interval may be derived by considering the finite range of the measurement/aspect of reality that is modelled as a response variable of the GLM. By considering the range of these quantities, one may derive useful bounds for the range of parameters which in turn may inform the selection of a suitable "non-informative" prior. To illustrate this procedure, consider a single outcome variable $y_1$ bounded in the interval $[y_1^{min}, y_1^{max}]$ and a single parameter $\beta_1$, corresponding to the intercept term of the GLM. Then:

$$y_1^{min} \leq y_1 \leq y_1^{max} \Rightarrow \beta_1^{min} = g^{-1}(y_1^{min}) \leq \beta_1 \leq g^{-1}(y_1^{max}) = \beta_1^{max}$$

Restrictions on the range of other parameters for the GLM may be similarly derived by considering the finite range, $[X_i^{min}, X_i^{max}]$, of each of the variables $X_i$ in the design matrix for the GLM and solving the inequalities:

$$\beta_i^{min} = \frac{g^{-1}(y_1^{min})}{X_1^{min}} \leq \beta_i \leq \frac{g^{-1}(y_1^{max})}{X_1^{max}}$$

Having bounded the range of each parameter, in the case it was the only one appearing in the design matrix of the model, one then introduces the uniform prior:

$$p(\beta_1, \beta_2, \ldots, \beta_p) = p^{-1} \prod_{i=1}^{p} p(\beta_i) = p^{-1} \prod_{i=1}^{p} \frac{[\![\beta_i \in [\beta_i^{min}, \beta_i^{max}]]\!]}{\beta_i^{max} - \beta_i^{min}}$$

Scaling by the number of parameters in the design matrix ensures that the linear predictor of the GLM, which is a function of all parameters will always be bounded in the interval $\left[g^{-1}(y_1^{min}), g^{-1}(y_1^{max})\right]$. Such scaling also avoids internal inconsistencies e.g. violations of the bounds of the interval $[y_1^{min}, y_1^{max}]$ if the variables are considered only one at a time. In particular, there is no guarantee that $\sum_i X_i^{min} \beta_i^{min} \geq y^{min} \wedge \sum_i X_i^{max} \beta_i^{max} \leq y^{max}$, and in fact both statements will nearly always be simultaneously false, unless the prior is scaled by the number of parameters. The Iverson bracket notation $[\![\beta_i \in [\beta_i^{min}, \beta_i^{max}]]\!]$, signifies that the prior assumes the value of zero if any of the parameters are outside their bounds, and restricts the prior to be uniform in the hypercube. Though the restriction of the prior in the hypercube may seem to reduce the prior's generality, one should notice that this restriction is implicit in any numerical calculation due to the finite precision of numerical computations. In fact, one could simply invoke this argument to bound priors to the range of numbers that can be represented by whatever arithmetic library and floating point precision is used for calculations.

### 3.1.2. Functional uniform priors for GLMs

We can also arrive at the same uniform (over the hypercube) prior from a different perspective, namely that of seeking a uniform prior over the space of the functional shapes estimated through GLM regressions (Ghosal et al. 1997; Bornkamp 2012). Such priors are constructed by mapping the parameter $\boldsymbol{\beta}$ from a compact, subspace $\boldsymbol{B} \in \mathbb{R}^p$ to another compact metric space $(M, d)$ equipped with a metric $d$ by using a differentiable map $\varphi: \boldsymbol{B} \rightarrow M$. The metric $d(\varphi, \varphi_0) = d(\varphi(\boldsymbol{\beta}), \varphi(\boldsymbol{\beta_0})) = d^*(\boldsymbol{\beta}, \boldsymbol{\beta_0})$ is assumed to be sufficiently well behaved to be locally approximated via the Mahalanobis distance, i.e. :

$$d^*(\boldsymbol{\beta}, \boldsymbol{\beta_0}) = c_1\sqrt{c_2(\boldsymbol{\beta} - \boldsymbol{\beta_0})^T \boldsymbol{U}(\boldsymbol{\beta_0})(\boldsymbol{\beta} - \boldsymbol{\beta_0}) + O(\|\boldsymbol{\beta} - \boldsymbol{\beta_0}\|^k)} \qquad \text{Eq. 13.}$$

where $\boldsymbol{U}(\boldsymbol{\beta_0})$ a symmetric matrix with strictly positive eigenvalues, $k \geq 3$ and $c_1, c_2$ positive constants. Imposing a uniform distribution in $(M, d)$, induces a potentially non-uniform distribution on $\boldsymbol{B}$:

$$p(\boldsymbol{\beta}) = \frac{\sqrt{|U(\boldsymbol{\beta})|}}{\int_B \sqrt{|U(\boldsymbol{\beta})|}d\boldsymbol{\beta}}, \quad |U(\boldsymbol{\beta})| \equiv \det U(\boldsymbol{\beta})$$

Eq. 14.

For GLMs, we may ask for a *uniform* distribution in the scale of the linear predictor for a single observation, i.e., the map $\varphi$ is $\eta = x^T\boldsymbol{\beta}$ under the Euclidean distance metric $d^*(\boldsymbol{\beta}, \boldsymbol{\beta}_0) = \sqrt{(x^T\boldsymbol{\beta} - x^T\boldsymbol{\beta}_0)^T(x^T\boldsymbol{\beta} - x^T\boldsymbol{\beta}_0)}$. These choices lead to a constant $U(\boldsymbol{\beta}_0) = xx^T$ and thus a constant prior over the hypercube with arbitrary yet bound finite bounds. Note that the prior $\propto \sqrt{\det xx^T}$ though closely related to the Jeffrey's prior for GLM (Ibrahim and Laud 1991), only coincides with the latter for the Gaussian case. In fact, the Jeffrey's prior itself can be seen as a functional uniform prior on the space of the densities of the residuals of the statistical model, under the Hellinger or the Kullback-Leibler distance(Balasubramanian 1997; Ghosal et al. 1997; Bornkamp 2012). Extensions to the entire $\mathbb{R}^p$ are possible by taking limits of a sequence of growing compact spaces (Ghosal et al. 1997), yet the resulting prior density may not be integrable, since the convergence of the integral in the denominator is not guaranteed.

### 3.1.3. Likelihood dominated priors for communities of investigators.

A likelihood dominated prior, is one in which the likelihood of the data is more concentrated ("peaked") than the prior, which essentially changes very little ("flat") over the region the likelihood is appreciable and does not assume large values outside that region. This scenario corresponds to the "stable estimation" principle for scientific inference (Edwards et al. 1963).There are two main justifications(Box and Tiao 1973) for adopting such priors. First, scientific investigations are usually not undertaken unless they are likely to increase our knowledge relative to what is known prior to the studies. Hence, the prior adopted for such scenarios should have as small as possible of an impact on the final inference. Second, scientific discourse is never conducted in isolation: the scientist will use the data from the experiments to convince their colleagues about the state of the world. The latter will often hold different opinions from the scientist conducting the investigation. Such scattering of opinion may be so extensive that an informative prior directly elicited from the investigators will be diffusely spread out over the potential range of observed data.   How does one go about conducting Bayesian inference in these two scenarios? A straightforward answer is to choose a default *reference* prior, which represents the opinion of someone who a priori knows very little about the state of the world. This individual's posterior will be dominated by the likelihood ("data"), not the prior, and thus it is independent of what anyone may have thought before the data were available. Such priors are indeed necessary, because in certain situation (e.g. regulatory approval of drugs or devices), "the appearance of objectivity is often required"(Berger 2006). Outside regulatory contexts, scientists often call upon statistics to provide an objective validation for their findings, and in such situations Objective Bayesian priors achieves "readily understandable communication of the information in the observed data, as communicated through a statistical model", while simultaneously ensuring that the answers for *the particular problem* are conditional on the data at hand and "respecting the frequentist notion that the methodology must ensure success in repeated usage by scientists" (Berger 2006).

A *locally* uniform prior, i.e., one that is constant, or nearly so (Edwards et al. 1963)), over the range of the likelihood is appreciable, could serve as the basis for a default prior. Furthermore, if this prior does not strongly favor some other region of the likelihood, it satisfies the criteria for a prior dominated by the likelihood. The uniform prior over the hypercube fits both criteria, if the boundaries are set wide enough to not "cut-off" areas in which the likelihood is appreciable.  While these considerations, going back to the 1960s, suggest the use of the uniform prior as a reference choice for the community of investigators, the mathematical argument that leads to the uniform prior is not transparent. Such transparency can be gained by considering formal mathematical representations of the diversity of opinions held by scientists ("prior elicitation").

### 3.1.3.1. Prior elicitation and the locally uniform prior.

Prior elicitation for GLMs has been explored in numerous publications (Bedrick et al. 1996; Denham and Mengersen 2007; Garthwaite et al. 2013; Hosack et al. 2017) since the early work on the linear regression model(Kadane et al. 1980). In these *predictive* approaches, the experts are queried about their beliefs regarding a potential observable (e.g., the conditional mean of the response variable) for different design points. The opinions of the experts are then analyzed by a suitable chosen regression model e.g. (un)weighted least squares(Kadane et al. 1980; Denham and Mengersen 2007; Hosack et al. 2017) or even GLM regression(Bedrick et al. 1996). Such an approach bypasses the much more difficult cognitive task of the *structural* method in which one elicits beliefs about the values of the regression parameters $\beta_i$ themselves(Kadane and Wolfson 1998). Different procedures for inducing priors from an expert often end up expressing them as multivariate Gaussians (Racine et al. 1986; Grieve 1988; Bedrick et al. 1996; Denham and Mengersen 2007; Garthwaite et al. 2013; Hosack et al. 2017). Hence if one were to elicit a prior from the community, the prior would assume the form:

$$\boldsymbol{\beta}|\boldsymbol{\beta}_{prior}, \boldsymbol{\Sigma}_{prior} \sim MVN(\boldsymbol{\beta}_{prior}, \boldsymbol{\Sigma}_{prior}), \qquad \boldsymbol{\Sigma}_{prior} = \boldsymbol{\sigma}_{prior} \boldsymbol{C}_{prior} \boldsymbol{\sigma}_{prior} \qquad \text{Eq. 15.}$$

In the latter equation the hyperparameters, $\boldsymbol{\beta}_{prior}, \boldsymbol{\Sigma}_{prior}$, would be fixed to values obtained through elicitation. The covariance of this prior could be equivalently parameterized through the marginal variances $\boldsymbol{\sigma}_{prior} = \sqrt{diag(\boldsymbol{\Sigma}_{prior})}$ and the correlation matrix $\boldsymbol{C}_{prior}$.

In the absence of a formal elicitation activity only the mathematical form of the community prior is available. The problem has now been transformed into one of specifying the mode, marginal variances, and correlation matrix of the multivariate Gaussian for the latter to be as diffuse as possible over the range of potential data/likelihood. We use Eq. 15 to build such "locally flat" priors in a stepwise manner: first, we ask whether a mathematical definition can be given to the concept of local uniformity. We feel that this question can be answered in the affirmative by defining this concept through the

Radon-Nikodym derivative with respect to a σ-finite measure e.g., the Lebesgue measure over the Euclidean space $\mathbb{R}^p$. If the derivative of the probability measure implied by Eq. 15 with respect to the Lebesgue measure is constant on $\mathbb{R}^p$, except possibly for a set of probability zero(Campbell 1966), then it is uniform locally over all measurable sets (including $\boldsymbol{B}$) whose union makes up $\mathbb{R}^p$. A less obvious benefit of defining local uniformity through Radon-Nikodym derivatives, is that the latter may also be used as generalized measures of range of the probability measure corresponding to Eq. 15. Informally, a measure that "extends" further out than another one will cover large swaths of the parameter space $\boldsymbol{B}$; as the corresponding kernel of the density is integrable, spreading it over a large area will ensure that the prior assumes small numerical values throughout the space.

To specify the covariance matrix a locally flat Gaussian, we note that the normalization constant or the differential entropy(Campbell 1966; Peña and Rodríguez 2003; Chen et al. 2016) are readily available *scalar* measures of "spread/range/extent" of a distribution: the former is the Riemannian integral of the kernel of the density, while the latter is the Lebesgue integral of the negative logarithm of the Radon-Nikodym derivative of a probability measure with respect to the density corresponding to the probability measure. Both the normalization constant and the differential entropy are increasing functions of the determinant, i.e. the "generalized variance"(Wilks 1932), of the covariance matrix of the multivariate normal. Application of the Hadamard inequality for determinants shows that $|\boldsymbol{\Sigma}_{\text{prior}}| = |\boldsymbol{C}_{prior}||\boldsymbol{\sigma}_{prior}|^2 \leq |\boldsymbol{\sigma}_{prior}|^2$, i.e., the spread of a general multivariate normal will be smaller than that of another multivariate Gaussian with the same marginal variance, but independent components. Hence, a significant first step in defining a locally uniform prior through multivariate Gaussians is by choosing the correlation matrix to be the identity one, spreading the distribution as far out as possible. Furthermore, the Gaussian has only a single mode, hence if the mode for the prior is taken to lie inside the interior of the parameter space (where the peak of the likelihood is likely to occur), then prior will not only be dominated by the likelihood but will not assume large values as one moves toward the boundary of $\boldsymbol{B}$. Hence, setting the mode of the Gaussian somewhere, anywhere, inside $\boldsymbol{B}$ goes a long away to ensuring local uniformity.

### 3.1.3.2.  Priors for testing and priors for exploration.

As a result of the heuristic arguments of the last two paragraphs, we can now concentrate on defining priors locally in subsets of $\mathbb{R}$, considering each Gaussian component, $\beta_i$, of the parameter vector $\boldsymbol{\beta}$ independently of all others. At this point, we will distinguish between two possibilities: *priors-that-test* in which the mode of the Gaussian prior $\beta_{i,0}^{prior}$ is of special interest to be specified explicitly:

$$p_{test}(\beta_i|\beta_{i,0}^{prior},\sigma_{i,0}^{prior}) \sim N(\beta_i|\beta_{i,0}^{prior},\sigma_{i,0}^{prior})$$

and *priors-that-explore* in which the mode itself is treated probabilistically by assigning it a hyperprior over a closed interval $[\beta_i^{min},\beta_i^{max}]$ that contains, i.e. "is-larger", the interval in which $\beta_i$ is likely to lie in:

$$p_{explore}(\beta_i|\beta_{i,0}^{prior},\sigma_{i,0}^{prior}) = \int_{\beta_i^{min}}^{\beta_i^{max}} N(\beta_i|\beta_{i,0}^{prior},\sigma_{i,0}^{prior})p(\beta_{i,0}^{prior})d\beta_{i,0}^{prior}$$

The relevant interval itself may be specified through a "finite-world" argument as in Section 3.1.1 and the simplest hyperprior for $\beta_{i,0}^{prior}$ is the uniform one over the interval $[\beta_{i,0}^{min},\beta_{i,0}^{max}]$. For both testing and exploration priors, the standard deviation $\sigma_{i,0}^{prior}$ of the prior for $\beta_i$ should also be specified in a manner compatible with local uniformity. The most straightforward choice is for $\sigma_{i,0}^{prior}$ to be taken to be a fixed, large number, with the meaning of "large" operationalized as the largest value that does not compromise numerical computations. Alternatively, one could treat the standard deviation as a nuisance parameter and integrate it out using a suitable hyperprior. Adopting similar arguments to the prior of the scale parameter for the GLM (Section 3.2), we could set the hyperprior $p(\sigma_{i,0}^{prior}) = $

$[\![\sigma_{i,0}^{prior} \in [\sigma_{i,0}^{min}, \sigma_{i,0}^{max}]]\!]/(\sigma_{i,0}^{max} - \sigma_{i,0}^{min})$, to obtain a non-conventional prior-that-tests for $\beta_i$ that is proportional to the difference of two incomplete gamma functions[1]:

$$\left[\Gamma\left(0, \frac{(\beta_i - \beta_{i,0}^{prior})^2}{2\sigma_{i,0}^{max\,2}}\right) - \Gamma\left(0, \frac{(\beta_i - \beta_{i,0}^{prior})^2}{2\sigma_{i,0}^{min\,2}}\right)\right] / \left[2\sqrt{2\pi}(\sigma_{i,0}^{max} - \sigma_{i,0}^{min})\right]$$

A more conventional choice for the hyperprior of the variance would be the scaled inverse chi-square distribution leading to a Student-t prior-that-tests for $\beta_i$ with scale $s_i$ and $\nu_0$ degrees of freedom. The corresponding priors-that-explore can be derived by integrating out $\beta_{i,0}^{prior}$ from the priors-that-test. The un-normalized form for the six priors that test and explore is shown in Table 1. They are all unimodal and attain their maximum value when $\beta_i = \beta_{i,0}^{prior}$; the prior-that-explore are nonzero only in the subset $[\beta_{i,0}^{min}, \beta_{i,0}^{max}]$. The relative difference of the maximum from the value at a point $\beta_{i,0}^{min} < \beta_i < \beta_{i,0}^{max}$ can be made arbitrarily small by suitable choice of the hyperprior.

*Table 1 Unnormalized densities for priors that test or explore, according to the hyperprior for the variance parameter*

| Hyperprior for the variance $\sigma_{i,0}^{prior}$ | Priors-that-test | Priors-that-explore[†] |
|---|---|---|
| **Fixed value** | $N(\beta_i \mid \beta_{i,0}^{prior}, \sigma_{i,0}^{prior})$ | $0.5 \times \left(\Phi\left(\frac{\beta_i - \beta_{i,0}^{max}}{\sqrt{2}\sigma_{i,0}^{prior}}\right) - \Phi\left(\frac{\beta_i - \beta_{i,0}^{min}}{\sqrt{2}\sigma_{i,0}^{prior}}\right)\right)$ |
| **Uniform over** $[\sigma_{i,0}^{min}, \sigma_{i,0}^{max}]$ | $\Gamma\left(0, \frac{(\beta_i - \beta_{i,0}^{prior})^2}{2\sigma_{i,0}^{max\,2}}\right) - \Gamma\left(0, \frac{(\beta_i - \beta_{i,0}^{prior})^2}{2\sigma_{i,0}^{min\,2}}\right)$ | $\int_{\beta_{i,0}^{min}}^{\beta_{i,0}^{max}} \left[\Gamma\left(0, \frac{(\beta_i - \beta_{i,0}^{prior})^2}{2\sigma_{i,0}^{max\,2}}\right) - \Gamma\left(0, \frac{(\beta_i - \beta_{i,0}^{prior})^2}{2\sigma_{i,0}^{min\,2}}\right)\right] d\beta_i$ [‡] |
| **Inv-$\chi^2(\nu_0, s_i^2)$** | $St(\beta_i \mid \beta_{i,0}^{prior}, \sigma_{i,0}^{prior}, \nu_0)$ | $\int_{\beta_{i,0}^{min}}^{\beta_{i,0}^{max}} St(\beta_i \mid \beta_{i,0}^{prior}, \sigma_{i,0}^{prior}, \nu_0) d\beta_i$ [¶] |

*[†]These priors are identically zero $[\![\beta_i^0 \notin [\beta_{i,0}^{min}, \beta_{i,0}^{max}]]\!]$, [‡]this integral can be computed in closed form using algebraic expressions of error and incomplete gamma functions, [¶] this integral can be computed in closed form using the Gauss hypergeometric function 2F1*

We illustrate these points graphically in Figure *1* using the case of a single parameter for the six priors in Table 1. To generate this figure, we took the value of $\sigma_{i,0}^{prior} = 1000$ (a common "default" noninformative option for the fixed value priors), $\beta_{i,0}^{min} = -200$, $\beta_{i,0}^{max} = +200$, $\beta_{i,0}^{prior} = 0$, $\sigma_{i,0}^{min}=900$, $\sigma_{i,0}^{max} = 1100$, and the degrees of freedom of the Student t distribution to be equal to one. The resultant priors vary little (less than 0.25%) in the interval $[-50,50]$, and thus are locally uniform in that interval. Despite the very different functional form of these priors, their local behavior relative to the attained maximum is rather similar, i.e., they all resemble a "flat" parabola. In fact, selection of the hyperparameters can be done in such a way that the shape of the priors not only becomes identical in $[\beta_{i,0}^{min}, \beta_{i,0}^{max}]$, but the priors themselves can be taken as effectively constant over the same interval.

---

[1] This formula assumes that $\beta_i \neq \beta_{i,0}^{prior}$, otherwise the incomplete gamma functions are not defined. The value of the density for $\beta_i = \beta_{i,0}^{prior}$, can be obtained in closed form as $\log(\sigma_{i,0}^{max}/\sigma_{i,0}^{min})/[\sqrt{2\pi}(\sigma_{i,0}^{max} - \sigma_{i,0}^{min})]$.

*Figure 1 x2 Comparison of priors that test and priors that explore for a single parameter.*

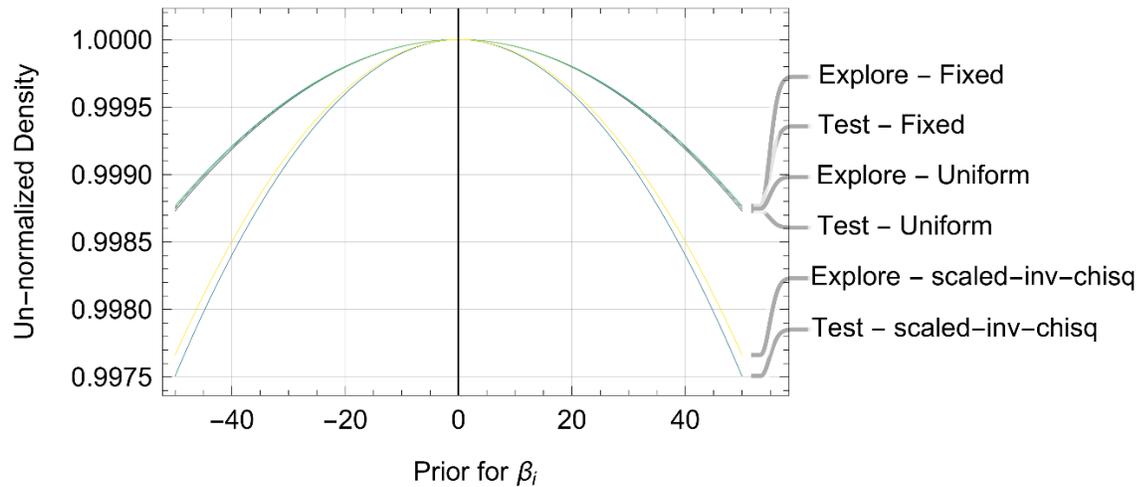

### 3.2. Prior for the scale parameter of the GLM

The prior for the scale parameter of the GLM may be derived by noting that this parameter serves as a bona-fide scale parameter for the response variable of the GLM. It also serves to scale the uncertainty in the likelihood, a point that is most evident by inspecting the saddlepoint approximation in Eq.9 and the ML solution given in Eq. 3. Non-informative priors for scale parameters in other contexts, e.g., the linear regression or the location-scale model are taken to be constant in the logarithm, or inverse to the untransformed value of the scale parameter: $p(\log(\phi)) \propto c \Rightarrow p(\phi) \propto \phi^{-1}$. This is also known as the

Jeffrey's prior for scale parameters. As with the case of the $\boldsymbol{\beta}$ we may restrict the range of the scale parameter to be a proper subset of the real line:

$$0 \leq \phi_{min} \leq \phi \leq \phi_{max} < \infty$$

For the purpose of this work we will take $\phi_{min} = 0$ and thus the uniform prior will be equal to $\phi = \phi_{max}^{-1}$. Alternatively, we may argue along the same lines as $\boldsymbol{\beta}$ and constrain the prior for $\phi$ to be uniform in the interval $[\phi_{min}, \phi_{max}], p(\phi) \propto c$. We justify this choice by noting that nothing, not even the variation, is infinite in a finite world.

### 3.3. The numerical equivalence between Bayesian and likelihoodist analyses

#### 3.3.1. The case of the known GLM scale parameter

It is straightforward to apply the Bayes theorem to GLM regression modeling under the uniform prior over the hypercube:

$$p(\boldsymbol{\beta}|\mathbf{y}, \mathbf{X}, \phi) = \frac{p^{-1} \exp\left(\sum_{i=1}^{n} \frac{y_i x_i' \boldsymbol{\beta} - b(x_i' \boldsymbol{\beta})}{\alpha_i(\phi)} + c(y_i, \phi)\right) \prod_{i=1}^{p} \frac{[\![\beta_i \in [\beta_i^{min}, \beta_i^{max}]]\!]}{\beta_i^{max} - \beta_i^{min}}}{p^{-1} \int \exp\left(\sum_{i=1}^{n} \frac{x_i' \boldsymbol{\beta} - b(x_i' \boldsymbol{\beta})}{\alpha_i(\phi)} + c(y_i, \phi)\right) \prod_{i=1}^{p} \frac{[\![b_i \in [\beta_i^{min}, \beta_i^{max}]]\!]}{\beta_i^{max} - \beta_i^{min}} d\boldsymbol{\beta}}$$

$$= \frac{\exp(ll(\boldsymbol{\beta}, \phi; \mathbf{y}|\mathbf{X}))[\![\beta_i \in [\beta_i^{min}, \beta_i^{max}]]\!]}{\int \exp(ll(\boldsymbol{\beta}, \phi; \mathbf{y}|\mathbf{X}))[\![b_i \in [\beta_i^{min}, \beta_i^{max}]]\!] d\boldsymbol{\beta}} \approx \frac{\exp(ll(\boldsymbol{\beta}, \phi; \mathbf{y}|\mathbf{X}))}{\int \exp(ll(\boldsymbol{\beta}, \phi; \mathbf{y}|\mathbf{X})) d\boldsymbol{\beta}}$$

The approximate equality of the last line will hold true if the boundaries of the hypercube extend past the region the likelihood assumes an appreciable value so that extending the limits of integration to infinity will incur exponentially small errors when evaluating these integrals via the Laplace approximation. There are a few notable observations about this calculation: a) the priors over the parameters are irrelevant as long as they are bounded and constant over the parameter space b) the bounds of the parameters are only relevant to the extent that they restrict the range of integration in the normalizing constant in the denominator and bounds each parameter within the hypercube used to define the prior.

To relate the Bayesian $\exp(ll(\boldsymbol{\beta},\phi;\boldsymbol{y}|\mathbf{X}))/\int \exp(ll(\boldsymbol{\beta},\phi;\boldsymbol{y}|\mathbf{X}))\,d\boldsymbol{\beta}$ to the likelihoodist analysis, we have to consider what makes the latter "tick". Subsequently we can shift attention to the likelihood (numerator) and the multivariate integral in the denominator of the Bayesian solution and derive expressions that are compatible with the assumptions of the likelihoodist analysis. This will ensure that the derivation of the two solutions is based on the same mathematical assumptions. Computationally, the likelihoodist analysis uses only the location of the maximum value of the log-likelihood and the curvature around that point. An approximate Bayesian solution that only uses this information may be obtained by a) expanding the numerator as multivariate Taylor series around the maximum of the likelihood and retain only the first two nonzero terms (the constant and the quadratic terms) b) evaluate the normalizing constant by the multivariate Laplace approximation in the denominator. Similar to the ML approach, the Laplace approximation uses the maximum value of the integrand and the curvature around the maximum to approximate the value of an exponential integral(Fulks 1960; Fulks and Sather 1961). As long as the maximum is located away from the boundary of the parameter space we can write:

$$\exp(ll(\boldsymbol{\beta},\phi;\boldsymbol{y}|\mathbf{X})) \approx \exp\left(ll(\widehat{\boldsymbol{\beta}},\phi|\boldsymbol{y}) - \frac{1}{2}(\boldsymbol{\beta}-\widehat{\boldsymbol{\beta}})'\boldsymbol{\Sigma}^{-1}(\boldsymbol{\beta}-\widehat{\boldsymbol{\beta}})\right)$$

$$\int \exp(ll(\boldsymbol{\beta},\phi;\boldsymbol{y}|\mathbf{X}))\,d\boldsymbol{\beta} \approx (2\pi)^{\frac{p}{2}}\sqrt{|\boldsymbol{\Sigma}|}\exp\left(ll(\widehat{\boldsymbol{\beta}},\phi;\boldsymbol{y}|\mathbf{X})\right)$$

$$p(\boldsymbol{\beta}|\boldsymbol{y},\mathbf{X},\phi) \approx MVN(\boldsymbol{\beta}|\widehat{\boldsymbol{\beta}},\boldsymbol{\Sigma})$$

Eq. 16.

In the second and third lines, we used the numerical equality of the expected and the observed Fisher information matrices for the canonical link of the GLM. For non-canonical links, the observed and expected Fisher information are only equivalent asymptomatically as the sample size increases. This derivation thus establishes the **numerical** equivalency between a Bayesian and likelihoodist approximate analyses. In both cases the curvature of the normal distribution (which determines the concentration of probability around the peak value) is identical and computable from the observed Fisher information. We should point out that this highly heuristic derivation can be made substantially more rigorous (Kass et al. 1990) by considering properties of the log-likelihood for GLMs. The relevant assumptions are that we are dealing with six-times continuously differentiable integrands, design matrices of full rank and variance parameters away from the boundary.

### 3.3.2. The case of unknown location and scale parameters

The posterior distribution in this case may be written as:

$$p(\boldsymbol{\beta}, \phi | \boldsymbol{y}, \mathbf{X}) = \frac{\exp(ll(\boldsymbol{\beta}, \phi; \boldsymbol{y}|\mathbf{X}))\, p(\phi)}{\iint \exp(ll(\boldsymbol{\beta}, \phi; \boldsymbol{y}|\mathbf{X})) p(\phi)\, d\boldsymbol{\beta}\, d\phi} \qquad \text{Eq. 17.}$$

Inferences about $\boldsymbol{\beta}$ which are usually the parameters of interest may be made by integrating out the nuisance parameter $\phi$:

$$p(\boldsymbol{\beta}|\boldsymbol{y}, \mathbf{X}) = \int p(\boldsymbol{\beta}, \phi|\boldsymbol{y}, \mathbf{X})\, d\phi = \int p(\boldsymbol{\beta}|\phi, \boldsymbol{y}, \mathbf{X}) p(\phi|\boldsymbol{y}, \mathbf{X})\, d\phi \qquad \text{Eq. 18.}$$

with $p(\phi|\boldsymbol{y}, \mathbf{X}) = \int p(\boldsymbol{\beta}, \phi|\boldsymbol{y}, \mathbf{X})\, d\boldsymbol{\beta}$. Before considering the Bayesian estimation in detail, we will first present a heuristic derivation that captures the essence of our calculations. In particular, one will note that in GLMs $\phi$ has many fewer dimensions than $\boldsymbol{\beta}$, i.e. it is a scalar quantity. Given substantial amount of data $y$, $p(\phi|\boldsymbol{y}, \mathbf{X})$ will often be relatively sharply peaked, implying that $p(\boldsymbol{\beta}|\boldsymbol{y}, \mathbf{X}) \approx p(\boldsymbol{\beta}|\boldsymbol{y}, \mathbf{X}, \hat{\phi})$, where $\hat{\phi}$ is the marginal mode for $\phi$ - more or less the ML estimate of $\phi$ from $p(\boldsymbol{y}|\mathbf{X}, \phi)$ provided the prior $p(\phi)$ is fairly uninformative or "flat".

The approximation $p(\boldsymbol{\beta}|\boldsymbol{y}, \mathbf{X}, \hat{\phi})$ has the form that would have arisen:

1. If $\phi$ had been estimated hierarchically as : $p(\boldsymbol{y}|\mathbf{X}, \phi) = \int p(\boldsymbol{y}|\mathbf{X}, \boldsymbol{\beta}, \phi) p(\boldsymbol{\beta}|\phi) d\boldsymbol{\beta}$

   and using the flatness of the prior and the sharpness around the mode in the Bayes' Theorem:

   $$p(\phi|\boldsymbol{y}, \mathbf{X}) \propto p(\boldsymbol{y}|\mathbf{X}, \phi) p(\phi) = p(\boldsymbol{y}|\mathbf{X}, \phi) \approx p(\boldsymbol{y}|\mathbf{X}, \hat{\phi})$$

2. The posterior for $\boldsymbol{\beta}$ was calculated in the normal way on the basis of a prior $p(\boldsymbol{\beta}|\hat{\phi})$

It is as if "the prior is estimated from the data" - this is what is often termed "empirical Bayes", an approximation to a proper Bayesian analysis.

The numerical equivalence between Bayesian and frequentist approximations is more formally obtained by variations of an algorithm in which Bayesian integrals are evaluated via either a) convolutions with Dirac delta functions for sharply peaked integrands, or b) via the Laplace method, if they involve "not-as-sharply" peaked exponentiated log-likelihoods, that can be approximated via multivariate quadratic polynomials:

1. First, we expand the log-likelihood around the ML estimate of $\hat{\boldsymbol{\beta}}$ for any given $\phi$ in both numerator and denominator to factorize the posterior density into two blocks: the value of the log-likelihood at $\hat{\boldsymbol{\beta}}$ as a function of $\phi$ and the kernel of the multivariate normal density that depends on $\boldsymbol{\beta}$ and $\phi$. Symbolically we have:

$$p(\boldsymbol{\beta}, \phi|\boldsymbol{y}, \boldsymbol{X}) \approx \frac{\exp\left(ll(\hat{\boldsymbol{\beta}}, \phi; \boldsymbol{y}|\boldsymbol{X}) - \frac{1}{2}(\boldsymbol{\beta} - \hat{\boldsymbol{\beta}})'\boldsymbol{\Sigma}^{-1}(\boldsymbol{\beta} - \hat{\boldsymbol{\beta}})\right)p(\phi)}{\iint \exp\left(ll(\hat{\boldsymbol{\beta}}, \phi; \boldsymbol{y}|\boldsymbol{X}) - \frac{1}{2}(\boldsymbol{\beta} - \hat{\boldsymbol{\beta}})'\boldsymbol{\Sigma}^{-1}(\boldsymbol{\beta} - \hat{\boldsymbol{\beta}})\right)p(\phi)\,d\boldsymbol{\beta}\,d\phi}$$

Eq. 19.

2. Subsequently, we carry out the integration in the denominator over $\boldsymbol{\beta}$ to recover the normalizing constant of the multivariate normal distribution: $(2\pi)^{\frac{p}{2}}\sqrt{|\boldsymbol{\Sigma}|} = (2\pi\phi)^{\frac{p}{2}}\sqrt{\left|\left(\boldsymbol{X}^T\boldsymbol{W}_{\hat{\boldsymbol{\beta}}}\boldsymbol{X}\right)^{-1}\right|}$. Next, we multiply both numerator and denominator by $\phi^{\frac{p}{2}}$:

$$p(\boldsymbol{\beta}, \phi|\boldsymbol{y}, \boldsymbol{X}) \approx \frac{\exp\left(ll(\hat{\boldsymbol{\beta}}, \phi; \boldsymbol{y}|\boldsymbol{X})\right)\phi^{\frac{p}{2}}p(\phi)}{\int \exp\left(ll(\hat{\boldsymbol{\beta}}, \phi; \boldsymbol{y}|\boldsymbol{X})\right)\phi^{\frac{p}{2}}p(\phi)\,d\phi} \times \frac{\exp\left(-\frac{1}{2}(\boldsymbol{\beta} - \hat{\boldsymbol{\beta}})'\boldsymbol{\Sigma}^{-1}(\boldsymbol{\beta} - \hat{\boldsymbol{\beta}})\right)}{(2\pi)^{\frac{p}{2}}\sqrt{|\boldsymbol{\Sigma}|}}$$

Eq. 20.

3. Then we integrate $\boldsymbol{\beta}$ out of the posterior density to obtain the marginal density for $\phi$ as:

$$p(\phi|\mathbf{y}, \mathbf{X}) \approx \frac{\exp\left(ll(\widehat{\boldsymbol{\beta}}, \phi; \mathbf{y}|\mathbf{X})\right) \phi^{\frac{p}{2}} p(\phi)}{\int \exp\left(ll(\widehat{\boldsymbol{\beta}}, \phi; \mathbf{y}|\mathbf{X})\right) \phi^{\frac{p}{2}} p(\phi) \, d\phi} \qquad \text{Eq. 21.}$$

4. Finally, we assume that this marginal density is sharply concentrated around its mode $\phi_{MAP}$, and thus can be approximated by a Dirac delta function: $p(\phi|\mathbf{y}, \mathbf{X}) = \delta(\phi - \phi_{MAP})$. In such a case, integrating out $\phi$ in Eq. 18. would be equivalent to setting $p(\boldsymbol{\beta}|\mathbf{y}, \mathbf{X}) \approx p(\boldsymbol{\beta}|\mathbf{y}, \mathbf{X}, \hat{\phi})$. This approximation step yields an empirical Bayes estimator as a numerical approximation to the marginal posterior density for $\beta$.

The only approximations that are made in the derivation are the second order, normal like expansion in step 1 and the empirical Bayes estimator in step 4. Existing non-Bayesian approaches for the estimation of GLMs may be interpreted as variations of this approximation schema, if the deviance formulation and the saddlepoint approximation to the GLM (Jorgensen 1987, 1997; Nelder and Pregibon 1987) are used in the intermediate steps.

➢ If the prior over $\phi$ is taken to be uniform, evaluation of the integral in Eq. 21 with the Laplace approximation requires that we compute the mode of the integrand. This leads directly to the modified profile likelihood estimator $\hat{\phi}_{mPL}$. The empirical Bayes estimator then follows with $\phi_{MAP} = \hat{\phi}_{mPL}$.

➢ Plugging the saddlepoint approximation (Eq. 9) and the non-informative Jeffrey's prior for the variance parameter $p(\phi) \propto \phi^{-1}$ modifies Eq. 21 to:

$$p(\phi|\mathbf{y}, \mathbf{X}) \approx \frac{\exp(-D(\mathbf{y}; \widehat{\boldsymbol{\mu}})/2\phi)\phi^{-\left(\frac{n-p}{2}+1\right)}}{\int \exp(-D(\mathbf{y}; \widehat{\boldsymbol{\mu}})/2\phi)\phi^{-\left(\frac{n-p}{2}+1\right)} d\phi} \qquad \text{Eq. 22.}$$

This density can be recognized as the scale – inverse chi-square with $n - p$ degrees of freedom, and scale parameter $D(\mathbf{y}; \widehat{\boldsymbol{\mu}})/(n - p)$. The marginal distribution $p(\boldsymbol{\beta}|\mathbf{y}, \mathbf{X})$ is thus a multivariate t-distribution with $n - p$ degrees of freedom and covariance matrix

$\frac{D(\mathbf{y};\hat{\boldsymbol{\mu}})}{n-p}(\mathbf{X}^T\mathbf{W}_{\hat{\boldsymbol{\beta}}}\mathbf{X})^{-1}$. This distribution convergences towards the empirical Bayes estimator, which in turn is identical to the frequentist result in Eq. 3 as the effective sample size, $n-p$ increases.

➤ When the saddlepoint approximation (Eq. 9) is combined with the uniform prior $p(\phi) \propto 1$ the marginal density of $\phi$ is still scale-inverse chi-square but with $n-p-2$ degrees of freedom, and scale parameter $D(\mathbf{y};\hat{\boldsymbol{\mu}})/(n-p-2)$. The mode of this distribution is equal to $\phi_{MAP} = D(\mathbf{y};\hat{\boldsymbol{\mu}})/(n-p)$ and thus the empirical Bayes estimate $p(\boldsymbol{\beta}|\mathbf{y},\mathbf{X},\phi_{MAP})$ is multivariate normal with covariance matrix equal to $\frac{D(\mathbf{y};\hat{\boldsymbol{\mu}})}{n-p}(\mathbf{X}^T\mathbf{W}_{\hat{\boldsymbol{\beta}}}\mathbf{X})^{-1}$. The exact marginal distribution for $p(\boldsymbol{\beta}|\mathbf{y},\mathbf{X})$ under the saddlepoint approximation is however multivariate t with $(n-p-2)$ degrees of freedom. The saddlepoint approximation is exact for the inverse Gaussian and Gaussian families for all values of the dispersion parameter and hence the results established under the constant and Jeffrey's priors are exact for these distributions (Blæsild and Jensen 1985; Reid 1988) for all $\phi$. This is also the case for the renormalized saddlepoint approximation to the gamma density (Blæsild and Jensen 1985; Jorgensen 1997).

➤ The EQL approach to GLMs is a modification of the algorithm 1-4 under the uniform prior for $\phi$. First, we substitute the log-likelihood implied by the saddlepoint density in Eq. 17:

$$p(\boldsymbol{\beta},\phi|\mathbf{y},\mathbf{X}) = \frac{\exp\left(-\frac{n}{2}\log(2\pi\phi) - \frac{1}{2}\sum\log(V(y_i)) - \frac{1}{2\phi}\sum d(y_i,\mu_i)\right)}{\iint \exp\left(-\frac{n}{2}\log(2\pi\phi) - \frac{1}{2}\sum\log(V(y_i)) - \frac{1}{2\phi}\sum d(y_i,\mu_i)\right)d\boldsymbol{\beta}\,d\phi} \quad \text{Eq. 23.}$$

Subsequently we apply the Laplace approximation to simultaneously handle the multiple integral for $\boldsymbol{\beta}$ and $\phi$. This results into a MVN approximation for the joint posterior distribution:

$$p(\boldsymbol{\beta},\phi|\mathbf{y},\mathbf{X}) \sim MVN\left(\begin{bmatrix}\hat{\boldsymbol{\beta}}\\\hat{\phi}_{EQL}\end{bmatrix}, \begin{bmatrix}\hat{\phi}_{EQL}(\mathbf{X}^T\mathbf{W}_{\hat{\boldsymbol{\beta}}}\mathbf{X})^{-1} & 0\\ 0 & \frac{2\,\hat{\phi}_{EQL}}{n}\end{bmatrix}\right)$$

Although asymptotically equivalent to the approximations considered so far, this is a substantially less satisfactory a computation for small to moderate effective sample sizes. To see why this is so, note that this approximation is equivalent to estimating the standard deviation of a normal sample without applying the Bessel correction.

- The shortcoming of the EQL can be addressed by resorting to Restricted Maximum Likelihood (REML) estimation, e.g. pages 87-91 in (Lee et al. 2017). In the REML approach the log-likelihood in Eq. 23 is maximized first with respect to $\boldsymbol{\beta}$, and the resulting adjusted profile log-likelihood in the frequentist perspective (equal, up to a constant, to the posterior marginal log-likelihood for $\phi$) is maximized with respect to $\phi$. This sequence of maximizations amounts to a hybrid iterated Laplace/symbolic integration procedure that recapitulates the essence of the algorithm : a) first one applies the Laplace method with respect to $\boldsymbol{\beta}$ (steps 1-2), effectively approximating $p(\boldsymbol{\beta}, \phi|\mathbf{y}, \mathbf{X}) \approx p(\boldsymbol{\beta} - \widehat{\boldsymbol{\beta}}|\mathbf{y}, \mathbf{X}, \phi) p(\phi|\mathbf{y}, \mathbf{X})$, followed by b) analytical integration over $\boldsymbol{\beta}$ (step 3) and concluding c) with a second application of the Laplace approximation in order to derive a normal like approximation to the posterior marginal for $\phi$ (step 4). Computed this way, and under the uniform prior, the mode of posterior marginal for $\phi$ occurs at $D(\mathbf{y}; \widehat{\boldsymbol{\mu}})/(n-p)$. Plugging-in this value, the empirical Bayes approximation to the posterior for $\boldsymbol{\beta}$ is similar to the one computed under the Jeffrey's prior for $\phi$. Thus, another way of thinking about a REML approach is that it is an approximate Bayesian approach in two stages: in the 1st stage the fixed parameters are integrated out with respect to the uniform prior using the Laplace approximation, while in the 2nd stage ML estimates – or their numerically identical Bayesian posterior modal estimates – are obtained.

- The method of moment estimators may also be seen as an approximate Bayesian calculation. To do so we expand the exponentiated kernel in Eq. 23. 23 to second order with respect to $y_i$

around $\mu_i$. For the deviance component, such an operation amounts to expanding to second order the term:

$$w_i\{y_i((b')^{-1}(y_i) - (b')^{-1}(\mu_i)) - b((b')^{-1}(y_i)) + (b')^{-1}(\mu_i)\} \qquad \text{Eq. 24.}$$

Using the identities $((b')^{-1})' = 1/b'' \circ (b')^{-1}$, $b'((b')^{-1}(\mu_i)) = \mu_i$, $b''((b')^{-1}(\mu_i)) = b''(\theta_i) = V(\mu_i)$ and after cancellation of terms with opposite signs, the second order expansion of the deviance is approximately equal to the Pearson residual. The leading term of the expansion for $V(y_i)$ is $V(\mu_i)$ which can be combined with $-\frac{n}{2}\log(2\pi\phi V(\mu_i))$ inside the exponential function; the remaining terms can be ignored since they do not depend on $\phi$ and will thus cancel out from both numerator and denominator. Subsequently we can adopt either the uniform or the Jeffrey's prior for $\phi$ and derive equivalent posterior expressions in which the sum of deviance residuals has been replaced by the sum of Pearson residuals.

*In summary*, likelihoodist analysis of a GLM regression problem with a known scale parameter (e.g. Poisson or logistic regression) is numerically identical to an approximate Bayesian analysis of the same data under a non-informative, uniform, prior that restricts all parameters to assume values in a finite interval. Bayesian counterparts of all flavors of frequentist methods for GLMs may be also recovered under two different non-informative priors for the scale parameters. These are also expected to yield nearly identical numerical results to the likelihoodist analyses for GLMs approximated by their saddlepoint densities, with the differences being in principle quantifiable by the discrepancy between the t and normal distributions of the appropriate degrees of freedom.

### 3.4. Magic formulas, non-quadratic likelihoods and locally uniform priors

The quadratic behavior of the likelihood around the maximum is a key requirement for the numerical equivalence between Bayesian and frequentist solutions. It is also an assumption that can be violated for two biomedically important GLMs, the binomial logistic and Poisson regression models. During the late 70s and early 80s, a series of papers(Barndorff-Nielsen and Cox 1979; Durbin 1980; Barndorff-Nielsen 1980, 1983; Hinkley 1980) investigated expressions for the distribution of the ML estimates of parameters in exponential family models with known scale parameter. This is precisely the case of the Poisson and Logistic GLMs. The essence of these investigations can be summarized in what many authors (Ghosh 1994; Efron 1998; Pawitan 2013; Lee et al. 2017) have previously called the likelihood p-formula :

$$p(\widehat{\boldsymbol{\beta}}|\boldsymbol{\beta},\mathbf{X},\phi) = \underbrace{\left(\frac{n}{2\pi}\right)^{\frac{p}{2}} |\bar{I}(\widehat{\boldsymbol{\beta}})|^{\frac{1}{2}} \frac{\exp\bigl(ll(\mathbf{X},\boldsymbol{\beta},\phi;\boldsymbol{y}|\mathbf{X})\bigr)}{\exp\bigl(ll(\widehat{\boldsymbol{\beta}},\phi;\boldsymbol{y}|\mathbf{X})\bigr)}}_{likelihood\ p-formula} \{1 + O(n^{-1})\} \qquad \text{Eq. 25.}$$

This approximation needs the same elements as the quadratic one made in conventional treatments of GLMs, i.e. the ML estimates and the mean observed information matrix $\bar{I}(\widehat{\boldsymbol{\beta}}) = (n\boldsymbol{\Sigma})^{-1}$ at the maximum, yet it achieves an error of order $n^{-1}$ rather than $n^{-1/2}$ typical of ML estimators (Barndorff-Nielsen and Cox 1979; Durbin 1980). The error order of the likelihood p-formula can be improved by renormalizing Eq.25 so that it integrates to unity. The resultant $p^*(\widehat{\boldsymbol{\beta}}|\boldsymbol{\beta},\mathbf{X},\phi) = R(\boldsymbol{\beta})p(\widehat{\boldsymbol{\beta}}|\boldsymbol{\beta},\mathbf{X},\phi)$ approximation has an error of $n^{-3/2}$ and is known as the magic, p* formula of Barndorff-Nielsen. The Bayesian equivalent to these formulas is derived as:

$$p(\boldsymbol{\beta}|\boldsymbol{y},\mathbf{X},\phi) = \frac{\exp\bigl(ll(\mathbf{X},\boldsymbol{\beta},\phi;\boldsymbol{y}|\mathbf{X})\bigr)p(\boldsymbol{\beta})}{\int \exp\bigl(ll(\mathbf{X},\boldsymbol{\beta},\phi;\boldsymbol{y}|\mathbf{X})\bigr)p(\boldsymbol{\beta})d\boldsymbol{\beta}} = \frac{\exp\bigl(n\bar{ll}(\mathbf{X},\boldsymbol{\beta},\phi;\boldsymbol{y}|\mathbf{X})\bigr)p(\boldsymbol{\beta})}{\int \exp\bigl(n\bar{ll}(\mathbf{X},\boldsymbol{\beta},\phi;\boldsymbol{y}|\mathbf{X})\bigr)p(\boldsymbol{\beta})d\boldsymbol{\beta}} \qquad \text{Eq. 26.}$$

$$= \left(\frac{n}{2\pi}\right)^{\frac{p}{2}} |\bar{I}(\widehat{\boldsymbol{\beta}})|^{\frac{1}{2}} \frac{\exp\bigl(ll(\mathbf{X},\boldsymbol{\beta},\phi;\boldsymbol{y}|\mathbf{X})\bigr)}{\exp\bigl(ll(\widehat{\boldsymbol{\beta}},\phi;\boldsymbol{y}|\mathbf{X})\bigr)} \times \frac{p(\boldsymbol{\beta})}{p(\widehat{\boldsymbol{\beta}})} \{1 + O(n^{-1})\}$$

by retaining the first term of the Laplace approximation to the integral of the denominator and forming the reciprocal. In the latter expression, $\bar{ll}(\boldsymbol{\beta},\phi|\boldsymbol{y})$ is the average log-likelihood of the statistical model.The resulting asymptotic expansion has the same leading term as the p formula and is also $O(n^{-1})$ accurate on account of the properties of the multivariate Laplace approximation (see Section

8.3, equation 8.3.50 in (Bleistein and Handelsman 2010). Under a locally, uniform prior $\frac{p(\boldsymbol{\beta})}{p(\widehat{\boldsymbol{\beta}})} \approx 1$ for all $\boldsymbol{\beta} \in \mathbf{B}$, and approximately $p(\boldsymbol{\beta}|\mathbf{y},\mathbf{X},\phi) = p(\widehat{\boldsymbol{\beta}}|\boldsymbol{\beta},\mathbf{X},\phi)\{1 + O(n^{-1})\}$. Like the likelihoodist result we can renormalize the approximation to the Bayesian result so that it integrates to unity. Renormalization can be used to "absorb" the error of $p(\boldsymbol{\beta}) \approx p(\widehat{\boldsymbol{\beta}})$ into the error of the approximation of the normalization constant leading to a $O(n^{-3/2})$ accurate Bayesian approximation which we denote as $p^*(\boldsymbol{\beta}|\mathbf{y},\mathbf{X},\phi)$.

Therefore, the renormalized Bayesian and p* approximations are related asymptotically as:

$$\begin{aligned} p^*(\widehat{\boldsymbol{\beta}}|\boldsymbol{\beta},\mathbf{X},\phi) &= p(\boldsymbol{\beta}|\mathbf{y},\mathbf{X},\phi)\{1 + O(n^{-3/2})\} \\ p^*(\boldsymbol{\beta}|\mathbf{y},\mathbf{X},\phi) &= p^*(\widehat{\boldsymbol{\beta}}|\boldsymbol{\beta},\mathbf{X},\phi)\{1 + O(n^{-3/2})\} \end{aligned}$$ Eq. 27.

To prove this, note that the asymptotic nature of the renormalized indirect Edgeworth(Daniels 1954; Barndorff-Nielsen and Cox 1979) and the Laplace approximation expansions underlying the likelihoodist and Bayesian solutions, implies the existence of positive real constants $M_F$ and $M_B$:

$$p(\widehat{\boldsymbol{\beta}}|\boldsymbol{\beta},\mathbf{X},\phi) = |p(\widehat{\boldsymbol{\beta}}|\boldsymbol{\beta},\mathbf{X},\phi)| \leq p^*(\widehat{\boldsymbol{\beta}}|\boldsymbol{\beta},\mathbf{X},\phi)\left(1 + \frac{M_F}{n^{\frac{3}{2}}}\right),$$

$$p(\boldsymbol{\beta}|\mathbf{y},\mathbf{X},\phi) = |p(\boldsymbol{\beta}|\mathbf{y},\mathbf{X},\phi)| \leq p^*(\widehat{\boldsymbol{\beta}}|\boldsymbol{\beta},\mathbf{X},\phi)\left(1 + \frac{M_B}{n^{3/2}}\right).$$

Therefore:

$$\begin{aligned} \frac{p(\boldsymbol{\beta}|\mathbf{y},\mathbf{X},\phi)}{p(\widehat{\boldsymbol{\beta}}|\boldsymbol{\beta},\mathbf{X},\phi)} &= \frac{|p^*(\boldsymbol{\beta}|\mathbf{y},\mathbf{X},\phi)|}{|p(\widehat{\boldsymbol{\beta}}|\boldsymbol{\beta},\mathbf{X},\phi)|} = \left(1 + \frac{M_B}{n^{3/2}}\right)\Big/\left(1 + \frac{M_F}{n^{3/2}}\right) \\ &= 1 + \frac{M_B - M_F}{n^{3/2}} \times \frac{1}{1 + \frac{M_F}{n^{3/2}}} \leq 1 + \frac{|M_B - M_F|}{n^{3/2}} \times \frac{1}{1 + \frac{M_F}{n^{3/2}}} \\ &\leq 1 + \frac{|M_B - M_F|}{n^{3/2}} \end{aligned}$$

Using a similar line of reasoning we can show that: $\frac{p(\widehat{\boldsymbol{\beta}}|\boldsymbol{\beta},\mathbf{X},\phi)}{p(\boldsymbol{\beta}|\mathbf{y},\mathbf{X},\phi)} = \left|\frac{p(\widehat{\boldsymbol{\beta}}|\boldsymbol{\beta},\mathbf{X},\phi)}{p^*(\boldsymbol{\beta}|\mathbf{y},\mathbf{X},\phi)}\right| \leq 1 + \frac{|M_B - M_F|}{n^{3/2}}$, thus proving the $n^{-3/2}$ asymptotic equivalency between the likelihoodist and Bayesian solutions. A practical

implication of this equivalency is that numerical procedures e.g. Monte Carlo Markov Chain simulation for calculating $p(\boldsymbol{\beta}|\mathbf{y},\mathbf{X},\phi)$ or quantities derived from it, simultaneously provide the corresponding calculations from $p^*(\widehat{\boldsymbol{\beta}}|\boldsymbol{\beta},\mathbf{X},\phi)$ to $n^{3/2}$ order for all models in which the p* approximation applies (models with sufficient statistics) and locally uniform priors. Obviously in the absence of renormalization the Bayesian and likelihoodist results are within $O(n^{-1})$, rather than $O(n^{-3/2})$. While we restricted attention to the nonquadratic case of known scale parameters for simplicity, the connection between Bayesian and likelihoodist results can be trivially extended to account for nuisance parameters in the saddlepoint approximations for GLMs. We outline the relevant steps, without carrying out the full derivation. First, one factorizes the integrand of the marginal density to obtain:

$$p(\boldsymbol{\beta}|\mathbf{y},\mathbf{X}) = \int p(\boldsymbol{\beta}|\mathbf{y},\mathbf{X},\phi)p(\phi|\mathbf{y},\mathbf{X})d\phi = \int p(\widehat{\boldsymbol{\beta}}|\boldsymbol{\beta},\mathbf{X},\phi)p(\phi|\mathbf{y},\mathbf{X})\{1+O(n^{-1})\}d\phi$$

In the second step, one appeals to the arguments in Section 3.3 to substitute the scaled-inverse-chi-square distribution with the appropriate degrees of freedom ($k$) for $p(\phi|\mathbf{y},\mathbf{X})$. Assuming that the mode of scaled-inverse-chi-square distribution is far from the boundary, a univariate application of the Laplace approximation, yields an asymptotic expansion $\propto p(\widehat{\boldsymbol{\beta}}|\boldsymbol{\beta},\mathbf{X},\hat{\phi})\{1+O(n^{-1})\}\left\{1+O\left(k^{-\frac{3}{2}}\right)\right\} \propto$

$p(\widehat{\boldsymbol{\beta}}|\boldsymbol{\beta},\mathbf{X},\hat{\phi})\{1+O(n^{-1})\}\left\{1+O\left(O(n)^{-\frac{3}{2}}\right)\right\} = p(\widehat{\boldsymbol{\beta}}|\boldsymbol{\beta},\mathbf{X},\hat{\phi})\{1+O(n^{-1})\}$.

# 4. The π value, the Bayesian alter-ego of the p value

Notwithstanding previous efforts to consider posterior predictive distributions when examining the reproducibility of two-sided p-values(Goodman 1992; Shao and Chow 2002; Senn 2002), to date there has been no obvious Bayesian justification to the p-value calculation per se. It is our intention in this section to present a Bayesian argument for the calculation and thresholding of p-values. To avoid confusion with the frequentist interpretation, we will designate these Bayesian quantities that are numerically identical to the p-values as *π value*.

## 4.1. The π-value quantifies the magnitude of a directional covariate effect

Many interpretations of estimated treated effects are couched in terms of the direction of the covariate effect, rather than its magnitude. For example, in the biological and social sciences the exact null hypothesis is almost always false because of the complexity and the multitude of the phenomena involved(Meehl 1967, 1978; Greenwald et al. 1996), and many theories are in fact assertions about the direction rather than the magnitude of a statistically determined effect(Greenwald et al. 1996; Gelman and Tuerlinckx 2000). In drug development, ascertainment that a drug is effective and safe, is a pre-requisite for drug registration. Both statements involve directional inferences that are not tied to generally agreed magnitudes for the effect size or the safety margin. In biomedical and social science application domains, measurement noise and experimental unit heterogeneity will often limit the translation of the magnitude, but not the direction of the effect observed to other settings.

Based on these considerations and the resulting need for directional inference, we now consider the posterior marginal mean of the direction of the covariate effect (or more generally the direction of the difference from a baseline level of $\beta_{i,0}$) in the linear predictor of the GLM. This direction is given by the sign function $sgn(\beta_i - \beta_{i,0})$, which assumes value of one if its argument is non-negative and minus one otherwise. Under the quadratic loss function, the best point estimate for directionality is the posterior mean:

$$\int sgn(\beta_i - \beta_{i,0})\, p(\beta_i|\mathbf{y}) d\beta_i = P(\beta_i \geq \beta_{i,0}|\mathbf{y}) - P(\beta_i < \beta_{i,0}|\mathbf{y}) = \quad \text{Eq. 28.}$$

$$sgn\big(P(\beta_i \geq \beta_{i,0}|\mathbf{y}) - P(\beta_i < \beta_{i,0}|\mathbf{y},\mathbf{X})\big) \times \left[1 - \overbrace{2 \times min(\{P(\beta_i \geq \beta_{i,0}|\mathbf{y},\mathbf{X}), P(\beta_i < \beta_{i,0}|\mathbf{y},\mathbf{X})\})}^{\pi-value}\right]$$

In Eq. 11 we established that the Wald test P-value for testing $H_0: \beta_i = \beta_{i,0}$ as twice the value of the smaller of the probabilities of the two complimentary random events: $\widehat{\beta}_i \geq \beta_{i,0}$ and $\widehat{\beta}_i < \beta_{i,0}$. Its Bayesian counterpart, the π value is equivalently defined as twice the value of the smaller of the two tail probabilities, $\{P(\beta_i \geq \beta_{i,0}|\mathbf{y},\mathbf{X}), P(\beta_i < \beta_{i,0}|\mathbf{y},\mathbf{X})\}$ of the posterior density for $\beta_i$. The π-value quantifies the absolute magnitude of a directional covariate effect, so that the smaller its value, the larger the effect in the direction of the sign of $\beta_i - \beta_{i,0}$. In summary, if we use the sign function to define a *map* of the treatment effect from $\mathbb{R} \to \{-1, 0, 1\}$, the π-value along with the sign of the difference of the two complimentary posterior tail probabilities to the left and the right of $\beta_{i,0}$ provide an estimator for the value of the sign function under the quadratic square loss through Eq. 28.

### 4.2. The numerical relation of p and π-values

In the case of GLMs of known scale parameter the p value and the π-value are identical since the posterior distribution and the MLE distribution are identical. When the scale parameter must be estimated by the data, the likelihoodist and the empirical Bayes posterior $p(\boldsymbol{\beta}|\mathbf{y},\mathbf{X},\phi_{MAP})$ are also identical if the likelihoodist ($\widehat{\phi}$) and Bayesian ($\phi_{MAP}$) point estimates are computed under the same assumptions. If one were to average over the uncertainty for $\phi$ and thus use the full proper marginal posterior for $\boldsymbol{\beta}$, the numerical difference between the p value (computed on the basis of the Gaussian distribution) and the π values is reduced the difference between the tail areas of a standardized normal distribution and that of Student t distributions ($St$) of the appropriate degrees of freedom: $n - p$ for the case of the $\phi^{-1}$ prior and $n - p - 2$ for the uniform one. In the case of the Gaussian and the $St_{n-p}$ distributions, the area probabilities correspond to the value of the standardized statistic $z_i$, while for $St_{n-p-2}$ the relevant statistic is equal to $z_i\sqrt{n-p/n-p-2}$. In

Figure 2, we plot the difference of the (twice) tail area probabilities corresponding to the three distributions, over a range of values of $z_i$ and degrees of freedom $n - p$. Differences among the tail areas of the three distributions (and thus between frequentist p and posterior marginal π-values) effectively disappear as the degrees of freedom increase. Since statistical packages base the Wald test on the Gaussian rather than the Student-t distribution, when the dispersion parameter is estimated from the data, the numerical discrepancies between Bayesian π and frequentist p will be minimal in practice, especially as the amount of data grows. Furthermore, even the choice of the prior for the scale parameter, does not matter much as long as the effective degrees of freedom exceed thirty.

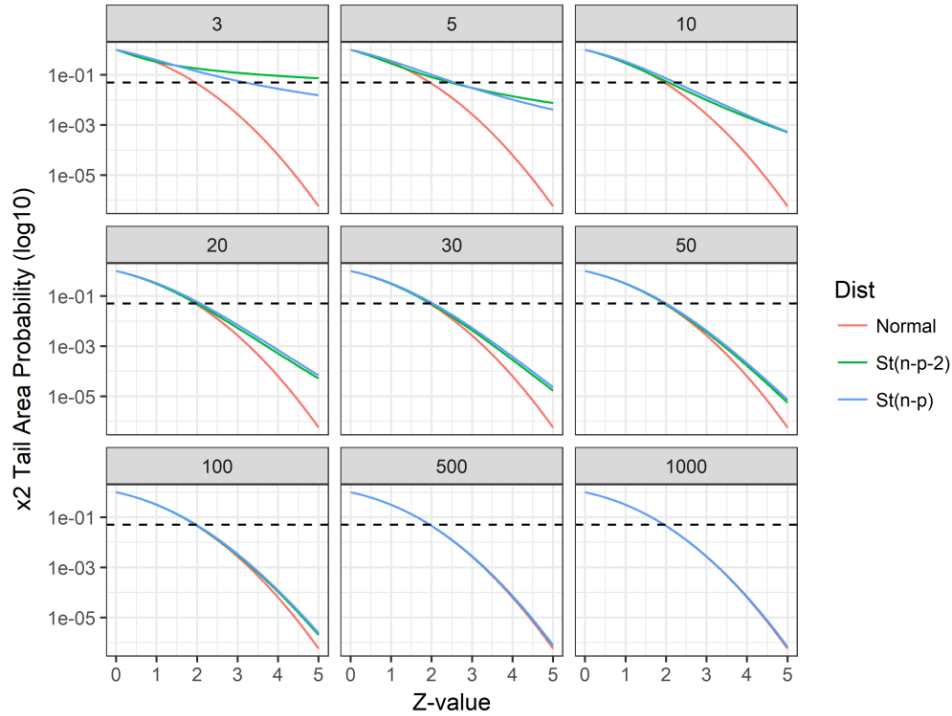

*Figure 2 Twice the tail area probabilities (π-values, in log10 scale) for the standardized normal distribution and the student t (St) distribution for different values of the degrees of freedom (individual panels) and values of the Wald statistic (Z-value). The dashed horizontal line corresponds to the 0.05 threshold.*

### 4.3. A decision analytic framework for thresholding π -values

There is nothing in the exposition so far that requires one to threshold π-values and in fact a Bayesian could simply use them as-is to quantify the magnitude of the treatment effect under the framework of Section 4.1. In this section we show that the justification for thresholding emerges from the interaction between two actors: an analyst, who *reports* the results of a GLM analyses and their clients. This interaction requires the application of a different loss function than the quadratic one used in Eq. 28. We develop this interaction in steps, by first considering the decisions (classifications of directional effects) of the analyst, followed by the decisions that the client must make after receiving the advice of the analyst(Hildreth 1963). In this section we also put forward an interpretation of π-values as measures of the value of the data for the analyst. On the other hand, the client's decisions create the need for a recalibration of the analyst's decisions to consider the context of use of scientific findings. The thresholding of π-values emerges naturally from this recalibration.

### 4.3.1. Decision problem for the "pure" analyst

The pure analyst's primary concern is with the reporting of the state of nature, conveniently identified as the difference between an estimated effect and a threshold $\delta_i = \beta_i - \beta_{i,0}$, as-is without *any* consideration to downstream applications. A pure analyst is thus only concerned with making wrong statements and thus his or her utility function $U_{an}(x)$ only considers gains or losses for correct or incorrect statements, respectively. To describe the decision problem for the analyst, we assume that an analyst has a baseline, not-necessarily monetary capital ($C$), e.g., reputation that can only be increased by correct statements but will surely decrease by erroneous ones. Furthermore, the analyst is indifferent with respect to the state of the world: the same gain or loss ($\alpha$) will be realized irrespective of the direction of the estimated effect as long as the analyst's call is aligned with the actual state of the world. The decision problem for the pure analyst is shown in Table 2:

Table 2 Decision table for a pure analyst with utility function $U_{an}(x)$, initial capital $C$ and symmetric gain or loss of $\alpha$ after having obtained data $y$

|  |  | Uncertain state of the world | |
|---|---|---|---|
|  |  | $\delta_i \geq 0$ | $\delta_i < 0$ |
| **Analyst's action** | State $\delta_i \geq 0$ | $U_{an}(C + \alpha)$ | $U_{an}(C - \alpha)$ |
|  | State $\delta_i < 0$ | $U_{an}(C - \alpha)$ | $U_{an}(C + \alpha)$ |
| **Probabilities** |  | $P(\delta_i \geq 0|y, X)$ | $P(\delta_i < 0|y, X)$ |

Then the utility of the two possible statements by the analyst are given by:

$$\begin{cases} U_{an}(C + \alpha) - P(\delta_i < 0|y, X) \times [U_{an}(C + \alpha) - U_{an}(C - \alpha)], & \text{state } \delta_i \geq 0 \\ U_{an}(C + \alpha) - P(\delta_i \geq 0|y, X) \times [U_{an}(C + \alpha) - U_{an}(C - \alpha)], & \text{state } \delta_i < 0 \end{cases}$$

Let us assume that conditional on the data, the two posterior tail probabilities are $P(\delta_i \geq 0|y, X) \geq P(\delta_i < 0|y, X) \Longrightarrow P(\delta_i < 0|y, X) \leq 0.5$. The optimal decision is to state $\delta_i \geq 0$, since the difference

between the two utilities of the two actions is equal to $[U_{an}(C + \alpha) - U_{an}(C - \alpha)] \times [1 - 2\, P(\delta_i < 0|\mathbf{y}, \mathbf{X})]$ which is a positive by hypothesis. On the other hand, if $P(\delta_i \geq 0|\mathbf{y}, \mathbf{X}) < P(\delta_i < 0|\mathbf{y}, \mathbf{X})$, the optimal decision is to state $\delta_i < 0$ with expected utility $U_{an}(C + \alpha) - P(\delta_i \geq 0|\mathbf{y}, \mathbf{X}) \times [U_{an}(C + \alpha) - U_{an}(C - \alpha)]$. The analyst should identify the state of the world as the direction with the maximum posterior tail probability, and the expected utility is:

$$U_{an}(C + \alpha) - min(\{P(\delta_i \geq 0|\mathbf{y}, \mathbf{X}), P(\delta_i < 0|\mathbf{y}, \mathbf{X})\}) \times [U_{an}(C + \alpha) - U_{an}(C - \alpha)] = \quad \text{Eq. 29.}$$

$$U_{an}(C + \alpha) - \frac{[U_{an}(C + \alpha) - U_{an}(C - \alpha)]}{2} \times \pi$$

Having observed the data, the analyst may wonder about the *expected value of perfect information* (EVPI) for this problem. This is the expected gain in utility that results from perfect information, the kind that can eliminate the possibility of making the wrong decision. The EVPI may be calculated as a weighted sum of the maximum utilities over the space of uncertain events *minus* the expected utility of the optimum action. Substituting the relevant quantities, we recover the EVPI as the product of the *central difference quotient* (CDQ) of the analyst's utility function at $C$, the gain $\alpha$ and the π-value:

$$U_{an}(C + \alpha) - \underbrace{\frac{\overbrace{[U_{an}(C + \alpha) - U_{an}(C - \alpha)]}^{\text{central difference}= \delta_{2\alpha}[U_{an}](C)}}{2}}_{} \times \pi = \underbrace{\boxed{\frac{\delta_{2\alpha}[U_{an}](C)}{2\,\alpha}}}_{CDQ \approx U'_{an}(C)} \times \alpha \times \pi \qquad \text{Eq. 30.}$$

If the expected gain is small relative to $C$, the CDQ becomes nearly equal to the derivative ("elasticity") of the utility function and thus the EVPI becomes equal to: $U'_{an}(C) \times \alpha \times \pi - value$. It thus follows that the EVPI is proportional to the π-value irrespective of the functional form of the analyst's utility function; the smaller the π-value, the less valuable the knowledge of the perfect information becomes for the analyst. *In this very specific sense, one may interpret the π-value as a measure of the strength of evidence for a directional effect through its relation to the value of perfect knowledge.* We cannot emphasize enough that despite them being numerically identical, no such interpretation is admissible for the p-value under the Nieman – Pearson perspective.

### 4.3.2. Decision problem for the client

The decision problem facing the client of the statistical report is somewhat different from that of the analyst generating that report. After receiving advice about a (directional effect) and a π-value, the client has to decide between acting on and "sleeping" on the analyst's statement about the state of the world. In the context of scientific investigation this action could simply be a declaration about the state of the world, while "sleeping" on the analyst statement, could be a summary as banal as "further studies are needed". In a regulatory context, action could be approval of an investigational drug and sleeping could be delaying approval until further data have been gathered. Similar to the pure analyst we assume that the client has a utility function $U_{co}(x)$ and "capital" ($M$, restricted to be non-negative) that can either be increased ("gain", if the analyst is proved to be correct about the direction of the effect) or decreased by a fraction ("loss", if the analyst is proved wrong) if the client acts on the statement of the analyst. If on the other hand the client decides to sleep on the analyst's statement, he or she may experience a capital decline because of the *opportunity cost* of inaction (Table 3):

Table 3 Decision table for a client of statistical analyses with utility function $U_{co}(x)$, initial capital M, asymmetric gains and losses and a relative opportunity cost c. The fractional gain ($\epsilon$) is restricted to be positive, while the relative loss ($\epsilon'$) and opportunity cost are restricted to lie in the interval $(0,1)$

|  |  | **Uncertain state of the world** | |
|---|---|---|---|
|  |  | *As analyst says* | *Contrary to analyst's claims* |
|  | Act on analyst's statements | $U_{co}(M \times (1 + \epsilon))$ | $U_{co}(M \times (1 - \epsilon'))$ |
| **Client's action** | Sleep on analyst's statements | $U_{co}(M \times (1 - c))$ | $U_{co}(M)$ |
|  | **Probabilities** | $1 - \frac{\pi - value}{2}$ | $\frac{\pi - value}{2}$ |

In this form, the decision problem for the client is too general to allow a clear glimpse of the behavioral implications of assuming different values for the fractional gain ($\epsilon$), fractional loss ($\epsilon'$), opportunity cost ($c$) and π-value. If we assume that client has a *declining marginal utility function*, which may be described reasonably well by the logarithmic function $U_{co}(x) = \log x$, then it is possible to evaluate the decision problem numerically. The client will decide to act on the analyst's statement if the utility of action is greater to that of inaction, i.e., iff:

$$\left(1 - \frac{\pi}{2}\right) \times \log(M \times (1 + \epsilon)) + \frac{\pi}{2} \times \log(M \times (1 - \epsilon')) > \left(1 - \frac{\pi}{2}\right) \times \log(M \times (1 - c)) + \frac{\pi}{2} \times \log(M)$$

Hence the risk averse client will choose to act (sleep) on the analyst's statement if the π-value is smaller (larger) than the critical value $\pi_{crit}$:

$$\pi_{crit} = min\left\{1, 2 \times \frac{\log(1 + \epsilon) - \log(1 - c)}{\log(1 + \epsilon) - \log(1 - c) - \log(1 - \epsilon')}\right\} \qquad \text{Eq. 31.}$$

The critical value is a non-decreasing function of the gain and opportunity cost but a non-increasing function of the loss. Decreasing thresholds for claiming significance map to rapidly shrinking volumes in the four-dimensional space $\pi_{crit}, \epsilon, \epsilon', c$ (Figure 3).. Eq. 31 and its graphical solution make it clear that threshold values cannot be considered as static immutable quantities, but rather intrinsic elements of the real-world problem. Different problems or different application domains may require adoption of different π threshold values.

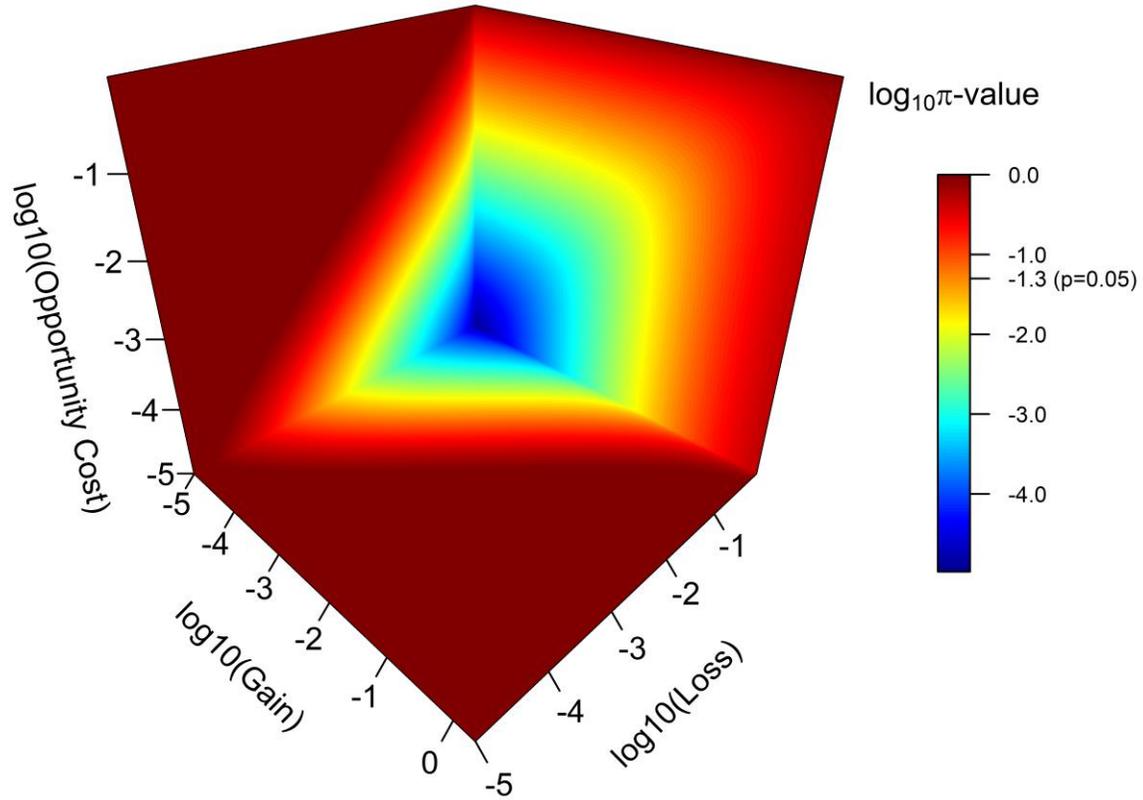

*Figure 3 Volumetric slice plot of the critical threshold against the fractional gain, loss and opportunity cost. The figure shows slices of the four-dimensional space $\pi_{crit}, \epsilon, \epsilon', c$*

### 4.3.3. Decision problem for the "recalibrated" analyst

The decision problem (Table 4) for an analyst "recalibrated" to the downstream actions of the client, expands the options of the pure analyst to include a statement about the non-actionability of $\delta_i$ conditional on the current evidence. In essence the recalibrated analyst will anticipate their client's decisions and adapt their own utilities to better align with the goals of their clients. The non-actionability action corresponds to the action of the client to "sleep" on the statement of the pure analyst. The utility of this action-statement by the analyst is a fraction of the utility of the optimal statement without thresholding, with the fractional loss quantified by the function $f(\pi_{crit}, \pi)$ at a given capital $C$ and potential gain $\alpha$.

*Table 4 Decision table for a pure analyst with utility function $U_{an}(x)$, initial capital $C$ and symmetric gain or loss of $\alpha$ after having obtained data $y$*

| | | Uncertain state of the world | |
|---|---|---|---|
| | | $\delta_i \geq 0$ | $\delta_i < 0$ |
| Analyst's action | State $\delta_i \geq 0$ | $U_{an}(C + \alpha)$ | $U_{an}(C - \alpha)$ |
| | State $\delta_i < 0$ | $U_{an}(C - \alpha)$ | $U_{an}(C + \alpha)$ |
| | State $\delta_i$ non-actionable on the current evidence | $U_{an}(C + \alpha) \times [1 - f(\pi_{crit}, \pi)]$ | $U_{an}(C + \alpha) \times [1 - f(\pi_{crit}, \pi)]$ |
| Probabilities | | $P(\delta_i \geq 0 | y, \mathbf{X})$ | $P(\delta_i < 0 | y, \mathbf{X})$ |

We require that that the utility of a non-actionability statement be somewhat less than the utility of a directional statement when $\pi \leq \pi_{crit}$. This is the meaning of recalibration, i.e., issuing a statement of non-actionability, thus advising the course of action that the client would have made for themselves after receiving the input of a pure analyst. The utility of a non-actionability statement is taken to be proportional to the utility of a correct directional statement, with the proportionality constant a function of $\pi, \pi_{crit}$. The corresponding function, $1 - f(\pi_{crit})$ is the recalibration factor of the pure analyst to the downstream decision their client will make. A restriction on the admissible forms of the recalibration factors follows from a basic requirement that the difference between the expected utility of a directional and a non-actionability statements to be a decreasing function $h$ of the difference $\pi - \pi_{crit}$. This requirement simply captures the desideratum that a non-actionability statement be of smaller utility, the larger the magnitude of the directional effect (Section 4.1) and the less valuable the value of perfect information (Section 4.3.1), and the further away the $\pi$ is from threshold of action $\pi_{crit}$. Therefore, we require:

$$U_{an}(C + \alpha) - \frac{\delta_{2\alpha}[U_{an}](C)}{2} \times \pi - U_{an}(C + \alpha) \times [1 - f(\pi_{crit}, \pi)] = h(\pi - \pi_{crit}) < 0$$

The simplest function that satisfies this requirement is $h(\pi - \pi_{crit}) = -k \times (\pi - \pi_{crit})$, $k = k(C, a, \pi, \pi_{crit}) > 0$. The fractional loss function $f(\pi_{crit}, \pi)$ is thus a function of the utility of a correct directional statement, the elasticity of the utility function, the observed π-value, the $\pi_{crit}$ and the proportionality constant $k$. If we choose $k = \frac{\delta_{2a}[U_{an}](C)}{2}$, $f(\pi_{crit}, \pi)$ becomes independent of $\pi$:

$$f(\pi_{crit}, \pi) = f(\pi_{crit}) = \frac{\delta_{2a}[U_{an}](C)}{2U_{an}(C+a)} \times \pi_{crit}$$

Furthermore, if the analyst's utility function in nearly linear, the fractional loss function becomes a simple function of the fractional capital gain $g = a/C$ of a correct directional statement:

$$f(\pi_{crit}) = \frac{g}{g+1} \times \pi_{crit} \approx \begin{cases} g \times \pi_{crit}, & g \downarrow 0 \\ \pi_{crit}, & g \uparrow \infty \end{cases}$$

This analysis shows that recalibration factors are smaller than, yet numerically close to, unity so that the analyst should be rewarded for a statement of non-actionability nearly as much as for a correct directional statement. We note that this is much more desirable situation for analysts than the one often encountered in the real world, when the p-value computed by the analyst is above the level of significance leading clients to request analysts embark on "p-value expeditions" or else! To complete the presentation, we note that the EVPI for the recalibrated analyst becomes equal to:

$$\frac{\delta_{2a}[U_{an}](C)}{2} \times min\{\pi, \pi_{crit}\} = CDQ \times \alpha \times min\{\pi, \pi_{crit}\} \qquad \text{Eq. 32.}$$

In a world of thresholds, the value of the perfect information for an analysis in which the π-value was above the threshold, is determined by the threshold itself.

## 5. Replication of scientific findings and reproducibility of p-values

The replication crisis in science is rooted to the failure of reported scientific findings to be reproducible when the same question is probed by two or more different research teams in repetitive experiments. While the recent re-emergence of the term(Pashler and Wagenmakers 2012) can be linked to specific incidents of scientific fraud (Stroebe et al. 2012) and psi phenomena such as precognition(Galak et al. 2012), in actuality concerns for the non-reproducibility in the scientific enterprise have been around for much longer(Goodman 1992; Sterne and Smith 2001; Ioannidis 2005; Harlow et al. 2013). The definition of a replicated finding is intimately linked to p-values obtained for tests of significance and this connection was eloquently stated by Goodman(Goodman 1992): "for many researchers, replication means the repetition of a statistically significant result in a study of statistical power" . The p-value focus persisted in subsequent commentaries about reproducibility (Sterne and Smith 2001; Wacholder et al. 2004; Ioannidis 2005) , even when authors criticized the use of p-values. In fact, in molecular epidemiology a "discovery" is often couched in terms of a thresholded p-value, leading to proposals to control the rate of False Discoveries by linking the p-value threshold to the number of tests performed during a given investigation (Benjamini and Hochberg 1995; Storey and Tibshirani 2003). Hence, even while replication is perceived to be more than p-values, in practice it is reduced to observing the reproducibility of a thresholded p-value. We now propose a general predictive probability framework for thinking about replication, which can be used to probe the replication of scientific findings. In this framework, reproducibility of p-values appears as a special and rather limited use case.

## 5.1. A predictive probability framework for replication of scientific findings

A framework for analyzing the replication potential of a given scientific investigation fulfills the needs of the final consumer of the results of that study, whether it is the public, a private organization or an official authority that commissioned, funded, or required that study in the first place. The consumer was not involved in the design of the study (other than perhaps posing the initial question and providing a general outline of acceptable ways to answer it), did not conduct the actual investigation or analyzed the results and has variable access to the source documentation for study procedures, protocol for measurements and raw data. Often, the only information available to the consumer is a highly simplified version of the research question and a topline summary of the study results. Based on this incomplete and partial knowledge, the consumer is faced with the task of modifying their behavior to align with the newly alleged state of the world, committing more resources to probe the question further (e.g., commissioning another study to confirm the finding), or simply move on to other questions. Faced with these alternatives which involve opportunity gains and costs, the consumer may ask whether the present study is sufficient to consider the research question resolved. In asking so, the consumer will be effectively conditioning on a) the entire body of data ($y_{init}$) and the design ($X_{init}$) of the initial study or b) sufficiently informative summaries of that data e.g., the sufficient statistics ($\widehat{\beta}$) the covariance matrix from the GLM fit and the deviance to 1) *predict* the data from a replicated study ($y_{rep}$), 2) make inferences about the parameters ($\beta_{rep}, \phi_{rep}$) in a replicate dataset with a potentially different design matrix ($X_{rep}$) and eventually 3) make decisions. Conditioning on the initial data, the consumer will be interested in inferences about:

$$p(\boldsymbol{\beta}_{rep}, \phi_{rep} | \boldsymbol{y}_{init}, \mathbf{X}_{init}, \mathbf{X}_{rep}) = \int \left\{ \begin{array}{l} \overbrace{p(\boldsymbol{\beta}_{init}, \phi_{init} | \boldsymbol{y}_{init}, \mathbf{X}_{init})}^{Inference\ from\ initial\ data} \times \\ \overbrace{p(\boldsymbol{\beta}_g | \boldsymbol{\beta}_{init}) p(\phi_g | \phi_{init})}^{Translation} \times \\ \overbrace{p(\boldsymbol{y}_{rep} | \mathbf{X}_{rep}, \boldsymbol{\beta}_g, \phi_g)}^{Replicate\ Data\ Generation} \times \\ \overbrace{p(\boldsymbol{\beta}_{rep}, \phi_{rep} | \mathbf{X}_{rep}, \boldsymbol{y}_{rep})}^{Replicate\ Data\ Inference} \end{array} \right\}$$
$$d\boldsymbol{\beta}_{init} d\phi_{init} d\boldsymbol{\beta}_g d\phi_g d\boldsymbol{y}_{rep}$$

Eq. 33.

In this predictive framework for replication, we allow for the *translation* of parameters from the initial study to *parameters* ($\boldsymbol{\beta}_g, \phi_g$) that generate the data for the replicate study. We distinguish notationally between the value of these parameters used to generate the replicate data and the targets of inference $\boldsymbol{\beta}_{rep}, \phi_{rep}$ using the replicate data. In an *exact translation* setup, $p(\boldsymbol{\beta}_g | \boldsymbol{\beta}_{init}) = \delta(\boldsymbol{\beta}_g - \boldsymbol{\beta}_{init})$ and $p(\phi_g | \phi_{init}) = \delta(\phi_g - \phi_{init})$, i.e. the parameters from the initial study translate faithfully to the generation process for the replicate study. In an *exact* replication $\mathbf{X}_{init} = \mathbf{X}_{rep}$, i.e., the design of the replicate study is identical to that of the initial study. While the meaning of reproducibility in existing studies is that of exact translation and replication, the framework of Eq.33 is more general and allows for imperfect translation and non-exact replication e.g., studies with different sample sizes, designs, or participant characteristics. It is not possible in general to carry out the integrations in Eq.33 analytically, and a conditional simulation framework is the most straightforward manner to approximate the predictive distribution $p(\boldsymbol{\beta}_{rep}, \phi_{rep} | \boldsymbol{y}_{init}, \mathbf{X}_{init}, \mathbf{X}_{rep})$.

While simulation based on Eq.33 is rather straightforward, the predictive distribution can be derived in closed form in the case of exact translation and replication under the assumptions of a) the saddlepoint approximation for GLMs b) a data likelihood that is well approximated by a multivariate quadratic polynomial and c) a known scale parameter $\phi$. The last assumption can be relaxed for models if there are sufficient data points to approximate the Student-t distribution with the Gaussian one. Models satisfying these assumptions include the Poisson and Logistic GLM, as well as survival models (e.g., the piecewise exponential, or the semiparametric proportional hazards model) that can be approximated by GLMs(Efron 1988; Argyropoulos and Unruh 2015; Bender et al. 2018). Under the assumptions a)-c) conditioning on the initial data and marginalization over the replicate data is replaced by conditioning on the sufficient statistics $\widehat{\boldsymbol{\beta}}_{init}$ and marginalization over $\widehat{\boldsymbol{\beta}}_{rep}$. Hence, the integral in Eq. 33 reduces to convolution of three multivariate normal distributions corresponding to a) the posterior distribution for $\boldsymbol{\beta}_{init}$ that conditions on $\widehat{\boldsymbol{\beta}}_{init}$ and $\boldsymbol{\Sigma}_{init} = \phi(\mathbf{X}_{init}^T \mathbf{W}_{\widehat{\boldsymbol{\beta}}_{init}} \mathbf{X}_{init})^{-1}$, b) the generative model for the replicate sufficient statistics $\widehat{\boldsymbol{\beta}}_{rep}$ and c) the posterior distribution for $\boldsymbol{\beta}_{rep}$ that conditions on $\widehat{\boldsymbol{\beta}}_{rep}$ and $\boldsymbol{\Sigma}_{rep} = \phi\left(\mathbf{X}_{init}^T \mathbf{W}_{\widehat{\boldsymbol{\beta}}_{rep}} \mathbf{X}_{init}\right)^{-1}$ over the space $\boldsymbol{\beta}_{init} \times \widehat{\boldsymbol{\beta}}_{rep}$:

$$p(\boldsymbol{\beta}_{rep}|\widehat{\boldsymbol{\beta}}_{init}, \mathbf{X}_{init}, \phi) = \int p(\boldsymbol{\beta}_{init}|\widehat{\boldsymbol{\beta}}_{init}, \mathbf{X}_{init}, \phi) p(\widehat{\boldsymbol{\beta}}_{rep}|\mathbf{X}_{init}, \boldsymbol{\beta}_{init}, \phi_g) p(\boldsymbol{\beta}_{rep}|\widehat{\boldsymbol{\beta}}_{rep}, \mathbf{X}_{init}, \phi) d\boldsymbol{\beta}_{init} d\widehat{\boldsymbol{\beta}}_{rep}$$
$$= \int \underbrace{MVN(\widehat{\boldsymbol{\beta}}_{rep}|\widehat{\boldsymbol{\beta}}_{init}, \boldsymbol{\Sigma}_{init} + \boldsymbol{\Sigma}_{rep})}_{p(\widehat{\boldsymbol{\beta}}_{rep}|\widehat{\boldsymbol{\beta}}_{init}, \mathbf{X}_{init}, \phi)} \times \underbrace{MVN(\boldsymbol{\beta}_{rep}|\widehat{\boldsymbol{\beta}}_{rep}, \boldsymbol{\Sigma}_{rep})}_{p(\boldsymbol{\beta}_{rep}|\widehat{\boldsymbol{\beta}}_{rep}, \widehat{\boldsymbol{\beta}}_{init}, \mathbf{X}_{init}, \phi)} d\widehat{\boldsymbol{\beta}}_{rep}$$

As $\boldsymbol{\Sigma}_{rep}$ is a function of $\widehat{\boldsymbol{\beta}}_{rep}$, integration over the latter cannot be carried out exactly in the expression above. Hence, we approximate $p(\boldsymbol{\beta}_{rep}|\widehat{\boldsymbol{\beta}}_{init}, \mathbf{X}_{init}, \phi)$ by plugging an analytic approximation to $p(\widehat{\boldsymbol{\beta}}_{rep}|\boldsymbol{\beta}_{rep}, \widehat{\boldsymbol{\beta}}_{init}, \mathbf{X}_{init}, \phi)$ into the Bayes theorem:

$$\begin{aligned} p(\boldsymbol{\beta}_{rep}|\widehat{\boldsymbol{\beta}}_{init}, \mathbf{X}_{init}, \phi) &= \frac{p(\boldsymbol{\beta}_{rep}, \widehat{\boldsymbol{\beta}}_{rep}|\widehat{\boldsymbol{\beta}}_{init}, \mathbf{X}_{init}, \phi)}{p(\widehat{\boldsymbol{\beta}}_{rep}|\boldsymbol{\beta}_{rep}, \widehat{\boldsymbol{\beta}}_{init}, \mathbf{X}_{init}, \phi)} \\ &\approx \frac{p(\widehat{\boldsymbol{\beta}}_{rep}|\widehat{\boldsymbol{\beta}}_{init}, \mathbf{X}_{init}, \phi) p(\boldsymbol{\beta}_{rep}|\widehat{\boldsymbol{\beta}}_{rep}, \widehat{\boldsymbol{\beta}}_{init}, \mathbf{X}_{init}, \phi)}{p_{approx}(\widehat{\boldsymbol{\beta}}_{rep}|\boldsymbol{\beta}_{rep}, \widehat{\boldsymbol{\beta}}_{init}, \mathbf{X}_{init}, \phi)} \\ &= \frac{p\left(\overline{\widehat{\boldsymbol{\beta}}}_{rep}|\widehat{\boldsymbol{\beta}}_{init}, \mathbf{X}_{init}, \phi\right) p\left(\boldsymbol{\beta}_{rep}|\overline{\widehat{\boldsymbol{\beta}}}_{rep}, \widehat{\boldsymbol{\beta}}_{init}, \mathbf{X}_{init}, \phi\right)}{p_{approx}\left(\overline{\widehat{\boldsymbol{\beta}}}_{rep}|\boldsymbol{\beta}_{rep}, \widehat{\boldsymbol{\beta}}_{init}, \mathbf{X}_{init}, \phi\right)} \end{aligned}$$

Eq. 34.

The approximate density $p_{approx}(\widehat{\boldsymbol{\beta}}_{rep}|\boldsymbol{\beta}_{rep}, \widehat{\boldsymbol{\beta}}_{init}, \mathbf{X}_{init}, \phi)$ may be derived by reorganizing the joint posterior $p(\widehat{\boldsymbol{\beta}}_{rep}, \boldsymbol{\beta}_{rep}|\widehat{\boldsymbol{\beta}}_{init}, \mathbf{X}_{init}, \phi) = p(\widehat{\boldsymbol{\beta}}_{rep}|\widehat{\boldsymbol{\beta}}_{init}, \mathbf{X}_{init}, \phi) p(\boldsymbol{\beta}_{rep}|\widehat{\boldsymbol{\beta}}_{rep}, \widehat{\boldsymbol{\beta}}_{init}, \mathbf{X}_{init}, \phi)$ in the

usual manner one prepares to carry out the convolution of the two Gaussians :

$$\overbrace{MVN\left(\widehat{\boldsymbol{\beta}}_{rep}|c(\widehat{\boldsymbol{\beta}}_{init},\boldsymbol{\beta}_{rep}),\left((\boldsymbol{\Sigma}_{init}+\boldsymbol{\Sigma}_{rep})^{-1}+\boldsymbol{\Sigma}_{rep}^{-1}\right)^{-1}\right)}^{p_{approx}(\widehat{\boldsymbol{\beta}}_{rep}|\boldsymbol{\beta}_{rep},\widehat{\boldsymbol{\beta}}_{init},\mathbf{X}_{init},\phi)} \times \overbrace{MVN(\boldsymbol{\beta}_{rep}|\widehat{\boldsymbol{\beta}}_{init},\boldsymbol{\Sigma}_{init}+2\boldsymbol{\Sigma}_{rep})}^{R(\boldsymbol{\beta}_{rep}|\widehat{\boldsymbol{\beta}}_{rep},\widehat{\boldsymbol{\beta}}_{init},\mathbf{X}_{init},\phi)}$$

$$c(\widehat{\boldsymbol{\beta}}_{init},\boldsymbol{\beta}_{rep}) = \left((\boldsymbol{\Sigma}_{init}+\boldsymbol{\Sigma}_{rep})^{-1}+\boldsymbol{\Sigma}_{rep}^{-1}\right)^{-1}\left(\boldsymbol{\Sigma}_{rep}^{-1}\boldsymbol{\beta}_{rep}+(\boldsymbol{\Sigma}_{init}+\boldsymbol{\Sigma}_{rep})^{-1}\widehat{\boldsymbol{\beta}}_{init}\right)$$

In the equation above, the factor on the left identified as $p_{approx}(\widehat{\boldsymbol{\beta}}_{rep}|\boldsymbol{\beta}_{rep},\widehat{\boldsymbol{\beta}}_{init},\mathbf{X}_{init},\phi)$ is an approximation, because the remainder factor on the right is not free of $\widehat{\boldsymbol{\beta}}_{rep}$ as the matrix $\boldsymbol{\Sigma}_{rep}$ is a function of this quantity.

To complete the derivation in Eq. 34, we must apply a heuristic argument to choose a value for $\overline{\widehat{\boldsymbol{\beta}}}_{rep}$ to ground the approximation: First, we apply the Laplace Approximation to integrate $p_{approx}(\widehat{\boldsymbol{\beta}}_{rep}|\boldsymbol{\beta}_{rep},\widehat{\boldsymbol{\beta}}_{init},\mathbf{X}_{init},\phi) \times R(\boldsymbol{\beta}_{rep}|\widehat{\boldsymbol{\beta}}_{rep},\widehat{\boldsymbol{\beta}}_{init},\mathbf{X}_{init},\phi)$ over $\boldsymbol{\beta}_{rep}$ by selecting the value of the latter variable that maximizes the remainder term, i.e. $\boldsymbol{\beta}_{rep} = \widehat{\boldsymbol{\beta}}_{init}$. After integration we obtain the marginal density $MVN\left(\widehat{\boldsymbol{\beta}}_{rep}\Big|\widehat{\boldsymbol{\beta}}_{init},\left((\boldsymbol{\Sigma}_{init}+\boldsymbol{\Sigma}_{rep})^{-1}+\boldsymbol{\Sigma}_{rep}^{-1}\right)^{-1}\right)$. The approximate mode of this density is found at $\widehat{\boldsymbol{\beta}}_{init}$, which is thus the most likely value for $\widehat{\boldsymbol{\beta}}_{rep}$. Since we would like the approximation in Eq. 34 to be as accurate as possible near the most likely value of $\widehat{\boldsymbol{\beta}}_{rep}$, we set $\overline{\widehat{\boldsymbol{\beta}}}_{rep} = \widehat{\boldsymbol{\beta}}_{init}$ in Eq.34 and after simplification of the product of the square roots of the determinants and cancelation of common terms we obtain:

$$p(\boldsymbol{\beta}_{rep},|\mathbf{y}_{init},\mathbf{X}_{init},\phi) = p(\boldsymbol{\beta}_{rep},|\widehat{\boldsymbol{\beta}}_{init},\mathbf{X}_{init},\phi) \approx MVN(\widehat{\boldsymbol{\beta}}_{init}, 3 \times \boldsymbol{\Sigma}_{init}) \qquad \text{Eq. 35.}$$

Hence, under the conditions of exact translation and replication with a known scale parameter, the predictive density for the inferences in the replicate study is a multivariate normal that is centered on the point estimate from the present study but has a covariance matrix that is a scalar multiplicate of the covariate matrix obtained in the initial study. This predictive density can be used to anticipate the decisions made in the replicate study by calculating π-values and subjecting them to the type of threshold analysis discussed in Section 4. The predictive π-value for an exact replication is related to the observed π-value of the initial study according to the formula: $\pi_{rep} = 2\Phi(\Phi^{-1}(\pi_{init}/2)/\sqrt{3})$. Using the 0.05 value to threshold p and π-values, we see that only an initial p-value of 0.0006 or less implies a predictive π-value of less than the threshold.

While Eq.33 assumes that both the initial and the replicate data will be analyzed from a Bayesian perspective it also allows mutatis mutandis a non-Bayesian framework for the analysis of replicability. From a non-Bayesian viewpoint, a predictive distribution for inferences in replicate datasets is given by $p(\widehat{\boldsymbol{\beta}}_{rep}|\mathbf{X}_{init}, \mathbf{X}_{rep}, \boldsymbol{\beta}_{init}, \hat{\phi}_{init}, \boldsymbol{\beta}_{rep}, \hat{\phi}_{rep})$ and both translation and multivariate integration now involve $\widehat{\boldsymbol{\beta}}_{init}$ rather than $\boldsymbol{\beta}_{init}$. The product $p(\boldsymbol{\beta}_g|\widehat{\boldsymbol{\beta}}_{init})p(\phi_g|\hat{\phi}_{init})$, provides the mechanism for translating estimators from the initial study to parameters of the generative model of the replicate data. Like the Bayesian case, setting each of these two probability density functions to a Dirac delta function corresponds to *exact* translation. Based on the material presented in Sections 2 and 3, under the (locally) uniform priors $p(\boldsymbol{\beta}_{init}, \phi_{init}|\boldsymbol{y}_{init}, \mathbf{X}_{init}) \approx p(\widehat{\boldsymbol{\beta}}_{init}, |\mathbf{X}_{init}, \boldsymbol{\beta}_{init}, \hat{\phi}_{init})$ and $p(\boldsymbol{\beta}_{rep}, \phi_{rep}|\mathbf{X}_{rep}, \boldsymbol{y}_{rep}) \approx p(\widehat{\boldsymbol{\beta}}_{rep}|\mathbf{X}_{rep}, \boldsymbol{y}_{rep}, \boldsymbol{\beta}_{rep}, \hat{\phi}_{rep})$, with exact equality for quadratic likelihoods and known scale parameters. Hence Eq.33 and the computations presented in Eq. 34 and 35 can be used to analyze replication under either Bayesian or non-Bayesian predictive probability framework given the close numerical agreement of the probabilities used for inference under these two different schools of statistical thought.

### 5.2. P and π– values are poorly reproducible

The predictive framework of Section 5.1 utilizes the joint predictive distribution of replicate data and replicate parameters. An alternative is to use a *marginal predictive framework*, in which the posterior probability for the parameters of *each* replicated dataset is used as input for downstream applications e.g. to calculate π-values. This marginal framework is implicitly used in all prior (Bayesian) analyses of the reproducibility probability of p-values(Goodman 1992; Shao and Chow 2002). We now describe a simulation-based approach for the analysis of reproducibility of the p and π-values under this framework before providing approximate formulas for models with known or precisely estimated scale parameters. This hierarchical simulation strategy will be used in the analysis of the clinical trial datasets considered in Section 6 and provide a straightforward way to check the saddlepoint and quadratic likelihood approximations used in this paper by sampling from the relevant "exact" (MCMC generated) or approximate distributions as shown in the schema below:

$$
\begin{aligned}
\boldsymbol{\beta}_{init}, \phi_{init} | \mathbf{y}_{init}, \mathbf{X}_{init} &\sim p(\boldsymbol{\beta}_{init}, \phi_{init} | \mathbf{y}_{init}, \mathbf{X}_{init}) \approx \\
&\quad MVN(\boldsymbol{\beta}_{init} | \widehat{\boldsymbol{\beta}}_{init}, \boldsymbol{\Sigma}_{init}) \times \text{Inv-}\chi^2\left(n-p, \frac{D(\mathbf{y}_{init}; \widehat{\boldsymbol{\mu}}_{init})}{n-p}\right) \\
\boldsymbol{\beta}_g, \phi_g | \boldsymbol{\beta}_{init}, \phi_{init} &\sim p(\boldsymbol{\beta}_g | \boldsymbol{\beta}_{init}) p(\phi_g | \phi_{init}) \\
\mathbf{y}_{rep} | \mathbf{X}_{rep}, \boldsymbol{\beta}_g, \phi_g &\sim f(\mathbf{y}_{rep} | \mathbf{X}_{rep}, \boldsymbol{\beta}_g, \phi_g) \\
\boldsymbol{\beta}_{rep}, \phi_{rep} | \mathbf{X}_{rep}, \mathbf{y}_{rep} &\sim p(\boldsymbol{\beta}_{rep}, \phi_{rep} | \mathbf{X}_{rep}, \mathbf{y}_{rep}) \\
&\approx MVN(\boldsymbol{\beta}_{rep} | \widehat{\boldsymbol{\beta}}_{rep}, \boldsymbol{\Sigma}_{rep}) \times \text{Inv-}\chi^2\left(n-p, \frac{D(\mathbf{y}_{rep}; \widehat{\boldsymbol{\mu}}_{rep})}{n-p}\right) \\
\pi_{rep,i} &= 2 \times min(\{P(\beta_{rep,i} \geq \beta_{i,0} | \mathbf{y}_{rep}, \mathbf{X}_{rep}), P(\beta_{rep,i} < \beta_{i,0} | \mathbf{y}_{rep}, \mathbf{X}_{rep})\}) \\
&\approx p_{rep,i} \\
p_{rep,i} &= 2 \times \Phi\left(-\frac{|\hat{\beta}_{rep,i} - \beta_{i,0}|}{\sqrt{\boldsymbol{\Sigma}_{rep}[i,i]}}\right) = 2 \times \Phi(-|\hat{z}_{rep,i}|)
\end{aligned}
$$
Eq. 36.

An analytical approximation for the distribution of the p and π-values in the replicate datasets under the exact translation and replication setup considered in Section 5.2 may be derived by first computing the posterior marginal (over $\boldsymbol{\beta}_{init}$) distribution $\widehat{\boldsymbol{\beta}}_{rep} | \mathbf{X}_{rep}, \mathbf{y}_{rep}, \phi_{rep} \sim MVN(\widehat{\boldsymbol{\beta}}_{init}, 2\boldsymbol{\Sigma}_{init})$, followed by integrating out all components except the i[th] one to obtain: $\hat{\beta}_{rep,i} \sim N(\hat{\beta}_{rep,i}, \sqrt{2 \times \boldsymbol{\Sigma}_{rep}[i,i]})$. The distribution of the corresponding test statistic is $\hat{z}_{rep,i} \sim N(\hat{z}_{init,i}, \sqrt{2})$. A straightforward change of random variables from $\hat{z}_{rep,i} \rightarrow \pi_{rep,i}$ then yields the distribution of the π-value. Previous work (Lambert and Hall 1982; Shao and Chow 2002; De Martini 2008; Boos and Stefanski 2011; De Capitani and De Martini 2016) has shown that the natural scale to consider the variability of p and thus values is a logarithmic one. Furthermore, our decimal system suggests that the base of the logarithmic scale to use is the base 10; in that scale, $\pi_{rep,i} = x \cdot 10^{-k} \rightarrow -\log_{10} \pi_{rep,i} = k - \log_{10} x$. Hence:

$$-\log_{10} p_{rep,i} = -\log_{10} \pi_{rep,i} \sim 10^{-\log_{10} \pi_{rep,i}} \exp\left(\mathrm{erfc}^{-1}\left(10^{-\log_{10} \pi_{rep,i}}\right)^2\right) \log 10 \sqrt{\frac{\pi}{2}} \times$$
$$\left[ N\left(\Phi^{-1}\left(\frac{10^{-\log_{10} \pi_{rep,i}}}{2}\right) \middle| \Phi^{-1}\left(\frac{10^{-\log_{10} \pi_{init,i}}}{2}\right), \sqrt{2}\right) + \right.$$
$$\left. N\left(-\Phi^{-1}\left(\frac{10^{-\log_{10} \pi_{rep,i}}}{2}\right) \middle| \Phi^{-1}\left(\frac{10^{-\log_{10} \pi_{init,i}}}{2}\right), \sqrt{2}\right) \right]$$

Eq. 37.

The expectation of the replicate p-values from Eq. 37 in the logarithmic and untransformed scale are compared to the predictive p-values from Section 5.2 in Figure 3A. The predictive approach leads to a more conservative assessment of the replication probability. The RPD of Eq. 37 is graphed in Figure 3B: the distribution gradually changes to a skewed "lognormal" shape as the initial p-value gets smaller ($-\log_{10} \pi_{rep,i}$ becomes more positive). Even after a "lognormality" has been attained, the relevant density will be rather diffuse, e.g., for p-values = 0.00001, there will be a considerable proportion of replicated p-values that are > 0.05. In Figures 3C and 3D we graph the relationship between the standard deviation and the mean in either the log10 or the untransformed scale. We observe that the standard deviation of the RPD is a sizeable fraction of the mean in log10 scale and is larger than the mean in the untransformed scale. The large variance of the RPD clearly illustrate that P and π-values exhibit excessive variation as random variables and their numerical values are poorly reproducible even when the initial p-value would suggest that findings should replicate. In the next section we will demonstrate the use of these concepts for the analysis of reproducibility in clinical trials, with R code that illustrates the programming of the relevant analyses provided in the Appendix.

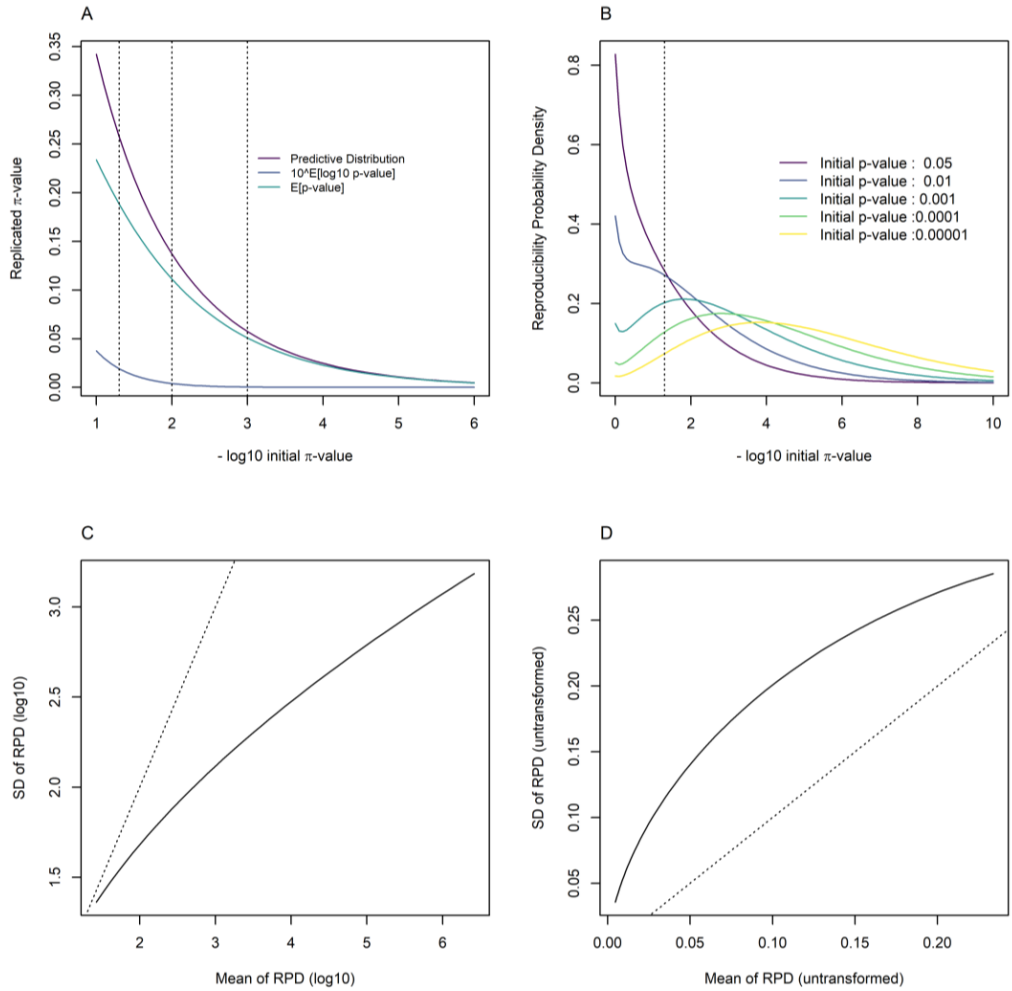

*Figure 4 (A) Relation between the predictive π-value in the replicate dataset and the π-value in the initial experiment (-log10 scale) for the predictive distribution and the expected values of the of the RPD computed in the log10 and untransformed scale. The vertical lines from left to right correspond to conventional levels of statistical significance, 0.05, 0.01, 0.001 (B) Replication probability density for different initial p values (vertical line corresponds to p=0.05) (C) Standard deviation vs mean of the RPD in the log10 scale (D) Standard deviation vs mean of the RPD in the untransformed scale. In (C) and (D) the dotted line is the line of identity*

## 6. Reproducibility in the SGLT2i kidney outcome clinical trials

Sodium Glucose Co-transporter 2 inhibitors (SGLT2i) were initially introduced as a class of modest antiglycemic agents that could affect a reduction in Glycated Hemoglobin A1c of 0.81 to 1.02% in newly treated patients with Diabetes type 2 and 0.57 to 0.63% when added to a metformin background (Tsapas et al. 2020). The target of these drugs is a protein expressed inside the kidney (the SGLT2 cotransporter) that is normally tasked with reclaiming the glucose molecules that are non-selectively filtered by the glomeruli, i.e., the small filtration units inside the kidneys, and returning the glucose to the circulation along with sodium. In the presence of diabetes, the excessively circulating glucose amounts leads to overactivation of the SGLT2 system, pathologic increase in the filtration function of the kidney which initiates the development of Chronic Kidney Disease (CKD) as the glomeruli inside the kidney are exhausted. Experiments over the last two and half decades involving naturally occurring SGLT2i (phlorizin(Vallon et al. 1999)) and diabetic mice lacking the SGLT2 transporter(Vallon et al. 2012) suggested that SGLT2i may in fact be beneficial in slowing the progression of CKD. Two human studies were undertaken with to test the effects of specific SGLT2i CKD: the CREDENCE (Perkovic et al. 2019) trial of canagliflozin involving patients with diabetes type 2 and the DAPA-CKD (Heerspink et al. 2020) of dapagliflozin that included patients with and without diabetes 2. The primary outcome in both studies was a composite end point of time to a) reduction of kidney function from baseline (estimated glomerular filtration rate, eGFR (reference range 90-120 ml/min/1.73m2) decline by 50% in DAPA-CKD or doubling of serum creatinine in DAPA-CKD ), b) the onset of End Stage Kidney Disease (need for dialysis, transplantation or eGFR < 15 ml/min/1.73m2) or c) death from cardiovascular or renal causes. The results about the primary outcome and a rare side effect (Diabetic Ketoacidosis, DKA) from these two trials are summarized in Table 5. In this table we also calculate a relative risk from a Poisson GLM fit to the summary data from these trials. The Poisson model estimates are near identical to those provided by the proportional hazards model used in the primary analyses.

Table 5 Primary outcome and Diabetic Ketoacidosis (DKA) events in the CREDENCE (Canagliflozin) and DAPA-CKD (dapagliflozin) kidney outcome trials of Sodium Glucose Cotransporter Two inhibitors (SGLT2i). Table shows the hazard ratio reported in the trials, number of events and event rates (not reported in the DAPA-CKD trial for the DKA outcome)

| Outcome | Trial | Relative Risk* | Hazard Ratio | SGLT2i N / Total N | Placebo N / Total N | SGLT2i Events/ 1000 PYs | Placebo Events/ 1000 PYs |
|---|---|---|---|---|---|---|---|
| Primary (1o) | CREDENCE | 0.71 (0.60 – 0.83) | 0.70 (0.59 – 0.82) | 245/2202 | 340/2199 | 43.2 | 61.2 |
| DKA | CREDENCE | 7.79 (1.48 – 41.09) | 10.8 (1.39 – 83.65) | 11/2200 | 1/2197 | 2.2 | 0.2 |
| Primary (1o) | DAPA-CKD | 0.61 (0.51 – 0.73) | 0.61 (0.51 – 0.72) | 197/2152 | 312/2152 | 46 | 75 |
| DKA | DAPA-CKD | 0.00 (0.00 – ∞) | – | 0/2149 | 2/2149 | – | – |

* Calculated by the authors using a Poisson regression model using as offset the a) follow up implied by the events/event rates reported by the study investigators (primary outcome in CREDENCE/DAPA-CKD and DKA in CREDENCE) b) the size of each study arm (DKA in DAPA-CKD). The point estimate and the standard error for the latter regression in log scale was -23 (42247.17), yielding the exact point estimate $1.03 \times 10^{-10}$ and 95% confidence interval of $3.16 \times 10^{-35972}$ - $3.36 \times 10^{35951}$. PY: Person-Years

In Figure 5 we show the contours of the log-likelihood around the ML estimate for the coefficients (log-relative risks) of the Poisson regression. For the primary outcome in both CREDENCE and DAPA-CKD, these contours are ellipsoids in the plane, at least up to 4 standard errors in each axis. For the DKA outcome different patterns emerge: a) in CREDENCE, the ellipsoid pattern does not quite hold true: the distance between the isocontours of probability differs above and below the main axis of the ellipsoid running from top left to bottom right corner b) in DAPA-CKD, the isocontours are nowhere near to being elliptical: in fact there appears to be a sharp discontinuity corresponding to a relative risk of 1.0 for those assigned to the SGLT2i arm. When the actual likelihoods are visualized in 3D (Figure 6), we see that those for the 1o outcome in either study assume the familiar form of a bivariate Gaussian density. The likelihood for the DKA outcome in CREDENCE clearly deviates from the bivariate Gaussian form as it has an asymmetric tail evident in the bottom right corner of the plot. Finally, the likelihood for the DKA outcome in CREDENCE resembles univariate Gaussians "pasted" together. In terms of the material presented in Sections 2 and 3, we would we expect the quadratic approximation to hold very well for the analyses of the primary outcome in either study, but less so for the DKA outcome in CREDENCE and not at all for the DKA in DAPA-CKD.

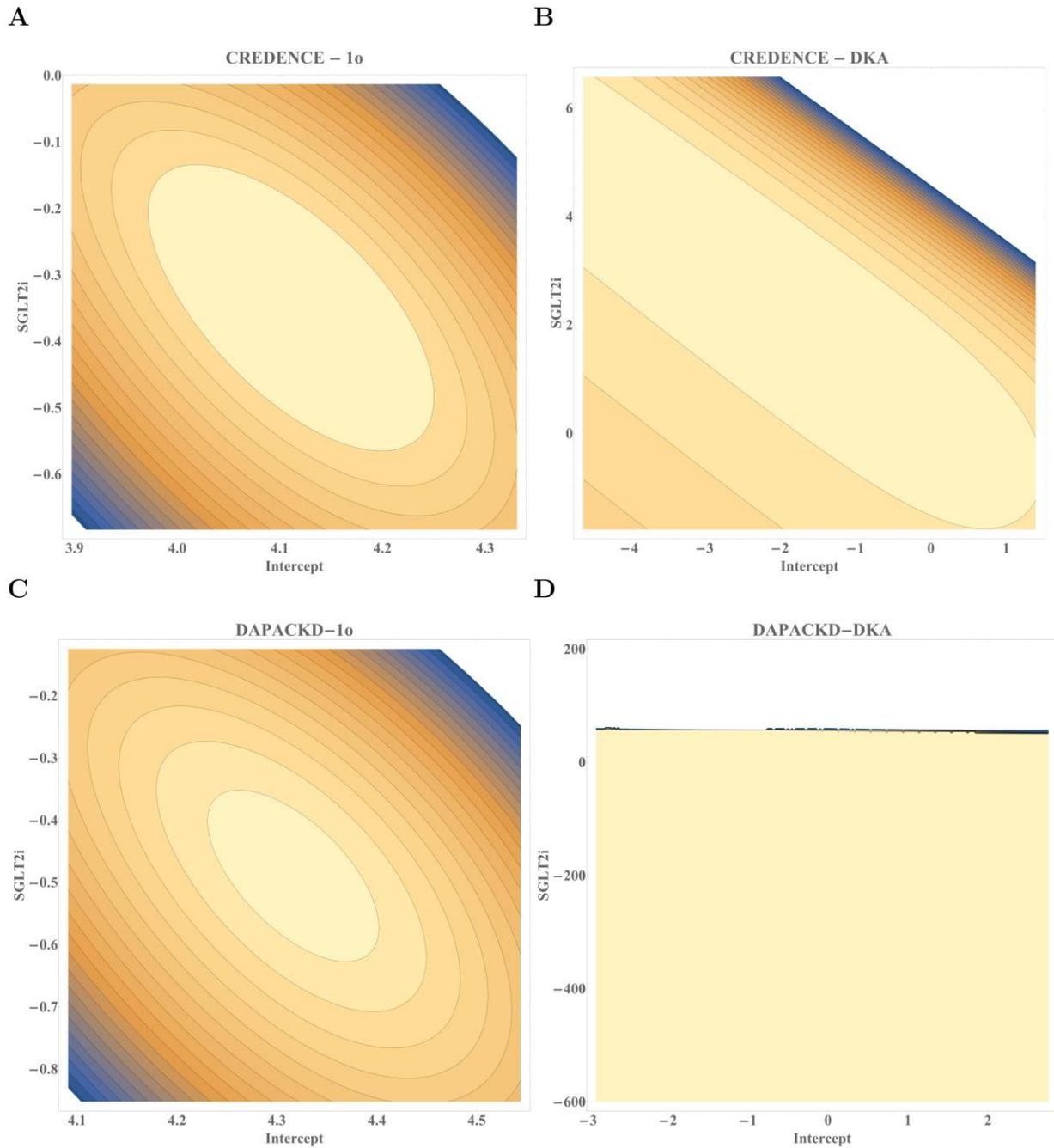

*Figure 5 Loglikelihood contours in the CREDENCE and DAPA-CKD trials for the primary (1o) outcome and Diabetic Ketoacidosis (DKA). X-axis : intercept of the Poisson model, y-axis treatment of SGLT2i therapy*

The corollary is that the Bayesian analysis will be insensitive to the choice of prior for primary outcome in either study. However, the analysis of DKA will be partially sensitive to the choice of prior for CREDENCE, and entirely dependent on the prior to yield something more useful than the entire set of the positive reals in DAPA-CKD.

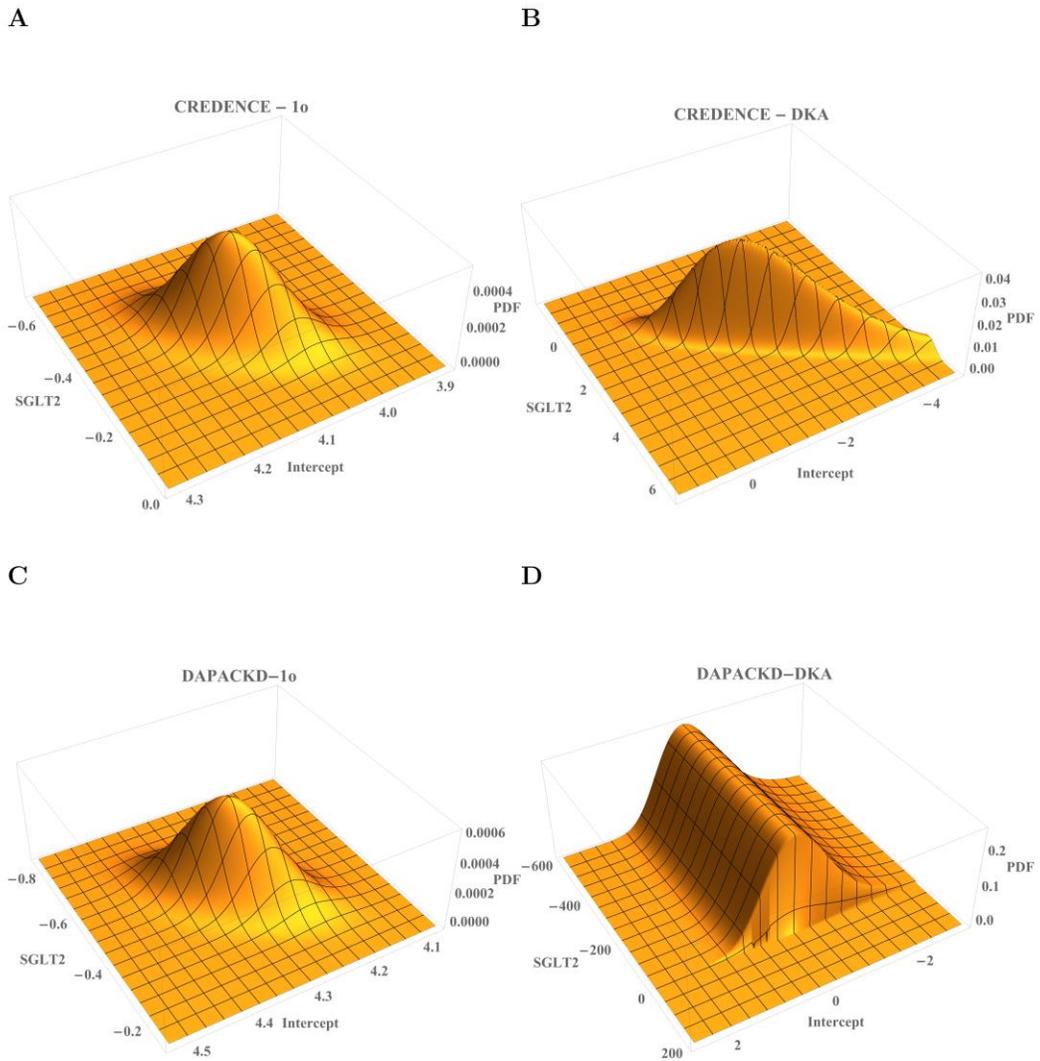

*Figure 6 Likelihoods in the CREDENCE and DAPA-CKD trials for the primary (1o) outcome and Diabetic Ketoacidosis (DKA). X-axis : intercept of the Poisson model, y-axis treatment of SGLT2i therapy*

For the Bayesian re-analyses, we adopted either a constant prior or the informative set of independent Cauchy (intercept) and Student-t with 2.5 degrees of freedom and scale parameter of 1.0 for the effect of SGLT2i. The latter prior is more informative (less diffuse) than the "weakly informative" prior of a Student-t with one degree of freedom and a scale parameter of 2.5 introduced for GLMs (Gelman et al. 2008), but less so than the standard Cauchy (nominally a Student-t with one degree of freedom and a scale of one) that is used for the intercept in that paper. Bayesian analyses were then carried out through the "No U Turn Sampler"(Hoffman and Gelman 2014) as implemented in the R package "rstanarm"(Goodrich et al. 2020). We also carried out the Bayesian analyses under the Student t prior using an approximate Expectation – Maximization (EM) algorithm as implemented in the R package "arm" (Gelman et al. 2008). For the MCMC based analyses, we simulated four independent chains for 10,000 iterations after discarding the first 10,000 samples generated by the sampler (burn-in). To compute π-values, we smoothed the MCMC samples by fitting them to finite Gaussian mixtures via the EM algorithm, and then computed tail area probabilities using the estimated mixing proportions and the component means and standard deviations. The ML and Bayesian estimates (summarized as posterior mean and posterior standard deviations are shown in Table 6 and Figure 7. As we comment below, a constant prior led to an improper posterior for the DKA outcome in DAPA-CKD and a diffuse Gaussian prior was used instead.

*Table 6 Analysis of the primary outcome and Diabetic Ketoacidosis (DKA) in the CREDENCE (Canagliflozin) and DAPA-CKD (dapagliflozin) kidney outcome trials of Sodium Glucose Cotransporter Two inhibitors (SGLT2i). Table shows the point estimate and its standard error for the maximum likelihood method (which is also the approximate posterior for quadratic log-likelihoods) and Bayesian analysis under the flat and Student-t informative priors estimated by the NUTS sampler. Also shown in an approximate Bayesian analysis based on Expectation Maximization (EM) for the Student-t prior.*

| Outcome | Trial | Maximum Likelihood | Flat Prior (NUTS) | Student-t Prior (EM) | Student-t Prior (NUTS) |
|---|---|---|---|---|---|
| **Primary (1o)** | CREDENCE | -0.3483 (0.0838) | -0.3483 (0.0835) | -0.3478 (0.0837) | -0.3449 (0.0834) |
| **DKA** | CREDENCE | 2.3990 (1.0440) | 2.9159 (1.3047) | 1.9633 (0.8084) | 1.7433 (0.8697) |
| **Primary (1o)** | DAPA-CKD | -0.4889 (0.0910) | -0.4896 (0.0904) | -0.4880 (0.0909) | -0.4831 (0.0912) |
| **DKA** | DAPA-CKD | -22.9957 (42247.1657) | -71.3650 (54.2870) | -1.2011 (1.1782) | -1.0364 (1.4168) |

For the analysis of the primary outcome in either study, the quadratic behavior of the log-likelihood takes over and its relative sharpness dominates any inferential differences between the Bayesian and non-Bayesian estimation methods, computational differences between the EM and the MCMC algorithms and the priors themselves: point estimates and standard errors are all identical up to the third decimal point. For the DKA outcome in CREDENCE, the moderate departure of the log-likelihood from the quadratic form leads to ML and EM estimates, that deviate from the estimates obtained by MCMC. The most extreme deviations are noted for the DKA outcome in CREDENCE: both the ML and the MCMC analysis under the flat prior lead to nonsensical answers. The Student-t prior effectively regularized this analysis, leading to posterior estimates that were realistic but imprecisely estimated.

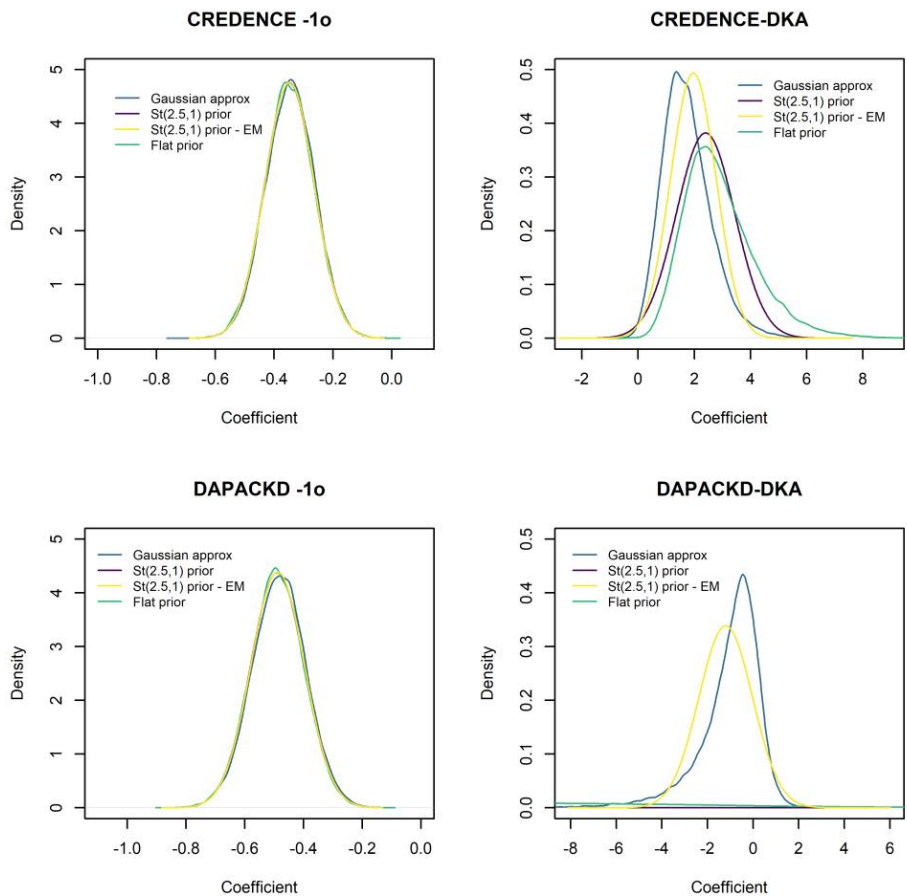

*Figure 7 Posterior probabilities for the treatment effect of SGLT2i in the CREDENCE and DAPA-CKD trials for the primary kidney specific outcome and Diabetic Ketoacidosis (DKA). Gaussian Approximation is based on the material in Sections 2-3 and coincides with the maximum likelihood distribution, Flat Prior = MCMC analysis under the constant prior, EM=Expectation Maximization.*

P and π-values are reported in Table 7 and illustrate the insensitivity of these values to the specific prior accepted and the method for carrying out the Bayesian computations. One notable disagreement is the π-value computed for the DKA outcome in CREDENCE under the flat prior; the π-value computed from the MCMC chain is 0.015, reflective of a small number of samples in which the log-hazard ratio was estimated to be greater than 0, yet the π-value from the quadratic approximation is >0.999.

Table 7 Analysis of the primary outcome and Diabetic Ketoacidosis (DKA) in the CREDENCE (Canagliflozin) and DAPA-CKD (dapagliflozin) kidney outcome trials of Sodium Glucose Cotransporter Two inhibitors (SGLT2i). Table shows the π-value for the maximum likelihood method (which is also the approximate π-value) and π-values from Bayesian analysis under the flat and Student-t informative priors estimated by the NUTS sampler. Also shown in an approximate Bayesian analysis based on Expectation Maximization (EM) for the Student-t prior.

| Outcome | Trial | Maximum Likelihood | Flat Prior (NUTS) | Student-t Prior (EM) | Student-t Prior (NUTS) |
|---|---|---|---|---|---|
| **Primary (1o)** | CREDENCE | 3.23 $10^{-5}$ | 3.05 $10^{-5}$ | 3.28 $10^{-5}$ | 3.5 $10^{-5}$ |
| DKA | CREDENCE | 0.022 | 0.002 | 0.015 | 0.016 |
| **Primary (1o)** | DAPA-CKD | 7.79 $10^{-8}$ | 6.16 $10^{-8}$ | 7.97 $10^{-8}$ | 11.7 $10^{-8}$ |
| DKA | DAPA-CKD | 0.999 | 0.015 | 0.308 | 0.343 |

To probe these discrepancies, we analytically integrated the likelihood of the DKA outcome in DAPA-CKD for the intercept parameter of the Poisson GLM, leading to the function $\left(1 + e^{\beta_1}\right)^{-2}$. Integration over $\beta_1$ (the treatment effect for dapagliflozin) will recover the normalization constant of the posterior; however, the integral $\int_{-\infty}^{\infty}\left(1 + e^{\beta_1}\right)^{-2} d\beta_1$ is divergent, and hence the posterior is *improper*. When the Student-t prior is introduced, the integral of the likelihood × prior no longer diverges, and univariate numerical integration yields the π-value of 0.334, which interpolates the corresponding entries for the DKA in DAPA-CKD in Table 7. However, the precise numeric value of these π-values is determined by the scale of the Student-t distribution and can be made arbitrarily small by selection of the value of the scale (not shown): when the data is limited in quantity, the Bayesian posterior is strongly affected by the prior, while non-Bayesian analysis and Bayesian analysis with non-regularizing priors will yield numerical nonsense.

Referring to Section 5.1, the predictive π-values for the primary outcome in CREDENCE and DAPA-CKD are 0.0164 and 0.0019 respectively, both below the nominal threshold of 0.05. We also undertook an analysis of replication under the marginal predictive framework of Section 5.2, utilizing the simulation schema in Eq 36. Specifically, we conducted a Bayesian analysis under the flat prior and generated samples from the posterior via MCMC. Each of these posterior samples were then used to generate Poisson distributed counts, which were then analyzed from a ML or a Bayesian perspective under a flat or a Student-t prior. A single draw from the posterior for each of these synthetic datasets was then sampled and the corresponding p-value was computed for each synthetic dataset. Non-parametric kernel density estimates of the treatment effect for the SGLT2i and the p/π-value in $-\log_{10}$ space were then compared against the approximations given in Eq. 35 and 37.

These analyses shown in Figure 8, illustrate that the closed formulas for the predictive distribution of the treatment effect and the p/π-value introduced in Eq. 35 and 37 are rather good approximations to the actual predictive density of these quantities (simulated exactly using Eq. 36) when the log-likelihood is well approximated via a quadratic function (primary outcome in CREDENCE/DAPA-CKD). For the DKA outcome in CREDENCE, the empirical predictive distribution of the treatment effect obtained from 5,000 replicate simulations had a mean (SE) of -7.15 (30,000) as many of those datasets had two few observations to analyze and estimates were often at the boundary of the parameter space. As a result, the p-values in the replicated datasets were rather diffusely distributed around the initial p value. The analytical approximation in. Eq 37 did not capture the tall "spike" of p-values close to one (Figure 8, middle row, right column). Bayesian analysis with either the flat or the Student-t prior could effectively regularize the analysis of these replicate datasets with these replicate π-values closely distributing around the initial p-value.

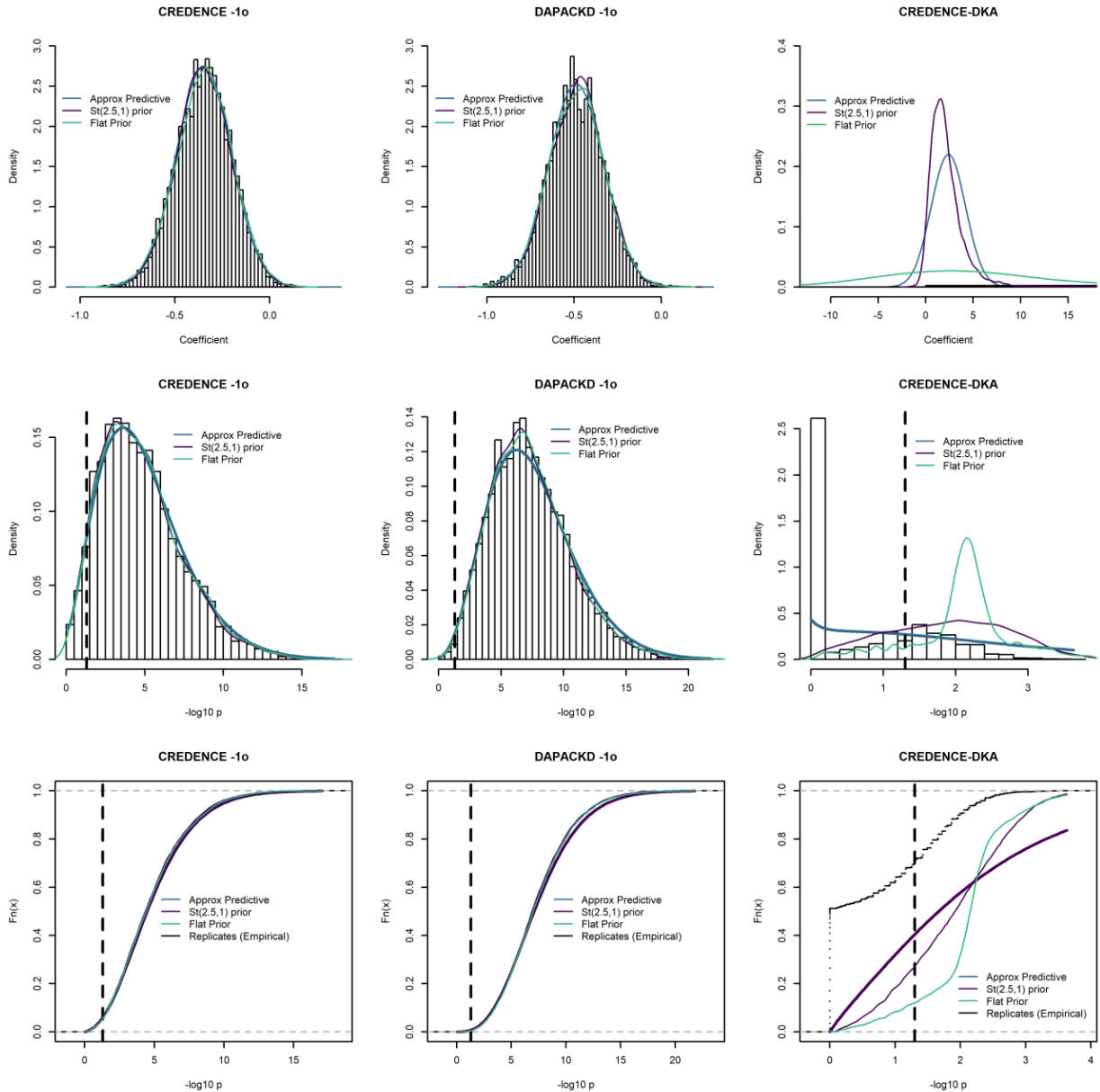

*Figure 8 Replication analysis of treatment effect in the SGLT2 kidney outcome trials. Top row , histograms of the predictive distribution of the treatment in the exact translated and replicated trials, approximate predictive density (Eq. 35) and non parametric kernel density estimates of replicated Bayesian analyses under two different priors. Middle row: histogram of observed p-values in the replicated trials, approximate predictive density of p and π-values (Eq. 37) and π-values obtained in replicate Bayesian analyses (MCMC – based on the NUTS sampler) using the schema in Eq. 36. For the Bayesian analysis, posterior samples from in replicated study were fit to a finite Gaussian mixture model and the parameters of the letter were used to calculate tail area probabilities. Bottom row cumulative density function of the p and π value based in Eq. 37. The vertical dashed lined in the two rows indicate the threshold of p=0.05.*

# 7. A Bayesian / Likelihoodist synthesis aiming for replication

The use of statistical significance testing is frequently and repeatedly e.g. every 10-15 years identified as the culprit underlying both real and imaginary failures of the scientific enterprise of the *users* of statistical tests(Carver 1978; Chow 1988; Harris 1997; Nickerson 2000; Shao and Chow 2002; Wasserstein et al. 2019; Gibson 2021). In the most current iteration, reliance of p-values thresholded at the "0.05" level have been implicated as a particular important cause of the replication crisis in science(Siegfried 2010; Trafimow and Marks 2015; Gibson 2021), followed by suggestions to lower the threshold of significance to "<0.005"(Benjamin et al. 2018), more than 130 years after Edgeworth's use of this value (Kennedy-Shaffer 2019), and eventually by the proposal to ban the use of the term "statistically significant" (Wasserstein et al. 2019) in the rejoinder of 43 articles published by the American Statistical Association about the misuses of p-values in statistical inference. This proposal has yet to be implemented and considering the pushback it received in editorials published in medical(Harrington et al. 2019) and clinical trial journals(Cook et al. 2019) is unlikely to ever be adopted, possibly because "complicated mathematical procedures lend an air of scientific objectivity to conclusions"(Carver 1978). During this debate the defenders of the p-values have not remained silent but identified the misuse of tests of statistical significance, rather than the tests per se as the real culprits. Such misuses stem from "failure of textbooks and tutorials to describe correctly the inferential meaning of p-values"(Greenland 2019), essentially a "human-factors problem" that could be addressed by "developing better evidence-based education and user-centered statistical software"(Lakens 2021). Short (Senn 2001; Greenland 2019; Wasserstein et al. 2019) and much longer lists of proscriptive don'ts (Nickerson 2000; Greenland et al. 2016) have also pointed out two main themes of user failures: the habitual misinterpretation of p-values as hypothesis or error probabilities, and the focus on repeated sampling properties of the p-value as a random variable, rather than the single value observed in a given analysis. It may very well be that "the key problem is that there are no interpretations of these concepts that are at once simple, intuitive, correct and foolproof"(Greenland et al. 2016) and that the "high cognitive demand has led to an epidemic of shortcut definition and interpretations". However, the mere fact that these issues with the p-values keep recurring with predictable frequency every 10-15 years, suggests that the p-value as an inferential apparatus is alien to human intellect and may indeed be quite an un-natural way to conduct inference for most scientists. Notwithstanding these observations, a significant amount of scientific research has been achieved using p-values, so there is apparent merit in the *numerical* calculation of these values when conducting scientific inference despite their disastrous abuse. The apparent discrepancy between the utility of the numerical computation and the systematic abuse of the relevant concept begs for an answer. We feel that the answer is obtainable if we discard the construct of the frequentist p-value and attach a drastically different, Bayesian interpretation to the number computed and reported as p in the Generalized Linear Model (GLM). We show that a Bayesian working under the same set of mathematical assumptions as a non-Bayesian to analyze data via a GLM will arrive at posterior distributions that are either identical or asymptotically equivalent to each other. We then establish the rationale for computing and a decision analytic framework for thresholding posterior tail areas ($\pi$-values) which for any given Bayesian analysis will be numerically identical to the non-Bayesian p-values. We conclude by proposing a predictive probability argument to reason about reproducibility that not only is more general than the replication probabilities of p and $\pi$-values but illustrates the severe shortcomings of relying on the latter to gauge reproducibility of research findings. As is customary in this area, we will issue a list of intertwined "dos" and "don'ts" for users of statistics afforded by the Bayesian $\pi$-value and their relationship to prior uses of the p-value while highlighting the technical requirements to reframe a reported p value as a $\pi$-value. We conclude by discussing opportunities to improve reporting of statistical analyses and the statistical exploration of reproducibility in publications and scientific discovery.

1. **DO use π-values to quantify directional (univariate) effects.** A Bayesian can quantify directional effects by calculating the posterior expectation of the *sign* of the parameter of interest, effectively mapping the range of the latter to the interval $[-1,1]$. The smaller the π-value, the larger the magnitude of the directional effect. Practitioners of the medical arts will be quick to recognize that probing directional effects and acting on them is aligned with the ethical principle of "to help, or at least to do no harm" the original formulation of the "primum non nocere" principle in medical ethics (Smith 2005). In a drug development context, a regulatory authority may be content to know that a novel therapy is more effective than a comparator treatment before granting its approval. Hence continuing to report p-values (which the audience should be free to interpret as π) in medical and clinical trial journals should continue.

2. **DO use π-values to make ordinal claims about directional effects.** This has been a major market for the p-value in applied research as has been repeatedly pointed out by previous commentators (Chow 1988; Harris 1997; Nickerson 2000; Lakens 2021). To see why, note that while one tests for absence of effects, no user has ever interpreted a statistically significant test in isolation of the *observed* direction of the effect. For example, when a significant p-value is reported in a *superiority* controlled clinical trial, no commentator will ever conclude that the tested intervention yielded data are just different (in a statistical sense) than those obtained in the controlled arm. They are expected to use the observed p-value to comment on the directionality of the observed effect and this is what is invariably done in scientific publications and communications. However, null hypothesis testing does not formally allow such directional statements to be made, only divisional hypothesis tests do. Aligning what scientists report, with the statistical procedures they actually perform, would require a radically different scheme for the reinterpretation of p-values via application of a three-valued logic(Harris 1997). In this approach the observed p-value is attached to statements about the magnitude of the parameter of interest as smaller, too close to call(Hunter 1997) or larger than the value specified in the null hypothesis. Such a schema was put forward more in the early 1960s (Kaiser 1960) as an extension of the Neyman-Pearson approach (in passing note the Bayesian literature has been a relative late-comer in the formal analysis of errors of direction, labelled as "S" errors (Gelman and Tuerlinckx 2000; Gelman and Carlin 2014; Lu et al. 2019)). In the non-Bayesian three value logic/directional inference framework, one can make three different errors: claiming a difference when there is none (alpha error), not finding a difference when one exists (beta error) or claiming a difference in the wrong direction (gamma error). Implementation of such an approach leads to split tailed significance tests (Braver 1975; Harris 1997) that mathematically combine the two possible one-sided tests that test divisional hypotheses. While adoption of a three valued logic could re-align interpretation and statistical procedure, we feel that our proposal for a framework that involves the "recalibrated" analyst provides a Bayesian decision theoretic justification for the behavior of scientists who use p-values (suitably interpreted as $\pi$-values) to claim a directional effect: what the analyst provides the client with, is a summary of functionals , i.e. "parcels of information", based on the posterior and an actionability statement that directly addresses the decision problem of the client(Hildreth 1963). If the *sign* functional is used, then according to Eq. 28 the parcels communicated is a sign (direction of effect) and a magnitude attached to the sign (the absolute value of the test statistic in non-Bayesian analysis or a p-value). Importantly such claims can be made, irrespective of whether the $\pi$-value is smaller ("subthreshold") than the threshold of the decision procedure or not.

3. **DON'T use automatic thresholds for the decision problems involving $\pi$-values.** Typical choices for level of significance (e.g. 0.05 or 0.01) in frequentist hypothesis testing are products of the era in which manual computation was the norm, and publication costs limited the size of the tables that could be reproduced and disseminated to scientific workers (Stigler 2008; Kennedy-

Shaffer 2019). The choice of 0.05 corresponds roughly to an estimated effect that is two standard deviations away from the mean of random Gaussian noise: multiplying by two can easily be done without a computer. Neither Fisher, nor Neyman considered the threshold to be fixed and universal, but rather application and domain specific. In the words of the latter (Neyman 1957): "There are weighty arguments against this automatism. In fact, it appears desirable to determine the level of significance in accordance with quite a few circumstances that vary from one problem to the next ". These warnings against the automatism mutatis mutandis carry over to π-values.  It is worth considering the relation between the critical p (frequentist) and π (Bayesian) thresholds. According to Neyman (Neyman 1957): "One is the consequences of committing an error (so-called of the first kind) when the hypothesis tested is in fact true and is rejected. Depending upon the problem, these consequences may be grave and, in such a case, a more stringent level of significance, perhaps 0.001, will be indicated. Another circumstance to be considered when fixing the level of significance is the consequences of failing to reject the hypothesis tested when, in fact, it is false (error of the second kind), and a specified alternative hypothesis is true. true. Also, the probability of this error (or, equivalently, the power of the test) must be taken into account." It thus seems that setting the threshold of significance (the probability of type I error if the null hypothesis were true) of p, involves a balancing act between the gain of getting the state of nature right, vs. the loss of getting it wrong. On the other hand, the value of the threshold for π-values is grounded on a richer decision theoretic framework that considers the opportunity cost of deferring the decision to act, in addition to the gains and losses of the decision itself. As this text is written, the world is going through the COVID19 pandemic, and the need to consider gains, losses, and opportunity costs about statistical decisions e.g., the size of the pediatric vaccine trials or the deployment of therapies that appear to yield a desirable clinical effect but with a not-to-small p-value in trials has never been greater. We proposed Eq. 31 as a quantitative instrument to start the discussion about how to set thresholds for statistical decisions about specific problems and application domains. While thresholds for p-values should also vary, we feel the semiquantitative balancing act implicit in the theory provides less guidance about how to tie their numerical values to downstream applications.

4. **DO use a subthreshold π-value to act as if a directional effect is present.** The only reason we work under a decision analytic framework is in the words of Neyman to take a "calculated risk, to an act of will to behave in the future (perhaps until new experiments are performed) in a particular manner, conforming with the outcome of the experiment"(Neyman 1957). The decision framework of Section 4 is designed to yield an actionability statement in the spirit of Neyman to be made if a subthreshold π-value is obtained, with an expiration date the day of arrival of new data.

5. **DON'T interpret a suprathreshold π-value as evidence for the lack of a directional effect.** The decision analytic framework depicted in Table 4 does not include an action/statement that the effect is zero, only that the effect is not actionable on the existing data. There is a "maligned cult of null-hypothesis significance testing" (Greenland 2019) who "has developed a strong (and unhealthy) focus on the idea that the main aim of the study should be to test null hypothesis"(Greenland et al. 2016) and the proposed framework protects against falling into the trap of the cult. In fact, our proposal can be seen as a return to the historical foundations of statistical significance testing (Edgeworth 1885) . It was Edgeworth who provided the first formal mathematical description of a significance test , in which the observed p-value was used to determine the strength of *particular* body of evidence before a matter is to be taken seriously(Stigler 2008; Kennedy-Shaffer 2019; Gibson 2021), i.e. acted upon. According to Edgeworth it "might be desirable to have a method not only uniform, and as it were impersonal, but also capable of distinguishing the degrees of evidence between reasonable probability and certainty almost mathematical." In the examples analyzed by Edgeworth almost mathematical certainty amounts to modern threshold of 0.005, yet values of 0.12 or even 0.20 provide "some evidence, but much weaker than that which we have considered in previous examples, in favour of a real difference". In our decision framework, failure of the π-value to fall below the threshold only defers the decision to act as if the effect was there. However, the decision to defer further decision is made *conditional* on the threshold for actionability. Changing the threshold only changes the strength of evidence before the scientific finding is to be considered further, but does not necessarily rule out the presence of an effect. Further exploration of the posterior using the effect size can be used to probabilistically answer the question of the presence of a directional effect when a suprathreshold π-value is obtained.

6. **DO interpret the π-value and/or its distance from the threshold as a measure of the strength of evidence for a directional effect.** This immediately follows from the relation between the expected value of the perfect information for the pure and recalibrated analysts. Informally, the smaller the π-value, the less valuable any additional information becomes for ordinal claims about directional effects. The historical connection to Edgeworth's intended use of p-values as distinguishing degrees of evidence should also be apparent.

7. **DO interpret a published p-value as a π-value when the mathematical assumptions for the numerical equivalency of Bayesian and non-Bayesian analyses are verified.** While we worked with the GLM for practical reasons, the technical assumptions for the numerical equivalency apply to a wider class of models. These assumptions can be conveniently grouped to those pertaining to the likelihood and the prior. On the likelihood side we require that the latter be concentrated away from the boundary of the parameter space, that the log-likelihood behaves like a multivariate quadratic locally around its maximum and that the parameter vector is of finite dimension. The key assumption on the likelihood side is that of "quadraticity" of the log-likelihood(Cam and Yang 2000; Geyer 2013) which simultaneously makes the ML estimator tend to a multivariate normal distribution, while allowing one to carry out the integrations for the Bayesian analyses with the Laplace approximation. On the prior side, the major assumption is that of an exact, or approximate locally uniform prior, an *objective* choice that is justified from three different perspectives that all relate to the needs of an analyst working with parameters that assume values in compact subsets of the real hypercube. It is a favorite pastime of both Bayesians and non-Bayesians to look down such uniform priors as relics of the era of Laplace, but we feel that the arguments regarding inference in a finite world, the induction of a uniform prior for linear functionals and the limiting form for a community of investigators who are not to be bothered to elicit an informative prior justify these priors for GLMs. Once these assumptions are made, one can apply the Laplace approximation and derive the numerical equivalence between non-Bayesian and Bayesian approaches. There exist some rather sophisticated presentations and analyses of the consistency of the posterior and convergence of the latter to the ML estimator are available in the literature for the exponential family and models with quadratic log-likelihoods (Locally Asymptotic Normal, Locally Asymptotic Mixed Normal) as considered by Le Cam (Johnson and Ladalla 1979; DiCiccio and Stern 1993; Ghosal et al. 1995; Ghosal 1997). Rather than appealing to these results we carried out the relevant derivations for pedagogical purposes. To our knowledge this material has never been put together in a manner that allows the reader to clearly see that some of the most elegant GLM results in likelihood theory are really nothing more than low precision approximations to Bayesian calculations under non-informative, locally uniform priors. The serious student of asymptotic analysis can go even further and verify that this equivalency will hold even for non-uniform, weakly informative priors such as the Student-t. The relevant mathematical machinery is based on the elegant and compact formulas regarding the higher order terms in the multivariate Laplace approximation , e.g. the implicit differentiation formula 8.3.50 in (Bleistein and Handelsman 2010) , the explicit representation in Theorem 1.2 by Kirwin (Kirwin 2010) and the more statistically focused presentations for the exponential family (Ghosal et al. 1995; Ghosal 1997, 1999, 2000) in particular Theorem 2.1 in(Ghosal 1997). The only requirement is that log-likelihood is sharper than the log-prior, so that the derivatives of the former dominate the derivatives of the latter in the higher order terms of the asymptotic approximation. We feel that not acknowledging, or simply paying lip service to this numerical equivalence(Senn 2001; Meng 2009; Lee et al. 2017) between Bayesian and non-Bayesian analyses under (locally) uniform or weakly informative priors, has done a considerable disservice to researchers who would like to be Bayesians but don't feel comfortable with MCMC, but also to non-Bayesians who are forced to work with the multivariate normal approximation, even when the log-likelihood deviates from the quadratic. The former can and should be able to "lift" ML computations to the Bayesian realm when the relevant assumptions are verified, while the latter should feel non-guilty if they fire off MCMC and use the output to stand for the magic p* formula which is not widely implemented in "off-the-shelf" statistical software. The analyses of the outcomes in the SGLT2i trials provide real world examples to absolve Bayesians who compute as non-Bayesians and vice versa of these sins. These examples also show that if the data presented to models whose log-likelihood behaves like a quadratic, the "asymptotia" will be reached even if for finite sample sizes and thus the prior will not play a role in the inference.

8. **DON'T conflate reproducibility of π-values with the replication of the relevant scientific finding.** The marginal predictive distribution we derived in section 5 for GLMs should make it rather clear that p and π-values are only weakly reproducible in the sense of being tightly clustering around the value obtained in the initial study. In practical terms while an inferential meaning is to be attached to a tail area probability computed from a single study, using a sequence of thresholded π-values from replicate studies to draw conclusions about the replication of scientific findings is a fool's game. A similar comment applies to the p-values under the non-Bayesian predictive framework briefly discussed in Section 5. The arguments in that section should address a major implicit technical objection against examining the reproducibility probability (Senn 2001, 2002; Greenland et al. 2016; Greenland 2019) of p-values, namely that the test statistic obtained in the initial study is also exactly in the subsequent studies. Eq. 37 provides a relatively simple formula to compute the RPD for p- and π-values in the setting of exact translation of treatment effects and exact replication of the initial study design. On the other hand, Eq. 36 can be used to explore more general setups that involve either an imperfect translation (e.g., bias) and under-resourced replication (e.g., reduced sample size of the replicate study) while accounting for violations of "quadraticity". Such simulations will invariably show that the RPD of the p- and π-values will be substantially more spread out in the presence of bias and/or compromises on the resources used to replicate study findings. These considerations make it even more compelling to do away with the tendency to pit significant against non-significant p-values obtained in replicate studies to gauge replicability of findings. Most studies in biomedical and psychological applications are barely powered to detect their assumed effect, so p and π-values will bounce quite a bit to the left and right of the significance threshold. It is thus best if one does not over-react to such apparent inconsistencies because they will be quite common in practice (Senn 2002).

9. **DON'T use the sequence of publicly available π-values to gauge the totality of evidence in a given field.** This point is somewhat related to the previous but is not identical. Practitioners in many fields, will often reduce voluminous amounts of data to summaries of the proportion of studies exploring a finding that yielded a significant π-value and use this proportion to infer the replicability of the finding. Examples of such statements that one may casually here include: "drug X works, because all studies found a statistically significant reduction in outcome Y", or "the effect of factor W is rather weak, because only half of the studies in the field have documented a statistically significant effect". This approach effectively treats the reported π-values as quantities that can be used to conduct a meta-inference in an unbiased manner. If the latter assumption was verified, then one could in principle use the absolute values of π-values and fit them to a lognormal distribution to Eq. 37 attempt to synthesize π-values as is done in the Fisher's method for a combined probability test (Mosteller and Fisher 1948) or Stouffer's z-score method(Whitlock 2005) used in meta-analysis. However, the published p-values about a topic are likely to be severely biased, because of a the "file-drawer" problem: if experiment 1 is significant, we replicate in experiment 2 and if significant again then all is well, but if not, we have a potential case of non-reproducibility, that we address statistically be appealing to the distribution of the p-values. But what happens in the real world, is that if experiment 1 is non-significant, then the experiment will be filed in the drawer and never published. If more than one groups work competitively in a given area, only those who get the subthreshold π-value are likely to see their result published before they move on to greener pastures. Furthermore, the chances of getting additional resources to replicate a suprathreshold π-value is low, and the large sequence of π-values that one hopes to utilize in this approach to meta-analysis may never be generated, only the subthreshold ones will be retained in the publication record. This will create an even more skewed, hyper-optimistic assessment of the reproducibility of the scientific finding. Hence, (non-)reproducibility is by definition directionally biased to include only the sub-threshold π-values, while also being extremely path-dependent based on the early history (who published what and when) of a finding in a given field. Formal meta-analytic techniques that explore the possibility of selection effects, e.g. funnel plots(Sterne et al. 2011) of treatment effects and their precisions, prior to evidence synthesis are more likely to be informative for assessing reproducibility.

10. **DON'T expect likelihood ratio tests to rescue testing for dichotomous existence ("lack of an effect") claims when the mathematical assumptions that render Bayesian and non-Bayesian analyses numerically equivalent are verified.** Dichotomous statistical tests for existence of effects appear to be necessary from the perspective of methodological falsificationism (see (Tunç et al. 2021) for an overview and defense of this perspective). This thesis which encapsulates the idea of a critical experiment ("experimentum crucis"), one that is decisively capable of determining whether a particular theory is superior to all other possible hypotheses, is unfortunately not a tenable one if the outcome of the crucial experiment is to be judged based on a thresholded p-value. In the latter case, the corresponding statistic would be judged against the value of a $\chi^2$ with degrees of freedom equal to the dimensionality of the hypothesis tested, i.e., the likelihood ratio. It would seem an uncontroversial position that such a test could be used to decide on the absence of effect if the hypothesis tested is one in which one or more components of the treatment effect vector are zero. However, someone working entirely within the framework of Section 4, in which treatment effects are always present, could arrive at precisely the same value of the likelihood ratio test by replacing the sign function with another discretizing function which maps all points in the posterior within the squared Mahalanobis distance $(\widehat{\boldsymbol{\beta}} - \boldsymbol{\beta_0})^T \boldsymbol{\Sigma}^{-1} (\widehat{\boldsymbol{\beta}} - \boldsymbol{\beta_0})$ from the hypothesis tested ($\boldsymbol{\beta} = \boldsymbol{\beta_0}$) to zero, and one otherwise. Under the square loss function the posterior expectation of this function is the complement of the integral of the multivariate normal function over an ellipsoid region, which coincides with the tail area (p-value) of a $\chi^2$ distribution with the appropriate degrees of freedom. The decision problem for the pure analyst/client and recalibrated analyst can similarly be updated to reflect the discretizing map introduced. Informally, this map discretizes the continuous values of the parameter to zero ("less distant from the hypothesis than what was observed") and one ("further away from the point hypothesis than what was observed") and the decision problem is transformed to make a statement about the magnitude of this derived variable. These considerations clearly show that one person's test for the existence of an effect, is another person's statement about the discretized magnitude of an effect. Unless the Mahalanobis distance used in the discretization is infinite (in which case the multivariate normal posterior would coincide with a multivariate delta Dirac function) the interpretation of absence of effect statements based on likelihood ratio tests with suprathreshold p-values is thus contestable. In such situations it is more accurate to state that the analysis failed to detect a signal within the resolution power(=discretization) of the experiment/analysis used to analyze the data.

So how could one apply the list of do's and don'ts in practice when reporting a scientific study, while considering replicability? A straightforward solution would involve the following elements:

1. **Be a likelihoodist in the results section of your paper or report but a Bayesian in the discussion.** Scientific papers are effectively a way to communicate data AND inferences to external audiences. The numerical equivalency of Bayesian and frequentist analysis, implies that any summary table constructed for a non-Bayesian GLM analysis is identical to one that would have been constructed for a Bayesian analysis of the same data if the relevant assumptions of quadraticity and likelihoods away from the boundary are verified. Presenting results from a likelihood perspective but discussing the same results from a Bayesian vantage point, frees the user from various mathematically irrelevant and thus irrational prohibitions of interpretation. Working entirely from within a Bayesian framework for example p-values (which are interpreted as π-values) are non-controversial, but relatively uninteresting posterior summaries of treatment effects based on crude discretizations, which can be discussed by applying the list of do's and don'ts.

2. **Switch to formal meta-analysis when discussing the present finding in the context of prior findings rather than compare p-values.** Given the poor numerical reproducibility of p-values, the tendency to compare p-values (and π-values!) against previously observed p-values relative to the threshold of significance is simply "feeding the strong cognitive bias of dichotomania: the compulsion to replace quantities with dichotomies ("black-and-white thinking"), even when such dichotomization is unnecessary and misleading for inference" and "blinds the user to actual data patterns, thus invalidating conclusions and sometimes rendering them ludicrous" (Greenland 2017). A simple meta-analysis may offer a much more productive way to see what the present study adds to the preexisting knowledge. In that regard, acting as a likelihoodist in the results section, an objective Bayesian in the first part of the discussion and as a meta-analytic Bayesian in the middle part of the discussion clearly separates the task of communicating what the study found, what are the inferences that can be drawn in isolation from the pre-existing knowledge and what the analyses adds to the totality of evidence.

3. **Pre-emptively address concerns about the replication of key findings of the present study using the predictive framework.** This can be done rather effortlessly for exact replication and translation through the approximate formula of Eq. 35. The argument will be valid only for this rather limited situation; it will be particularly hard to justify this calculation except in the context of highly regulated scientific activities with rather rigorous quality control and audit mechanisms in place. Examples of such research includes high energy physics and many (but not all) drug and device registrational trials. In these domains application of Eq.35 leads to simple heuristics such as the 5-sigma rule, or even drug approvals based on the result of a single trial if the p-value is appropriately small. Most research is unfortunately not well resourced to fulfill these requirements, and we feel that an exploration of reproducibility that considers imperfect translation and replication using the Monte Carlo schema of Equation 36 should be preferred over the formulas that assume exact replication and translation. An appropriate place to include such analyses is the late part in the discussion section when limitations are considered. Limitations of translation ("generalizability"), numerical approximation ("quadraticity") and replication ("small samples") can be explored rather rigorously via a simulation framework and presented in the manuscript.

4. **Include verifications of quadraticity of the log-likelihood in the methods section and the supplementary results.** This is a numerical assumption that is seldom verified in practice, even though it is implicit in non-Bayesian GLM calculations and is required for the interpretation of the latter as approximate Bayesian computations. Quadraticity can be used to operationalize the notion of having enough data for the inferential likelihood task at hand(Geyer 2013), and we have seen verification of this assumption can render the choice of prior irrelevant for posterior inference. Tests for the quadratic behavior of log-likelihood based on the curvature of the log-likelihood around the maximum were introduced for non-linear least squares(Bates and Watts 1980) as an extension of the nonlinearity measures of Beale(Beale 1960) and the bias expressions for nonlinear regression introduced by Box(Box 1971). These measures are applicable to GLMs (Kass and Slate 1994) and can be computed from the IWLS output if one would like to supplement the direct visualization of the likelihood. Failure for the quadraticity to be verified will in general invalidate likelihoodist results, unless one switches to methods based on the p-formula. Since numerical implementations of the latter are not widely available, but the asymptotically equivalent MCMC with the uniform prior is, switching to Bayesian numerical methods may be the only way for a non-Bayesian to conduct their analyses in a numerically correct fashion. However, once one switches to a Bayesian fitting procedure it would be advisable to try multiple alternative priors to get an idea of whether the inference is prior dominated vs. data dominated(Kass and Wasserman 1996). The analyses of the rare side effects in the SGLT2i trials provide examples of "how-to-do-this", with the CREDENCE illustrating a successful application and DAPACKD providing evidence that one should sometimes not even conduct this analysis, until further data from other studies become available.

In summary, we established a numerical equivalence between Bayesian and non – Bayesian analysis for GLMs and provide some suggestions to leverage this equivalence to improve reporting of analyses while moving away from unsound cognitive practices (e.g. p-value dichotomania) when discussing and inferring reproducibility of study findings. We hope that the material presented in the short monograph meaningfully contribute to the discussion about the role of p-values and provide a roadway towards the full adoption of the Bayesian perspective when appraising evidence based on GLMs.

# Appendix : Programming Reproducibility Analyses in R

In this Appendix we will provide software code in the R analysis for the analysis of reproducibility in Section using the SGLT2i clinical trial data (Section 6) to illustrate the relevant program snippets. A familiarity with the R programming language is assumed. A working installation of the STAN program, a versatile facility for fitting Bayesian models with Hamiltonian Monte Carlo and the No U Turn Sampler (NUTS) should also be available in your system, if one is interested in fitting GLMs with MCMC.

## A1. Reproducibility Calculations

*Setting up the environment in R*

We first load the libraries that are needed to fit the models under various priors and post-process the results. These libraries should be installed before their first use:

```r
library(arm)        ## for Expectation Maximization Bayesian fitting
library(rstanarm)   ## for fitting using the No U Turn Sampler (NUTS)
library(viridis)    ## for colorblind friendly palette during plotting
library(pracma)     ## for the complementary inverse error function
library(mclust)     ## clusters MCMC samples and is used to compute p-values
                    ## through a model-based integration approach
```

*Enter and preprocess the data for these analyses*

We now enter the data found in Table 5 in R, and select the primary outcome in CREDENCE for illustration purposes (the same code can be used for the analyses of the other three outcomes, but to conserve space only the primary outcome in CREDENCE will be considered here. In the dataframe below, the number of events in the SGLT2i trials for the active treatment (either canagliflozin or dapagliflozin ) and the Placebo (PBO) and the total duration of follow up in person years is given by the PYTrt, PYPBO respectively.

```r
## provide the data
data<-data.frame(
    Outcome=c("Composite Cardio Renal", "DKA",
              "Composite Cardio Renal", "DKA"),
Drug=c("Canagliflozin","Canagliflozin","Dapagliflozin","Dapagliflozin"),
```

```
           Study=c("CREDENCE","CREDENCE","DAPA-CKD","DAPA-CKD"),
           NGSLT2=c(2202, 2200, 2152, 2149),
           NPBO=c(2199, 2197, 2152, 2149),
           NEvTrt=c(245,   11, 197,    0),
           NEvPBO=c(340,    1, 312,    2),
           PYTrt=c(5671.296, 5000.000, 4282.609, NA),
           PYPBO=c(5555.556, 5005.000, 4160.000, NA)
           )

## select the primary (Composite Cardio Renal) outcome in CREDENCE for
## illustration
datExample<-subset(data, Outcome=="Composite Cardio Renal" &
Study=="CREDENCE")
## this will yield the design matrix for the initial study, whose replication
## potential we will assess. Named xinit to identify with the material
## and mathematical notation adopted in Section 5
xinit<-data.frame(Count=c(datExample$NEvTrt, datExample$NEvPBO),Treat=c(1,0),
                  FUT=c(datExample$PYTrt, datExample$PYPBO)/1000)
```

*Fitting of the GLM using Bayesian and likelihoodist methods*

We used a Poisson GLM to analyze the outcomes (number of events, NEvTrt & NEvPBO) in the SGLT2i trials are undertaken, using the person years (PYTrt, PYPBO) as offsets. The variable Treat gives the treatment assignment (Treat = 1 , assignment to the SGLT2i arm in the clinical trial, Treat = 0 in the Placebo, PBO arm). We fit and extract the z-values from the likelihoodist fit to this Poisson GLM via the commands:

```
## fit the GLM via Maximum Likelihood and extract the z values which are used
## to compute the p-values
finit<-glm(Count~Treat+offset(log(FUT)),data=xinit,family="poisson")
z<-abs(finit$coefficients/sqrt(diag(vcov(finit)))) ## to calc p formulas
```

We fit the Student-t prior using the EM approach by the following command (the function *bayesglm* is available from the package *arm*):

```
## EM fit
finit.Bayes.EM<-bayesglm(Count~Treat+offset(log(FUT)),data=xinit,
                         family="poisson",  prior.df=2.5,scale=1)
```

We can also fit the Poisson GLM via MCMC using the NUTS sampler. Flat and Student-t priors require separate fittings. For each prior we fit 4 independent Markov chains and obtain 20,000 samples. While we will randomly subsample the chains during the assessment of reproducibility, a large number of iterations ensures that the Monte Carlo error for the estimates remains low and we can compare coefficients to three or four significant digits.

```r
## NUTS fit with a flat prior - set seed for reproducible results
finit.Bayes.flat<-stan_glm(Count~Treat, offset = log(FUT),
                    data=xinit, family=poisson(link="log"),
                    prior = NULL, chains=4,cores=4,seed=4321,
                    prior_intercept = NULL, iter=20000)

## NUTS fit with Student t priors- set seed for reproducible results
finit.Bayes<-stan_glm(Count~Treat, offset = log(FUT),
                  data=xinit, family=poisson(link="log"),
                  prior = student_t(df = 1, 0, 1), chains=4,seed=1234,
                  prior_intercept = student_t(df = 2.5, 0, 1),iter=20000)
```

To obtain the results by each method we simply call the relevant summary method in R:

```r
summary(finit)                          ## maximum likelihood results

Call:
glm(formula = Count ~ Treat + offset(log(FUT)), family = "poisson",
    data = xinit)

Deviance Residuals:
[1]  0  0

Coefficients:
            Estimate Std. Error z value Pr(>|z|)
(Intercept)  4.11415    0.05423  75.861  < 2e-16 ***
Treat       -0.34831    0.08380  -4.156 3.23e-05 ***
---
Signif. codes:  0 '***' 0.001 '**' 0.01 '*' 0.05 '.' 0.1 ' ' 1

(Dispersion parameter for poisson family taken to be 1)

    Null deviance: 1.7517e+01  on 1  degrees of freedom
Residual deviance: 2.7756e-14  on 0  degrees of freedom
AIC: 19.007

Number of Fisher Scoring iterations: 2

summary(finit.Bayes.EM)                 ## EM Bayesian results

Call:
```

```
bayesglm(formula = Count ~ Treat + offset(log(FUT)), family = "poisson",
    data = xinit, prior.df = 2.5, scaled = 1)

Deviance Residuals:
[1]  0  0

Coefficients:
            Estimate Std. Error z value Pr(>|z|)
(Intercept)  4.11382    0.05422  75.868  < 2e-16 ***
Treat       -0.34780    0.08374  -4.153 3.28e-05 ***
---
Signif. codes:  0 '***' 0.001 '**' 0.01 '*' 0.05 '.' 0.1 ' ' 1

(Dispersion parameter for poisson family taken to be 1)

    Null deviance: 1.7517e+01  on 1  degrees of freedom
Residual deviance: 4.3978e-05  on 0  degrees of freedom
AIC: 19.007

Number of Fisher Scoring iterations: 4

summary(finit.Bayes.flat,digits=4)   ## NUTS - flat prior results

Model Info:

 function:     stan_glm
 family:       poisson [log]
 formula:      Count ~ Treat
 algorithm:    sampling
 priors:       see help('prior_summary')
 sample:       40000 (posterior sample size)
 observations: 2
 predictors:   2

Estimates:
                 mean      sd      2.5%      25%       50%       75%       97.5%
(Intercept)     4.1125   0.0541   4.0051    4.0762    4.1132    4.1496    4.2168
Treat          -0.3485   0.0832  -0.5113   -0.4049   -0.3485   -0.2922   -0.1858
mean_PPD      292.5517  17.1682 260.0000  281.0000  292.5000  304.0000  327.0000
log-posterior  -8.5040   0.9897 -11.1636   -8.8986   -8.2028   -7.7934   -7.5297

Diagnostics:
                mcse   Rhat   n_eff
(Intercept)    0.0003 0.9999 28495
Treat          0.0005 1.0001 26253
mean_PPD       0.0962 1.0000 31856
log-posterior  0.0077 1.0001 16609

For each parameter, mcse is Monte Carlo standard error, n_eff is a crude
measure of effective sample size, and Rhat is the potential scale reduction
factor on split chains (at convergence Rhat=1).

summary(finit.Bayes,digits=4)        ## NUTS - Student t prior results

Model Info:

 function:     stan_glm
```

```
family:       poisson [log]
formula:      Count ~ Treat
algorithm:    sampling
priors:       see help('prior_summary')
sample:       40000 (posterior sample size)
observations: 2
predictors:   2

Estimates:
                mean      sd      2.5%      25%       50%       75%       97.5%
(Intercept)    4.1097   0.0543   4.0023   4.0733    4.1103    4.1470    4.2135
Treat         -0.3439   0.0840  -0.5092  -0.4001   -0.3442   -0.2872   -0.1797
mean_PPD     292.1201  17.1153 259.5000 280.5000  292.0000  303.5000  326.5000
log-posterior -13.9318  0.9958 -16.6100 -14.3169  -13.6291  -13.2240  -12.9579

Diagnostics:
              mcse   Rhat   n_eff
(Intercept)  0.0003 1.0001  27925
Treat        0.0005 1.0000  24755
mean_PPD     0.0967 1.0000  31300
log-posterior 0.0075 1.0002 17582
```

For each parameter, mcse is Monte Carlo standard error, n_eff is a crude measure of effective sample size, and Rhat is the potential scale reduction factor on split chains (at convergence Rhat=1).

π-values for the Bayesian models are best calculated by smoothing the output of the MCMC, rather than using the samples directly in Eq. 28. To see why, note that the floor of a π-value estimated from the empirical distribution of the MCMC sample is 2/(Number of Samples in the MCMC). It would thus take >1,000, 1,0000, 100,000 samples to compute the π-value to 3, 4, 5 significant digits, while an estimate derived from a smoothing procedure may have greater precision. A smoothing-based estimate for the p-value may be obtained by fitting a low dimensional Gaussian mixture to the output of the chain, and then using the observed mixing proportions and z-values of the components of the mixture to compute the p-value algebraically. The following code will fit a univariate mixture of Gaussians to an MCMC sample and carry out the relevant computations; we fit each parameter independently, because in high dimensions there may not be enough samples in the MCMC output to fit a high dimensional mixture and capture the behavior at the tails, because of the curse of dimensionality. Formally, we take advantage of the fact that a joint MCMC simulation for multiple parameters, also samples from the marginal distribution for each parameter. Hence, we can substantially simplify the calculation of the π-values by working marginally, i.e. univariately.

```r
## approximate the tail area from a MCMC posterior using a Gaussian mixture
pval.MCLUST<-function(mcmc.df)
{
  kde.df<-densityMclust(mcmc.df, G=1:5,modelnames="V", verbose=F)
  G<-kde.df$G
  pro<-kde.df$parameters$pro
  mn<-kde.df$parameters$mean
  sd<-sqrt(kde.df$parameters$variance$sigmasq)
  z<-abs(mn/sd)
  sum((2*pnorm(-z))*pro)
}
```

We first cast the MCMC samples into a matrix, before applying the aforementioned function column-wise:

```r
## flat prior
bayes.flat<-as.matrix(finit.Bayes.flat)         ## get samples into matrix
p.bayes.flat<-apply(bayes.flat,2,pval.MCLUST)   ## calculate p-values

## Student - t prior
bayes<-as.matrix(finit.Bayes)                   ## get samples into matrix
p.bayes<-apply(bayes,2,pval.MCLUST)             ## calculate p-values
```

Contrast to the direct calculation of π-values from the empirical distribution:

```r
## empirical p-values
## flat prior
p.bayes.flat.emp<-apply(bayes.flat,2,function(x) 2*min(mean(x>=0),mean(x<0)))
## Student - t prior
p.bayes.emp<-apply(bayes,2,function(x) 2*min(mean(x>=0),mean(x<0)))
```

As there are only 80,000 samples (4 chains with 20,000 samples each) in the MCMC chain, the smallest non-zero π-value we can obtain with the empirical approach is 2/80,000 = 5 10$^{-5}$. However, the actual π-value for the intercept is smaller than this value, so the empirical calculation simply underflows to zero. On the other hand, the π-value is larger than the floor of the calculation and the empirical π-value

```r
rbind(p.bayes.flat.emp, p.bayes.flat)
                 (Intercept)        Treat
```

```
p.bayes.flat.emp 0.000000e+00 0.001050000
p.bayes.flat     1.751247e-05 0.001895736

rbind(p.bayes.emp, p.bayes)
             (Intercept)      Treat
p.bayes.emp 0.000000e+00 0.00860000
p.bayes     2.295234e-06 0.01161849
```

To implement the marginal predictive framework for reproducibility (Eq. 36), we must simulate from the posterior obtained under the flat prior, and use the sampled coefficients to generate a new dataset that is then used for estimation. The sampled coefficients from the posterior of the initial sample may be modified to emulate an imperfect translation and the size of the synthetic generated may be different than the initial sample to emulate imperfect replication. Note that if one were to analyze the replicate datasets with Bayesian methods, one would have to invest a considerable number of computational resources for the relevant MCMC sampling. To reduce the time required for these analyses, one must cut down on the amount of MCMC sampling for each replicate dataset, which in turn will reduce the precision of the π-value calculations, unless a smoothing technique is used to post-process the samples before computing the π-value.

```
## simulate Poisson counts
## note that changing the FUT (Follow Up Time) from the
## value observed in the initial study could be used to
## model imperfect translation, e.g. a study with smaller
## follow up time in each study arm
Count<-rpois(2,exp(lp + log(xinit$FUT)))
dsim<-as.data.frame(cbind(Treat=c(0,1),Count=Count,FUT=xinit$FUT))
## obtain Maximum Likelihood estimates
f1<-glm(Count ~ Treat + offset(log(FUT)),data=dsim, family="poisson")
coef<-mvrnorm(1,coef(f1),vcov(f1))
z<-abs(coef(f1)/sqrt(diag(vcov(f1))))

## fit Bayesian now
## NUTS - flat
finit.Bayes.flat<-stan_glm(Count~Treat, offset = log(FUT),
                    data=dsim, family=poisson(link="log"),
                    prior = NULL, chains=4,
                    prior_intercept = NULL, iter=2000)
bayes.flat<-as.matrix(finit.Bayes.flat) ## get samples
p.bayes.flat<-apply(bayes.flat,2,pval.MCLUST)
b.bayes.flat<-bayes.flat[sample(nrow(bayes.flat),1),]
## NUTS - Student t
finit.Bayes<-stan_glm(Count~Treat, offset = log(FUT),
```

```
                       data=dsim, family=poisson(link="log"),
                       prior = student_t(df = 1, 0, 1), chains=4,
                       prior_intercept = student_t(df = 2.5, 0,
                       1),iter=2000)
  bayes<-as.matrix(finit.Bayes) ## get samples
  p.bayes<-apply(bayes,2,pval.MCLUST)
  b.bayes<-bayes[sample(nrow(bayes),1),]
  ret<-c(2*pnorm(-z),p.bayes.flat,p.bayes,coef,b.bayes.flat,b.bayes)
  ## values returned are as follows: likelihoodist p-value for the intercept,
  ## likelihoodist p-value for the treatment effect, p-value for the
  ## intercept under a flat prior, p-value for the treatment effect under the
  ## flat prior p-value for the intercept under a Student-t prior,
  ## p-value for the treatment effect under a Student-t prior,
  ## likelihoodist intercept, likelihoodist treatment effect,
  ## intercept under the flat prior, treatment effect under the flat prior
  ## intercept under the Student-t prior,
  ## treatment effect under the Student-t prior
  names(ret)<-c("pB0","pB1","pB0.flat","pB1.flat","pB0.St","pB1.St",
                "B0","B1","B0.flat","B1.flat","B0.St","B1.St")
  ret
}
```

During the repeated application of the NUTS sampler in the replicate datasets, it is possible that the algorithm will fail to converge for one or more such datasets. We used R's exception handling facilities, to skip over datasets in which the MCMC failed to converge, by wrapping a function that catches such errors over the bona fide simulation function:

```
## wrapper to handle exceptions by STAN
faulttolerantsim<-function(b0,b1)
{
  tryCatch({ret <- sim(b0,b1);}, error = function(e) {ret <<- rep(NA,12)});
  ret
}
```

A limited number (e.g., 5000) number of replicate datasets are generated for the analysis of reproducibility. The following code will generate these datasets by sub-sampling the MCMC chain obtained by analyzing the data in the original dataset under the flat prior:

```
set.seed(12345)  ## set seed for reproducible results
Nsim<-5000       ## number of Monte Carlo samples to obtain
B<-bayes.flat[sample(nrow(bayes.flat),Nsim),]
colnames(B)<-c("b0","b1")
Best<-t(mapply(faulttolerantsim,B[,"b0"],B[,"b1"]))
```

The code for the analytical approximations to the π-value probability density (Eq.37) and its cumulative density function in logarithmic base 10 scale given the value of the statistic obtained in the initial dataset is the following:

```r
## approx. p value CDF
## T is the observed statistic in the initial sample,
## log10p the value of the replicate p-value we want to
## compute the CDF or the PDF at
pvalCDF<-function(log10p,T)
{
  c<-qnorm((10^log10p)/2)
  pnorm((T + c)/sqrt(2))+pnorm((-T + c)/sqrt(2))
}

## approx. p value PDF
pvalPDF<-function(log10p,T)
{
  c<-qnorm((10^log10p)/2)
  Jac<-abs(-2.0^(-0.5+log10p) * 5^(log10p) *sqrt(pi) *
log(10.0)*exp(erfcinv(10^log10p)^2))
  pdf<-(dnorm(-T+c,0,sqrt(2.0))+dnorm(T+c,0,sqrt(2.0)))
  pdf*Jac

}
```

## A2. Plotting the calculations

Contrasting the coefficient estimates obtained in the initial dataset by the different approaches (Gaussian approximation which coincides with the ML method, Student-t priors fit by EM, or the NUTS under the two different priors is done graphically to detect deviations from normality. The following code snippet can accomplish that:

```r
col=viridis(4) ## obtain a palette of four colour-blind friendly colors
main="CREDENCE -1o" ## overall plot label
x<-seq(quantile(Best[,"B1"],probs=0.01),quantile(Best[,"B1"],probs=0.99),
       length.out=100)
## Student-t prior,  NUTS
plot(density(bayes[,2]),col=col[2],ylab="Density",main="",xlab="Coefficient",
     xlim=c(-1.0,0.1),ylim=c(0,5))
## Gaussian approximation (from ML fit)
lines(x,dnorm(x,finit$coefficients[2],sqrt(diag(vcov(finit)))[2]),type="l",
      ylab="Density",main="Intercept",xlab="Coefficient",col=col[1])
## Flat prior NUTS
lines(density(bayes.flat[,2]),col=col[3])
```

```
## Student-p prior , EM algorithm
## some programming needed to get the coefficients out of the object
em.coef<-coefficients(finit.Bayes.EM)
em.cov<-summary(finit.Bayes.EM)$cov.scaled ## covariance matrix
em.se<-sqrt(diag(em.cov)) ## standard errors for intercept/treatment effect
lines(x,dnorm(x,em.coef[2],em.se[2]),col=col[4])
legend(-0.6,dnorm(finit$coefficients[2],
                  finit$coefficients[2],sqrt(diag(vcov(finit)))[2]),
       col=col[c(1,2,4,3)],legend=c("Gaussian approx","St(2.5,1) prior",
                                    "St(2.5,1) prior - EM",
                                    "Flat
prior"),bty="n",xjust=1,lty=1,cex=0.8)
```

which yields the following figure as output:

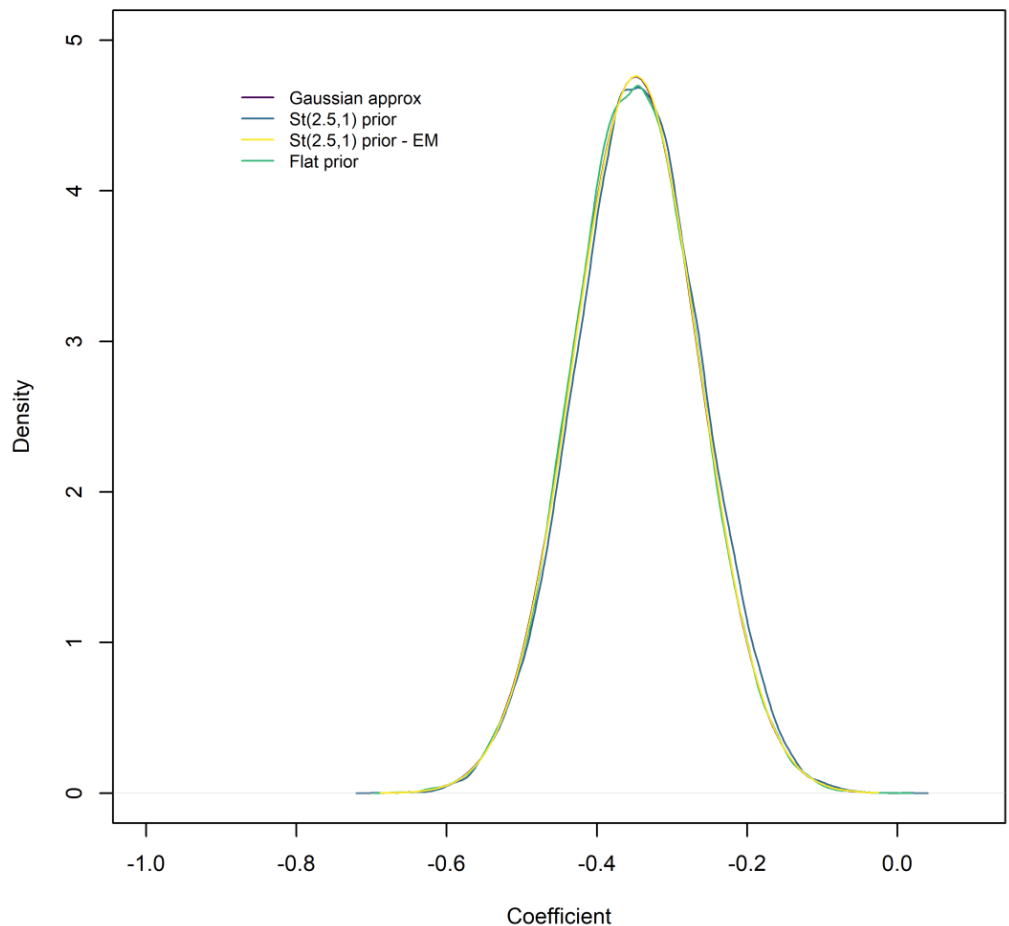

The following code summarizes graphically the analyses of the replicate datasets by ML, as well as the the MCMC estimates under the flat and Student-t priors.

```r
hist.algo<-"fd"   ## algorithm to use for the histogram evaluation
par(mfcol=c(3,1))
x<-seq(quantile(Best[,"B1"],probs=0.01),quantile(Best[,"B1"],probs=0.99),
       length.out=100)
x<-c(x,seq(finit$coefficients[2]-5*sqrt(3)*sqrt(diag(vcov(finit)))[2],
           finit$coefficients[2]+5*sqrt(3)*sqrt(diag(vcov(finit)))[2],
           length.out=100))
x<-sort(x)
## Distribution of coefficients obtain by ML estimation of replicate datasets
hist(Best[,"B1"],freq = F, breaks = hist.algo, add=F, main=main,
xlab="Coefficient",
     ylab="Density", xlim=range(x),ylim=c(0,3))
## approximate predictive based on the Gaussian approximation Eq.35
points(x,dnorm(x,finit$coefficients[2],sqrt(3)*sqrt(diag(vcov(finit)))[2]),
       col=col[2],type="l",main="SGLT2i",xlab="Coefficient",ylab="Density")
## MCMC estimate under the Student-t prior
points(density(Best[,"B1.St"]),type="l", col=col[1])
## MCMC estimate under the flat prior
points(density(Best[,"B1.flat"]),type="l", col=col[3])
legend(-0.6,2.5,
       col=col[c(2,1,3)],legend=c("Approx Predictive","St(2.5,1) prior",
                                  "Flat Prior"),bty="n",xjust=1,lty=1)

x<-seq(min(log10(Best[,"pB1"])),max(log10(Best[,"pB1"])),length.out=100)
## p-values from ML estimation of the replicate datasets
hist(-log10(Best[,"pB1"]),freq = F, breaks = hist.algo, main=main, xlab="-
log10 p")
## approximate predictive based on the Gaussian approximation Eq.35
points(-x, pvalPDF(x, z[2]),col=col[2],type="l", lwd=2)
## MCMC estimate of p-values under the Student-t prior (schema of Eq.36)
points(density(-log10(Best[,"pB1.St"])),col=col[1],type="l")
## MCMC estimate of p-values under the flat prior (schema of Eq. 36)
points(density(-log10(Best[,"pB1.flat"])),col=col[3],type="l")
abline(v=-log10(0.05),col="black", lwd=2,lty=2)
legend(14,.16,
       col=col[c(2,1,3)],legend=c("Approx Predictive","St(2.5,1) prior",
                                  "Flat Prior"),bty="n",xjust=1,lty=1)

## replicate p-value cumulative density function
x<-seq(min(log10(Best[,"pB1"])),max(log10(Best[,"pB1"])),length.out=100)
## p-values from ML estimation of the replicate datasets
plot(ecdf(-log10(Best[,"pB1"])),main=main, xlab="-log10 p",cex=0.1)
abline(v=-log10(0.05),col=1, lwd=2,lty=2)
## approximate predictive based on the Gaussian approximation Eq.35
points(-x,1-pvalCDF(x, z[2]),col=col[1],type="l", lwd=2)
## MCMC estimate of p-values under the Student-t prior (schema of Eq.36)
dummy<-ecdf(-log10(Best[,"pB1.St"]))
points(-x, dummy(-x),col=col[1],type="l")
## MCMC estimate of p-values under the flat prior (schema of Eq. 36)
dummy<-ecdf(-log10(Best[,"pB1.flat"]))
points(-x, dummy(-x),col=col[3],type="l")
legend(15,0.6,
       col=c(col[c(2,1,3)],1),
       legend=c("Approx Predictive","St(2.5,1) prior", "Flat Prior",
       "Replicates (Empirical)"),bty="n",xjust=1,lty=1)
```

The code will generate three plots for the coefficients, the probability density function of the replicate π-values and their cumulative density function. The output for the primary outcome in CREDENCE is shown below:

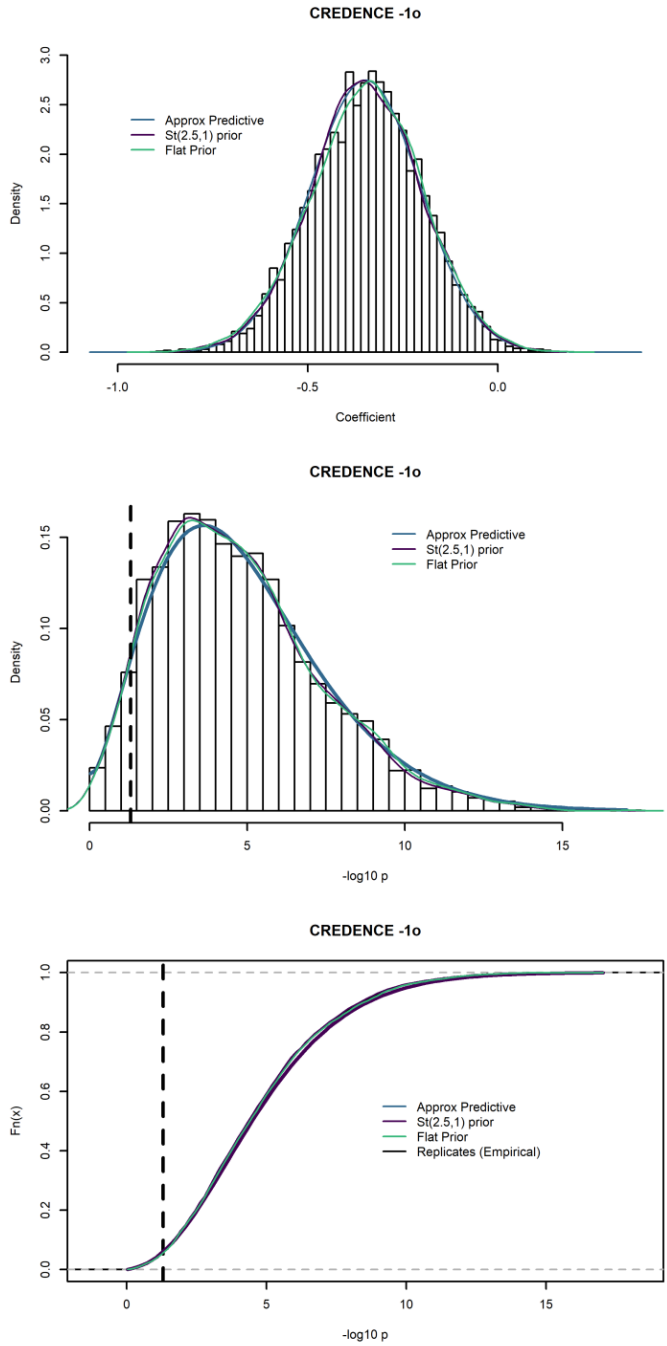

# Subject Index

**The index will be generated after the text has been finalized, i.e. after external and internal review.**